\documentstyle[12pt,aps]{revtex}
\catcode`@=11
\begin{document}
%\draft
\*
\eject
 
\hfill{CEBAF-TH-94-14}\par

\bigskip\bigskip

%\begin{title}
\centerline{\bf Semileptonic Meson Decays in the Quark Model:  An Update}
%\\
%\end{title} 

\bigskip\bigskip\bigskip\bigskip

\centerline{Daryl Scora}

\bigskip
\centerline{\it Department of Applied Mathematics}
\centerline{\it York University, 4700 Keele Street}
\centerline{\it North York, Ontario  M3J 1P3}

\bigskip\bigskip

\centerline{Nathan Isgur}

\bigskip
\centerline{\it Continuous Electron Beam Accelerator Facility}
\centerline{\it 12000 Jefferson Avenue, Newport News, Virginia 23606}

\bigskip\bigskip\bigskip\bigskip

\begin{abstract}
We present the predictions of ISGW2, an update of the ISGW quark model 
for semileptonic meson decays.  The updated model incorporates a number 
of features which should make it more reliable, including the constraints 
imposed by Heavy Quark Symmetry, hyperfine distortions of wavefunctions, 
and form factors with more realistic high recoil behaviors.
\end{abstract}

\newpage

\section{\bf Overview}

\bigskip
  It has been nearly ten years since the ISGW model [1] was introduced [2,3] 
so it is not surprising that the heavy quark semileptonic landscape now 
looks very different.  At that time, for both theoretical and experimental
reasons, inclusive decays were the main focus of attention, and the ISGW 
model, which studied exclusive decays and approximated the inclusive 
semileptonic spectra by summing over resonant channels, was considered 
quite eccentric.  Today, improvements in both theory and experiment have 
made exclusive semileptonic decays a main focus of attention.  Such 
decays seem very likely to provide the most accurate determinations of 
the weak mixing angles $V_{cb}$ and $V_{ub}$. They also provide 
excellent probes 
of hadronic structure via precision tests of Heavy Quark Symmetry (HQS) [4-7].

  The ISGW model was in many respects a stepping-stone to Heavy Quark 
Symmetry: it is a model which respects the symmetry in the heavy quark 
limit near zero recoil.  It also played a role in the discussion of 
the reliability of the free quark decay model (and its derivatives) 
for the endpoint region in $b \rightarrow u$ semileptonic decay.  Indeed, the model
had its origin in that discussion, and was {\it designed} to provide the minimum
reasonable prediction for the decay rate in this region for a fixed $V_{ub}$.
In this paper we present an updated version of ISGW, which (with the 
permission of the ISGW authors) we call ISGW2 to emphasize that it is {\it not} 
a new model but rather an improved version of an old one [8].  The 
new features are described in detail in Section III, but briefly 
they are:
\medskip

\begin{enumerate}
\item Heavy Quark Symmetry constraints on the relations between form 
factors away from zero recoil are respected,

\item Heavy Quark Symmetry constraints on the slopes of form factors 
near zero recoil are built in [9],

\item the naive currents of the quark model are related to the full weak currents
{\it via} the matching conditions of Heavy Quark Effective Theory (HQET) [6],

\item Heavy-Quark-Symmetry-breaking color magnetic interactions 
are included, whereas ISGW only included the symmetry-breaking 
due to the heavy quark kinetic energy,

\item the ISGW prescription for connecting its quark model form factors
to physical form factors is modified to be consistent with the constraints
of Heavy Quark Symmetry breaking at order $1/m_Q$,

\item relativistic corrections to the axial coupling constants 
(known to be important in the analogous coupling
$g_A$ in neutron beta decay) are taken into account, and

\item more realistic form factor shapes, based on the measured pion 
form factor, are employed.
\end{enumerate}

    The discovery of Heavy Quark Symmetry has not eliminated the need for 
models; it has 
rather provided a solid foundation for model-building and redefined the
role that models should play. Consequently, an updated version of the ISGW 
model that incorporates
the lessons of Heavy Quark Symmetry, and is designed with current usage in mind,
seems very worthwhile. Among other roles, models should:

\begin{enumerate}
\item provide predictions for the various universal form factors (``Isgur-Wise functions") 
of Heavy Quark Symmetry,

\item provide predictions for the form factors governing $b \rightarrow u$, 
$c \rightarrow s$, 
$c \rightarrow d$, and $s \rightarrow u$ transitions not directly 
governed by Heavy Quark 
Symmetry, and

\item give estimates for the sizes of Heavy-Quark-Symmetry-breaking
effects in the $b \rightarrow c$ decays determining $V_{cb}$, 
in the relations 
between $b \rightarrow u$ and $c \rightarrow d$ matrix elements which can be 
used to determine $V_{ub}$ 
from exclusive semileptonic decays [4,10], and in the relation between $c \rightarrow s$
and $b \rightarrow s$ matrix elements which enter into the prediction of
exclusive $b \rightarrow s \gamma$ decays [10]. 
\end{enumerate}

     In the next section we will give some of the background to ISGW and 
to the events leading up to ISGW2, as well as a quick review of 
the basic elements of the ISGW approach.  As already mentioned, 
Section III describes the new features of ISGW2 in detail.  In Section IV we 
present our results. Section V discusses their implications for Heavy
Quark Symmetry, while Section VI compares our results to experiment.
Section VII closes with a few 
comments.

\vfill\eject
\section{\bf   Background}
\medskip

\subsection{Some History}

    In 1985, when the model that was eventually published as the 
ISGW model [1] was introduced [2,3], its intended use was very 
different from its present use.  Moreover, much less was known 
about semileptonic $b$ and $c$ quark decays, both theoretically and 
experimentally.  The ISGW2 model presented here is designed to 
update the earlier version to address both of these shortcomings.
Ten years ago, the experimental study of the semileptonic decays of $b$ 
and $c$ quarks was in its infancy.  In particular, for $b$ quarks 
the main available data was on the inclusive lepton energy spectra 
for $\bar B \rightarrow X \ell \bar \nu_{\ell}$, generated by the quark level
$b \rightarrow c \ell \bar \nu_{\ell}$  and 
$b \rightarrow u \ell \bar \nu_{\ell}$ transitions.  
At that time the principal theoretical tool being used to analyze 
these spectra was the QCD-corrected parton model of ACCMM [11] 
and its relatives [12], with particular emphasis on extracting 
the Cabibbo-Kobayashi-Maskawa (CKM) [13] matrix elements $V_{cb}$ 
and $V_{ub}$  from inclusive lepton spectra.  Early fits to these 
spectra [14] near the $b \rightarrow c \ell \bar \nu_{\ell}$  
endpoint were leading to alarmingly 
small upper limits for the ratio ${\vert V_{ub}/V_{cb} \vert}^2$.  
Such results could of 
course simply be attributed to errors in the data.  Alternatively, 
they could be taken as serious limits on $V_{ub}$ which would indicate 
a failure of the Standard Model scenario for CP violation.  The 
ISGW model was introduced to explore a third possibility: that a 
partonic description of the
$b \rightarrow c \ell \bar \nu_{\ell}$  and 
$b \rightarrow u \ell \bar \nu_{\ell}$ transitions
in the endpoint
region, where the lepton energy is near its maximum, might be deficient.  
The basic motivation for this concern arises from the observation that 
the highest energy leptons in these decays are asssociated with the production 
of the lowest-mass hadronic final states $X$ in 
$\bar B \rightarrow X_c \ell \bar \nu_{\ell}$ and 
$\bar B \rightarrow X_u \ell \bar \nu_{\ell}$
respectively; the partonic description would only be expected to 
apply once the states $X_c$ and $X_u$ had masses above their 
respective resonance regions.

     We will discuss this issue in more detail below.  We raise it at 
this point to recall that one of the main goals of the ISGW model 
was the production of an alternative description of the endpoint 
region which intentionally represented an {\it extreme} example of how 
little $b \rightarrow u \ell \bar \nu_{\ell}$ could show up in the 
endpoint region.  The motivation was 
to illustrate the theoretical uncertainty which should be reflected 
in upper limits on ${\vert V_{ub}/V_{cb} \vert}^2$ extracted from 
inclusive endpoint spectra 
and to thereby place more realistic constraints on Standard Model 
CP-violation scenarios.
Along the path to this primary goal, the ISGW model produced a 
number of other results.  In retrospect, the most important of 
these were probably conceptual:  much of the framework for Heavy 
Quark Symmetry [4-7] was presented in these early papers [1-3], 
including the vital role of the zero recoil point 
(where $t=(p_{\ell}+p_{\bar \nu _{\ell}})^2$
is at its maximum value $t_m$), the 
insensitivity of $\bar B \rightarrow D \ell \bar \nu_{\ell}$ and 
$\bar B \rightarrow D^* \ell \bar \nu_{\ell}$ transitions to $m_b/m_c$, 
and the role 
of $D \rightarrow \bar K  \ell^+ \nu_{\ell}$ and 
$D \rightarrow \bar K^* \ell^+ \nu_{\ell}$ measurements in ``tuning" 
exclusive models to be 
used for the extraction of $V_{cb}$ and $V_{ub}$. ISGW also made a number 
of predictions.  For example, ISGW was the first exclusive model 
to calculate rates to channels other than the pseudoscalar and 
vector ground states and consequently to predict that in both
$b \rightarrow c \ell \bar \nu_{\ell}$ and $c \rightarrow s \ell^+ \nu_{\ell}$ 
decays the exclusive transitions $\bar B \rightarrow D, D^*$ and 
$D \rightarrow \bar K, \bar K^*$ 
would dominate.  This prediction (which is surprising since kinematically 
masses up to $m_B$ and $m_D$, respectively, are allowed), now has a 
firm basis in theory [15,16, 4-7].  They also pointed out that in the 
nonrelativistic limit (applicable to such exotic processes as 
$\bar B_c \rightarrow \psi \ell \bar \nu_{\ell})$, 
the weak transition form factors would be controlled by a set of 
universal functions given by the Fourier transforms of wave function 
overlaps and not by $t$-channel meson masses.  This point has since 
been explored by many authors [17].

      As mentioned in Section I, this update of ISGW has been prompted by 
a number of developments.  The most fundamental of these is the 
discovery and development of Heavy Quark Symmetry [4-7]. 
In particular, the development of Heavy Quark
Effective Theory  [6] as a tool for systematically treating both the
$1/ m_Q$ and perturbative QCD corrections to the extreme Heavy Quark
Symmetry limit has helped place models like ISGW in clear focus.  HQET divides
the calculation of current matrix elements into two steps:  matching the
currents of the full theory onto those of a low energy effective theory
associated with some relatively light renormalization scale $\mu$, and then
calculating matrix elements in the low energy effective theory.  From this
perspective, a quark model like ISGW or ISGW2 is presumed to be associated with a
quark model scale $\mu_{qm}
\sim {\cal O}$(1 GeV) where a valence constituent quark structure of hadrons
dominates the physics.

Since the constraints of Heavy Quark Symmetry for current matrix elements of the
low energy effective theory are consequences of QCD, 
every model 
should display these results (including 
an allowed symmetry-breaking pattern) in the appropriate 
limit.  In fact, in the low-recoil 
region where nonrelativistic dynamics apply, the ISGW model was 
already totally consistent with the Heavy Quark Symmetry limit.  Adding the 
constraints of Heavy Quark Symmetry in ISGW2 nevertheless has 
significant impact.  In high recoil $b \rightarrow c \ell \bar \nu_{\ell}$ 
transitions, some ISGW 
form factors have missing functions of $w \equiv v \cdot v'$ 
($v$ and $v'$ are the four-velocities of the initial and final hadronic systems; this
variable is called $w$ after the origin of the name of this letter in, {\it e.g.}, French) which are 
unity at zero 
recoil, $e.g.$, the $f$ form factor in $\bar B \rightarrow D^* \ell \bar \nu_{\ell}$
is missing a factor of ${1 \over 2}(1+w)$ 
which goes to unity at $w=1$.  A related issue is embedded in the 
recoil dependence of the ISGW form factors.  As discussed in ISGW, 
the slope of a quark model form factor consists of two terms: a 
normal ``transition charge radius" term and a relativistic correction (of 
order $1/m_jm_i$ in a $Q_i \rightarrow Q_j$ current matrix element) which 
is outside of 
the scope of a nonrelativistic quark model.  ISGW posited that 
such relativistic effects could be taken into account in an 
approximate way by replacing all factors of $(t_m-t)$ appearing in 
their nonrelativistic formulas for form factors by $\kappa ^{-2}(t_m-t)$, where 
$\kappa$ is the ratio of the nonrelativistic charge radius to the true 
charge radius.  Heavy Quark Symmetry [9] tells us that this prescription
(while fortuitously close numerically in the cases to which it was applied) 
is incorrect; the symmetry moreover 
dictates the correct result in the heavy quark limit.  This 
result, to be described below, is adopted in ISGW2.
Consideration of the allowed pattern of HQS-breaking at order $1/m_Q$ also
has an impact. Among other effects, it requires a change in the ISGW prescription
for relating the form factors of the weak binding limit calculated here to
physical form factors.   
Although such modifications to ISGW are only strictly required near 
the heavy quark limit, ISGW2 adopts the usual constituent quark model 
stance of treating all constituent quarks like heavy quarks, so 
the same changes are made, {\it e.g.}, to $c \rightarrow s$ transitions.

QCD also demands that the matrix elements of a low energy effective theory like
the quark model be corrected by the matching conditions which map them onto the
matrix elements of the full theory. At the level of the currents of the two theories,
these matching conditions take the generic
form
\begin{equation}
J^\mu_{ji}={\bf C}_{ji}{\bf J}^\mu_{ji}+{\alpha_s\over \pi}\Delta {\bf
J}^\mu_{ji}+ {1\over m_{Q_j}}
\delta_j{\bf J}^\mu_{ji}+{1\over m_{Q_i}} \delta_i {\bf J}^\mu_{ji}~.
\end{equation}
In ISGW2, we explicitly calculate the $1/ m_{Q_j}$ and $1/ m_{Q_i}$ corrections
in the quark model, so only the factor ${\bf C}_{ji}$ mapping the naive vector
$\left(\bar {\bf Q}_j \gamma^\mu {\bf Q}_i \right)$ and axial vector
$\left(\bar {\bf Q}_j \gamma^\mu \gamma_5 {\bf Q}_i \right)$ currents of the quark model onto
the true currents $\left(\bar Q_j \gamma^\mu Q_i \ {\rm and}\  \bar Q_j
\gamma^\mu \gamma_5 Q_i \right)$ and the expansion in terms of the new naive currents
appearing in
$\Delta {\bf J}^\mu_{ji}$ in order ${\alpha_s / \pi}$ are needed.  We will
give these matching factors in Section III.A below.

     There are other reasons why an update of the ISGW model is warranted.  
In the period since the publication of ISGW, its role in providing a 
very conservative upper limit on  ${\vert V_{ub}/V_{cb} \vert}^2$ 
has become antiquated; 
ISGW2 attempts to modernize ISGW so that its predictions become {\it best} 
estimates rather than {\it most conservative} estimates.  Consider, for 
example, the curve in Figure 1 showing the ISGW  
form factor $F_{\pi}(Q^2)$ with Gaussian wavefunctions.  
The charge radius of the pion was used to determine the value
of the parameter $\kappa$ which in turn determines
the rate of decrease of $F_{\pi}(Q^2)$ shown. Thus, instead of choosing 
a value that provided a best global fit to the data over the whole
kinematic range applicable to the 
$b \rightarrow u \ell \bar \nu_{\ell}$ transition, ISGW chose a value 
that fits at low $Q^2$ but, as a result of its
unrealistic Gaussian form, 
underestimates $F_{\pi}(Q^2)$ at high $Q^2$. This choice was driven by 
the ISGW goal of providing a minimum rate for 
$\bar B \rightarrow X_u \ell \bar \nu_{\ell}$ in the 
endpoint region.  In ISGW2 we attempt a more realistic description 
of the recoil dependence of all form factors.

     There have also been important experimental developments since 
1985!  In $\bar B$ decays [19], the inclusive spectra near the endpoint 
region show a definite $\bar B \rightarrow X_u \ell \bar \nu_{\ell}$ 
excess [20], although, for the reasons already 
mentioned, the resulting value of $V_{ub}$ is unclear.  The decays 
$\bar B \rightarrow D \ell \bar \nu_{\ell}$ and 
$\bar B \rightarrow D^* \ell \bar \nu_{\ell}$ 
have both been measured [21] in sufficient detail to extract 
the CKM matrix element $V_{cb}$ with some confidence since the observed
features of these decays are
consistent 
with the expectations of Heavy Quark Symmetry.  Preliminary evidence 
for $\bar B \rightarrow D^{**} \ell \bar \nu_{\ell}$ decays 
(here $D^{**}$ represents non-$D$ or $D^*$ decays) has been reported and searches have begun for the 
exclusive $b \rightarrow u \ell \bar \nu_{\ell}$ processes
$\bar B \rightarrow \rho \ell \bar \nu_{\ell}$ and 
$\bar B \rightarrow \omega \ell \bar \nu_{\ell}$ [22].  In 
D decays [23], where $V_{sc}$ is known, $D \rightarrow \bar K \ell^+  \nu_{\ell}$ and 
$D \rightarrow \bar K^* \ell^+ \nu_{\ell}$ decays have been measured [24]
in sufficient detail to extract the four $c \rightarrow s$ form factors 
contributing in the limit $m_{\ell} \rightarrow 0$, and rather tight limits on 
$D \rightarrow \bar K^{**} \ell^+ \nu_{\ell}$ have 
been set.  In  all cases the experimental results are qualitatively consistent 
with the  predictions of ISGW (despite some initial 
indications to the contrary [25]);  indeed, all results to date are 
consistent with ISGW within its anticipated ``quark model accuracy" 
of predicting matrix elements to $\pm 25\%$.  However, in the spirit 
of ``tuning" the quark model to higher accuracy, in ISGW2 we have 
taken note of a substantial failure of ISGW to predict the magnitude 
of the S-wave axial vector form factor $f$ in 
$D \rightarrow \bar K^* \ell^+ \nu_{\ell}$ decay.  In the 
quark model, this form factor is analogous to $g_A$ in neutron beta 
decay, where experiment is about 25\% below the quark model 
prediction of 5/3;  the data on $f$ indicate that it is also 
smaller than the quark model prediction.  There is a very natural 
explanation for the $g_A$ discrepancy within the quark model [26,27]:  
the matrix elements of the space components of the axial current in 
a relativistic S-wave spinor are reduced in proportion to the probability 
of lower components in that spinor.  We accordingly build this 
relativistic correction factor into ISGW2.

     With the predictions of Heavy Quark Symmetry to facilitate the 
extraction of $V_{cb}$ and $V_{ub}$ from the data, one of the main uses of 
models has shifted from predicting form factors to predicting the 
{\it deviations} of form factors, or relations between form factors,
from the predictions of Heavy Quark 
Symmetry.  In view of this changing role, we implement one further 
elaboration of ISGW in ISGW2:  we consider the effects of hyperfine 
interactions on meson wavefunctions.  ISGW already naturally took 
into account the other $1/m_Q$ effect in HQET
[6], the heavy quark kinetic energy, so this addition to 
the model completes the parallel with the most general symmetry-breaking effects allowed.
As we will see, the ``$g_A$ effect" and these hyperfine interactions, in concert with
matching corrections, eliminate the problem with the 
$D \rightarrow \bar K^* \ell^+ \nu_{\ell}$ form factor $f$.

     To summarize:  
ISGW2 is an updated version of ISGW 
designed to make ``best estimates" within the 
context of a constituent quark model that fully respects Heavy 
Quark Symmetry.

\bigskip
\subsection{ A Review of the Foundations of the ISGW Model}

\medskip

    In Section III we will describe in detail the new features which we 
incorporate in ISGW2.  Here, we review the basic ideas and methods 
of the ISGW model.

      ISGW breaks the problem of computing a current matrix element of a 
transition from a state $H$ of mass, momentum, and spin $m,p,s$ to 
$H'$ with $m',p',s'$ into kinematical and dynamical parts.  It first makes 
the usual mechanical Lorentz-invariant decomposition of the 
matrix element into Lorentz tensors and invariant form factors $f_i$ $(i=1,2,...N)$ 
which depend only on the four momentum transfer variable $(t_m-t)$ 
where $t=(p'-p)^2$ and where $t_m=(m'-m)^2$ is the maximum momentum transfer.  
The variable 
$(t_m-t)$ is used since it is zero at the ``zero recoil point" where $H'$ 
is left at rest in the rest frame of $H$; the importance of momentum 
transfers near $t_m$ will be made clear below.

     It should be noted that {\it any} specification of the functions $f_i(t_m-t)$ 
leads to a Lorentz invariant description of these weak decay processes.  
In this sense ISGW {\it is not} a nonrelativistic approximation.  
It is, however, a nonrelativistic estimate of the intercepts 
$f_i(0)$ and ``charge radii" $r_i \equiv [6{{df_i(0)} 
\over {d(t_m-t)}}]^{1 \over 2}$ (or more generally the shapes)
of the Lorentz invariant form factors 
$f_i(t_m-t)$.  These estimates are made by noting that there is a one-to-one 
correspondence between the $f_i$ and a partial wave expansion of the 
$\langle H' \vert j^{\mu}(0) \vert H \rangle$ matrix elements.  
For example, if $H$ and $H'$ are pseudoscalars 
$P$ and $P'$, then 

\begin{equation}
 \langle P'(p') \vert V^{\nu}(0) \vert P(p) \rangle = 
               f_+^{P'P}(p+p')^{\nu}+f_-^{P'P}(p-p')^{\nu}~~.
\end{equation}

\noindent This decay also has two partial wave 
amplitudes.  In the rest frame of $P$, $V^0(0) \vert P(0) \rangle$ is 
still a pseudoscalar, 
and so creates $P'$ in an S-wave; $\vec V(0) \vert P(0) \rangle$ is 
an axial vector and so must 
create $P'$ in a P-wave.  Thus $m(f_+^{P'P}+f_-^{P'P})+E'(f_+^{P'P}-f_-^{P'P})$ 
and $\vec p~'(f_+^{P'P}-f_-^{P'P})$ are proportional to the 
rest frame S-wave and P-wave amplitudes, respectively.

     A vital element of the ISGW model is that each partial wave 
amplitude is calculable in the nonrelativistic limit; the one-to-one 
correspondence with the $f_i$ then allows a calculation of each  
in this limit ($i.e.$, none of the $f_i$ are intrinsically
relativistic in character).  The ISGW model therefore calculates 
the $f_i$ in a  limit in which the model would in 
principle be exact, and then extrapolates these exact formulas 
to the physical regime.  It should be noted that the nonrelativistic 
limit requires more than  $\vec p/m$ and $\vec p~'/m'$ being small.  It also 
requires that the internal motion of the constituents of $H$ and 
$H'$ be nonrelativistic.  One of the essential assumptions of the 
ISGW model is that such ``mock meson" form factors
$\tilde f_i$ which are derived in the approximation 
that $m_u$, $m_d$, and $m_s$ are large compared to $\Lambda_{QCD}$ can 
be extrapolated 
down to their  actual constituent quark masses to estimate the $f_i$.

     In the heavy quark world in which the ISGW formulas would be exact 
in principle, the low-lying mesons would all be simple quarkonia.  
ISGW is therefore necessarily a model for matrix elements between 
resonances, {\it i.e.}, it does not directly address the issue of 
whether semileptonic meson decays are resonance dominated. 
The original ISGW paper argues that nonresonant contributions 
are likely to be small (their absence is correlated with the 
known success of the narrow resonance approximation), and there is 
some evidence from the data for this prediction.  Nevertheless, 
the issue remains a hotly debated one.  Note that this debate is relevant to our updating of 
the resonant matrix elements only once we use them to estimate 
the {\it inclusive} rates, {\it e.g.}, those in the 
$\bar B \rightarrow X_u \ell \bar \nu_{\ell}$ endpoint region.  At that 
point we will discuss this issue in more depth.

    The semileptonic decays of the $b \bar c$ meson $\bar B_c$ via the 
$\bar c \gamma^{\mu} (1- \gamma_5) b$ current 
provide a good illustration of a system in which the ISGW model 
would in principle be an excellent approximation.  Both $\bar B_c$ and 
the low-lying states of the $c \bar c$ system can be reasonably well-described 
as nonrelativistic, and matrix elements like 
$\langle \psi(p's') \vert A^{\nu} \vert \bar B_c(p) \rangle$ can be accurately 
calculated in the frame where $\vec p=0$ for small $\vec p~'$ as atomic-physics-type 
wavefunction overlap integrals.  This is the essence of the ISGW method.
However, serious model dependence can occur when these matrix elements are 
extrapolated to large recoils; moreover, it occurs even at small 
recoil when any quark mass is extrapolated down to the constituent 
masses of the $u$, $d$, or $s$ quarks.  Amongst the issues which must 
be addressed when a light quark plays a role are: 

\begin {enumerate}

\item	{\it the quarkonium approximation}:  It is a fundamental tenet of 
the constituent quark model that, up to ``small" corrections which 
arise from pair creation leading to resonance widths, systems 
containing a light quark can still be treated as quarkonia.  
{\it I.e.}, extra $q \bar q$ pairs and the gluonic degrees of freedom do not
have to be introduced explicitly.  ISGW adopts this approximation. 
In addition, for simplicity it uses harmonic oscillator 
wavefunctions to approximate the true quarkonium wavefunctions.

\item {\it the weak binding approximation}:  For heavy quarkonia, 
the quark masses and energies are approximately equal and as a result the 
hadron mass is approximately the sum of the constituent 
quark masses.  Once a quark becomes light, the failure of this approximation and other 
associated complexities make even a low velocity boost of the 
wavefunction problematic.  Moreover, when both constituents are 
light, there can be a great discrepancy between the mass of a 
hadron and the sum of the masses of its constituents; this in 
turn leads to ambiguities in the extrapolation of the 
nonrelativistic formulas.  ISGW adopts a specific prescription 
for dealing with such ambiguities.  

\item {\it relativistic corrections}:  Even if the extrapolation 
of the nonrelativistic ISGW formulas were straightforward, they would 
still suffer from their failure to incorporate important 
relativistic physics.  A simple example is the charge radius $r_i$:  
in general such a radius will receive both nonrelativistic 
contributions with a scale controlled by the radius of the 
quarkonium wavefunctions and relativistic contributions controlled
by the Compton wavelengths of the participating quarks.  
(The latter contributions are themselves of several types: 
relativistic corrections from the quarkonium wave equation, 
field-theory-induced pair creation effects, {\it etc.}  Such 
effects are simply lost in the nonrelativistic limit since 
$r \sim 1/p \gg 1/m$.)  Of course the form factor intercepts will also receive
relativistic corrections:  generically, we can write 
$f_i(0)=f_i^{nr}(0)[1+ {\cal O} (p/m)]$.  ISGW 
invokes the empirical success of the nonrelativistic 
quark model in assuming that $f_i^{nr}(0)$ will be a reasonable approximation 
to $f_i(0)$.  Finally, there is of course no guarantee that $f_i(t_m-t)$ won't 
contain $(t_m-t)$ dependence that is intrinsically relativistic.  For 
example, an additional ``kinematic" factor of 
$[1+{{t_m-t} \over {4mm'}}]$ would be ``seen" 
as unity by a nonrelativistic calculation.

\end {enumerate}

    All of these shortcomings, and others left unmentioned, make 
it surprising that the nonrelativistic constituent quark model 
works as well as it does.  It may be that its successes
are based on one crucial fact:  that ``it 
is better to have the right degrees of freedom moving at the 
wrong speed than the wrong degrees of freedom moving at the 
right speed" [28].  Given that the quark model would be correct 
if all the quarks were heavy quarks, its utility
may reside in its ability to {\it parameterize} the evolution of the 
properties of these correct degrees of freedom 
from heavy to light systems.

\vfill\eject

\section{\bf ISGW2:  the new features}

\medskip

     As already repeatedly emphasized, ISGW2 is {\it not} a new model:  it 
is a slightly 
improved version of ISGW [1].  In this section we describe one by 
one the differences between ISGW and ISGW2. For closely related studies 
of the marriage of the quark model with the physics of
Heavy Quark Symmetry, see the work cited in Ref. [7].

\medskip

\subsection{the constraints of Heavy Quark Symmetry}

    Although ISGW is completely consistent with the constraints of Heavy 
Quark Symmetry at maximum recoil $t_m$ (or $w=1$) in the symmetry limit, HQS 
also determines various aspects of the behavior of the form factors 
at finite recoil and at nonleading order in the $1/m_Q$ expansion.  
For example, the six form factors of 
$\bar B \rightarrow D \ell \bar \nu_{\ell}$ and 
$\bar B \rightarrow D^* \ell \bar \nu_{\ell}$
are required by Heavy Quark Symmetry to have, in the low energy
effective theory, the form [4-7]

\begin{equation}
    \tilde f_++\tilde f_-=\tilde f_+-\tilde f_-
=\tilde g= \tilde f=\tilde a_-- \tilde a_+= \xi(w)
\end{equation}
\begin{equation}
    \tilde a_++ \tilde a_-=0
\end{equation}
where $\xi(w)$ is the Isgur-Wise function (we have adopted conventions for
defining the HQS form factors which lead to these simple
forms; see Section V for explicit formulas relating these $\tilde f^{(\alpha)}$
to the usual ISGW form factors).  In the heavy quark 
limit, the ISGW model 
respects all of these constraints at all $w$ except for that on $\tilde f$:  
it gives $\tilde f  =( {2 \over {1+w} })\xi(w)$, corresponding to the nonrelativistic approximation 
$1+w \simeq 2$.  Such effects, which correspond to $v^2/c^2$ corrections 
to a leading 
nonrelativistic prediction, lie outside of the dynamical framework 
of ISGW, but are easily appended to the model (see, e.g., the second of Refs. [7]).  
Using eq. (3) and 
the corresponding results of Ref. [29] on the 
$\bar B \rightarrow D_2^* \ell \bar \nu_{\ell}$,
$\bar B \rightarrow D_1^{3/2} \ell \bar \nu_{\ell}$,
$\bar B \rightarrow D_1^{1/2} \ell \bar \nu_{\ell}$, and
$\bar B \rightarrow D_0^* \ell \bar \nu_{\ell}$ 
decays, the required modifications to be incorporated into ISGW2 
are easily enumerated. They are all listed explicitly in Appendix C.
The simplest example is the axial vector form factor $f$ of eq. (B15) of ISGW,
which, as the form factor corresponding to $\tilde f$ in eq. (3), 
picks up an additional factor of ${1 \over 2}(1+w)$ in the HQS limit (additional
nonleading effects in the $1/m_Q$ expansion will be described below). In addition to 
these modifications,
Heavy Quark Symmetry tells us that in heavy quark systems the 
eigenstates with $J^P=1^+$
are {\it not} the $L-S$ coupled states $^3P_1$ and $^1P_1$,
but rather the $j-j$ coupled states $P_1^{3/2}$ and $P_1^{1/2}$ with 
$s_{\ell}^{\pi_{\ell}}={3 \over 2}^+$
and ${1 \over 2}^+$, respectively [30,29]. We therefore also list the new
form factors appropriate to semileptonic decays to such excited $P$-wave mesons
in Appendix C.

In addition to these constraints of Heavy Quark Symmetry on the matrix
elements of the low energy effective theory, HQET prescribes how to match these
matrix elements onto matrix elements of the full theory, as already
mentioned above.  The matching of a generic form factor $f^{(\alpha)}_{ji}$ of
type
$\alpha$ associated with the underlying $Q_i\rightarrow Q_j$ transition can be
written in the form [31]
\begin {equation}
f^{(\alpha)}_{ji}={\bf C}_{ji}(w)\left[f^{(\alpha)}+\tilde
\beta^{(\alpha)}_{ji}(w) {{\alpha_s\left( \mu_{ji}\right)}\over \pi}\right] \xi
(w)
\end {equation}
where the $f^{(\alpha)}$ are unity for 
$\tilde f_++\tilde f_-$,
$\tilde f_+-\tilde f_-$,
$\tilde g$,
$\tilde f$, and
$\tilde a_+-\tilde a_-$,
and zero for $\tilde a_++\tilde a_-$. Here 
\begin {equation}
{\bf C}_{ji}=\left[ {{\alpha_s \left(m_i \right)}\over {\alpha_s \left(m_j
\right)}}
\right]^{a_I} \left[{{\alpha_s \left(m_j \right)} \over {\alpha_s
\left(\mu_{qm} \right)}}\right]^{{a_{L}}(w)}
\end{equation} 
is independent of $\alpha$ and has
\begin{equation}
a_I=-~{{6}\over {33-2N_f}} 
\end{equation}
and
\begin{equation}
a_L(w)={{8}\over {33-2N^{\prime}_f}}\left[wr(w)-1 \right]
\end{equation}
with
\begin{equation} 
r \equiv {{1}\over \sqrt{w^2-1}}{\it ln} \left(w+\sqrt{w^2-1}\right),
\end{equation}
$N_f$ the number of active flavors below the scale $m_i$ (four for $i=b$) and
$N^\prime_f$ the number below $m_j$ (three for $j=c$).  In contrast, the
radiative correction functions $\tilde \beta^{(\alpha)}_{ji} (w)$ multiplying
$\alpha_s / \pi$ (evaluated at a scale $\mu_{ji}$ intermediate between
$m_i$ and $m_j$ which we take to be the geometric mean $\mu_{ji}=(m_jm_i)^{1\over
2}$) are $\alpha -$dependent.

The $\tilde \beta^{(\alpha)}_{ji}(w)$ associated with each of six form factors
$\tilde f_{+}, \tilde f_{-}, \tilde g, \tilde f, \tilde a_{+}-\tilde a_{-}, $
and $ \tilde a_{+}+\tilde a_{-}$ are known.  At $w=1$ they are simply
\begin {eqnarray}
\tilde \beta^{(f_{+}+f_-)}_{ji}(1)&=& \gamma_{ji}-{2\over 3} \chi_{ji} \\ \cr
\tilde \beta^{(f_{+}-f_-)}_{ji}(1)&=&\gamma_{ji}+{2\over 3} \chi_{ji}  \\ \cr
\tilde \beta^{(g)}_{ji}(1)&=& {2\over 3}+ \gamma_{ji}  \\ \cr
\tilde \beta^{(f)}_{ji}(1)&=& -{2\over 3}+ \gamma_{ji}  \\ \cr
\tilde \beta^{(a_{+}+a_{-})}_{ji}(1)&=& -1- \chi_{ji} +{4\over 3} {1\over {(1-z_{ji})}}+
{2\over 3} {{1+z_{ji}}\over {(1-z_{ji})}^2} \gamma_{ji} \\ \cr
\tilde \beta^{(a_{+}-a_{-})}_{ji}(1)&=&-{4 \over 3} {1 \over {(1-z_{ji})}}-\chi_{ji} +
{1 \over 3}+ \left[ 1-{2 \over 3} {{1+z_{ji}} \over
{(1-z_{ji})^2}} \right] \gamma_{ji} 
\end{eqnarray}
where
\begin{equation}
\gamma_{ji}\equiv {{2z_{ji}}\over {1-z_{ji}}} {\it ln} {1\over {z_{ji}}}-2 
\end{equation}
and
\begin{equation}
\chi_{ji}\equiv-~1~-~\frac{\gamma_{ji}}{1-z_{ji}}
\end{equation}
with
\begin{equation}
z_{ji} \equiv {m_j\over m_i}. 
\end{equation}
We deviate from the use of these matching conditions only in case of
transitions between light ($u$, $d$, and $s$) quarks. Since, as described in
Appendix A, we assume that $\alpha_s$ ``freezes out" at 
the quark model scale $\mu_{qm}$, the ``renormalization group improved"
matching conditions, as embodied in the ${\bf C_{ji}}$ factor, are inappropriate
for such transitions. For them we use the expansion of ${\bf C_{ji}}$ to 
lowest order in $\alpha_s$, {\it i.e.} we resort to ``lowest order matching".

In principle the $\tilde \beta^{(\alpha)}_{ji}$ are functions of $w$, but this
$w-$dependence is predicted [31] to be so 
weak relative to uncertainties in the $w$-dependence associated with nonperturbative
effects that we ignore it here. (This dependence
would, for example, correspond to a change in the predicted rates for the exclusive $b
\rightarrow c$ decays of the order 1\% {\it if} it could be
distinguished from the $w -$dependence in the preasymptotic nonperturbative Isgur-Wise
functions.)
On the other hand, there is nontrivial $w-$dependence contained in the factor
$\left[{{\alpha_s(m_{j}})\over {\alpha_s(\mu_{qm})}}\right]^{a_L(w)}:$ for $w$
near 1,
\begin {equation}
\left[{{\alpha_s(m_{j}})\over {\alpha_s(\mu_{qm})}}\right]^{a_L(w)} \simeq 1 -
{2\over 3}\left( {8\over {33-2N^{\prime}_f}}\right) {\it ln} \left[ {{\alpha_s(\mu_{qm})}\over
{\alpha_s(m_j)}}\right](w-1) 
\end {equation}
We will make use of this factor below.

    The preceeding HQS-and HQET-induced modifications to ISGW are consequences
 which emerge from considerations of the heavy quark limit.  There are
additional modifications which arise from restrictions on the form of $1/
{m_Q}$ corrections to this limit.  From the most general form of $1/ m_b$
and $1/ m_c$ corrections to the $\bar B \rightarrow D \ell \bar\nu_{\ell}$ and $\bar
B \rightarrow D^* \ell \bar\nu_{\ell}$ form factors (see Section V), it is possible
to resolve an ambiguity in the procedure for relating the form factors of the
ISGW weak binding nonrelativistic calculation to physical from factors.  Such a
calculation in principle only determines form factors up to factors like $\left(
m_H/ \tilde m_H \right)^n$ where $m_H$ is a physical hadron mass and
$\tilde m_H$ is the sum of its constituent quarks' masses.  HQET resolves this
ambiguity in a pleasing way:  it specifies that the form factors $\tilde
f_i^{qm}$ being calculated in such a quark model are (up to order $1/ m_Q$)
the dimensionless form factors $\tilde f_i$ of the heavy quark limit which
expand matrix elements in Lorentz invariants using the heavy quark four
velocities $v^\mu$, $v^{\prime \mu}$ and {\it not} the $f_i$ which expand them
in terms of their momenta.  (As far as we can determine, it is purely by
accident that the ISGW notation $\tilde f_i$ for the weak binding form factors
coincides with the notation for the HQS form factors).  Such considerations in
addition demand that the conventional form factors $f_i$ be obtained from the
$f_i^{qm}$ by mass scaling factors which differ from the physical masses by at 
most $1/ m_Q$ effects.  In ISGW2 we choose to resolve this residual
ambiguity by using the hyperfine-averaged physical masses $\bar m_{H_1H_2}
\equiv \bar m_{H_1} \equiv \bar m_{H_2}$ of a HQS spin doublet of hadrons $H_1$
and $H_2$ to relate the $\tilde f_i^{qm}$ to the $f_i$.  (For the $s_ \ell =
{1\over 2}$ ground state doublet this mass is just $\bar m_{VP} = {3\over 4} m_V +{1
\over 4} m_P$;  in general $\bar m_{s_\ell}= \left( {s_\ell +1} \over {2 s_\ell +1}
 \right) m_{j= s_\ell + {1\over 2}} + \left( { s_\ell} \over {2 s_\ell + 1} \right)
m_{ j= s_\ell - {1\over 2}}$ for an HQS multiplet with light degrees of freedom
having spin $s_\ell$).  As in ISGW, we use the $f_i$ to compute all rates;  this
may be viewed as a residual model-dependent choice of certain $1 / m^2_Q$
terms, and illustrates very clearly how HQS and HQET have reduced the 
model-dependence of the results of ISGW2 relative to ISGW.

There are two additional but clearly related elements to the correspondence between
the $\tilde f_i^{qm}$ and the $f_i$.  The $\tilde f_i^{qm}$ are functions of $\tilde w$
which is the weak-binding variable analogous to the physical variable $ w$.  In passing to the
physical form factors $f_i$ which depend on $t_m-t$ we identify
\begin {equation}
\tilde w -1 = {{t_m -t}\over {2\bar m_{P_Q} \bar m_{X_Q}}}
\end{equation}
for a transition $P_Q \rightarrow X_q \ell \bar \nu_{\ell}$ induced by an underlying
$Q \rightarrow q \ell \bar \nu_{\ell}$ transition.  In addition, we correct all form
factors for relativistic terms proportional to $t_m-t$ that are required by the form
of $1/m_Q$ corrections:  see Section V and Appendix C for details.

    We next note that consideration of sum rules for the 
$Q_i \rightarrow Q_j \ell \bar \nu_{\ell}$
transition in the heavy quark limit provides a
constraint on the slope of the Isgur-Wise 
function $\xi(w)$.  If we define $\rho^2$ by the expansion 

\begin{equation}
    \xi(w)=1-\rho^2(w-1)+ \cdot \cdot \cdot 
\end{equation}

\noindent about $w=1$, 
then [9] in the heavy quark limit

\begin{equation}
    \rho^2= {1 \over 4}+ {\rho^2_{sd}} + \Delta \rho^2_{pert}.
\end{equation}
As will 
be seen below, the structure-dependent term $\rho^2_{sd}$
dominates for a weakly
bound system.
The $\Delta \rho^2_{pert}$ term is that appearing 
in eq. (19) from the $w$-dependence of
the matching factors.
The ${1 \over 4}$  represents
a relativistic correction to the nonrelativistic limit.  
It corresponds to the relativistic correction 
found in Ref. [27]; its generalization to systems with different 
spins is discussed in Ref. [32]. The results of Ref. [27] actually 
dictate the subleading (order $1/m_Q$ and $1/m_q$) corrections to the $\frac{1}{4}$
of the heavy quark limit. In terms of a conventional 
charge radius $r^2$ defined by

\begin{equation}
    f(t)=f(t_m)[1-{1 \over 6} r^2(t_m-t)+ \cdot \cdot \cdot] 
\end{equation}

\noindent in an expansion of 
the generic $P_Q \rightarrow X_q$ form factor $f$ around $t=t_m$, the relation corresponding to
eq. (22) with subleading terms from Ref. [27] included is 

\begin{equation}
     r^2={3 \over 4m_Qm_q} + r_{wf}^2 + {1 \over {\bar m_{P_Q} \bar m_{X_q}}}
\left( \frac{16}{33-2N'_f} \right) {\it ln}\left[ \frac{\alpha_s(\mu_{qm})}{\alpha_s(m_q)} \right]
\end{equation}

\noindent where, for a ground state harmonic oscillator wavefunction (see Appendix A),

\begin{equation}
      r_{wf}^2={3m_{sp}^2 \over 2 \bar m_{P_Q} \bar m_{X_q} \beta_{PX}^2}~~ .
\end{equation}

\noindent The terms of eq. (24) are associated in order with the 
terms of eq. (22). Indeed, $r_{wf}^2$ in eq. (25) is the 
transition matrix element of the
square of the interquark separation between $P_Q$ and $X_q$; it would be
{\it four} times the squared charge radius of the pion in the case where $P_Q=X_q=\pi$.
Since $m_{sp}/\beta_{PX} \sim m_{sp}/p_{wf} \gg 1$ in the nonrelativistic
limit, the $1\over 4$ is 
indeed a ``relativistic correction", as stated earlier.  However, 
in the constituent quark model $m_{sp}/p_{wf} \sim 1$, so it 
could be a very significant 
``correction"!  ISGW recognized the {\it generic} possibility of 
$1/m_Qm_q$ corrections 
to $r^2$ and accordingly introduced a ``relativistic correction factor" $\kappa$ 
to compensate for them: they took 
$f(t_m-t) \rightarrow f({{t_m-t}\over {\kappa ^2}})$,
corresponding to enlarging $r$ 
by a factor $\kappa ^{-1}$.  It is now clear that the $3 \over 4m_Qm_q$ term 
in eq. (24) is a 
well-defined and necessary relativistic kinematic correction which 
should be added to the $r_{wf}^2$ term of a nonrelativistic model.  In ISGW2 
we note that this required correction is actually sufficient to 
achieve the same empirical effect as the multiplicative factor $\kappa$ of 
ISGW, which was fit to the low $t$ pion charge form factor.  While relativistic 
dynamics missing from the constituent quark model might in principle 
still affect $r_{wf}^2$, we assume that such effects were on the whole 
subsumed into the quark model once its parameters were chosen to 
give a good description of the meson spectra.  Thus, in ISGW2 we 
drop the {\it ad hoc} $\kappa$ factor in favor of the use of eq. (24).  This 
has the 
additional bonus of making ISGW2 consistent with the dynamical 
constraints of the Bjorken sum rule [9]: in the heavy quark limit, 
the ``structure-dependent terms" in eq. (22) are determined by the 
amplitudes to excite final states $X_q$ with $s_{\ell}^{\pi_{\ell}}={3 \over 2}^+$ 
and
${1 \over 2}^+$. Finally, we note that the $\Delta \rho^2_{pert}$ term
vanishes for decays to $s$, $d$, and $u$ quarks since their masses are
already below the quark model scale $\mu_{qm}$ where the running coupling constant has been assumed to saturate
(see Appendix A). As a result, it only comes into play for $b \rightarrow c$ transitions.

    In addition to this improvement in the way we deal with the slopes of the
form factors, in ISGW2 we also abandon the gaussian form factors of ISGW, which
are unrealistic at large recoils. This modification is described in Section III.C
below.

\subsection{  some relativistic corrections to the quark model}

     As stressed in Section II, the ISGW model was introduced to 
illuminate some basic issues surrounding semileptonic decays.  
It therefore used the simplest possible version of the 
quark model capable of addressing these issues.  It is, however, 
known that the predictive accuracy of the naive nonrelativistic 
quark model can be substantially improved by considering various 
 relativistic corrections to that model.
One of the simplest such corrections occurs in the matrix elements 
of the axial current.  The naive nonrelativistic quark model 
predicts that $g_A={5 \over 3}$ in neutron beta decay.  However, 
it has been known for twenty years that when the constituent quarks 
are given realistic momenta, $g_A$ is reduced by a factor of 
$1-{4 \over 3}P_{lower}$, 
where $P_{lower}$ is the probability of lower components in the quark 
spinors [26].  By taking this effect into account, most models 
[26,27] obtain values of $g_A$ about 25\% smaller than $5 \over 3$, close to 
its observed value of $1.257 \pm 0.003$ [33].
For the $P_Q \rightarrow V_q$ axial vector S-wave form factor (called $f$ in ISGW) 
the correction factor is

\begin{equation}
     C_f=J_{V_qP_Q}^{-1} \int d^3p \phi^*_{V_q}(p)
              \Bigl[{{E_q+m_q} \over {2E_q}}\Bigr]^{1/2} 
              \Bigl[1-{{p^2} \over {3(E_q+m_q)(E_Q+m_Q)}} \Bigr]
               \Bigl[{{E_Q+m_Q} \over {2E_Q}}\Bigr]^{1/2}\phi_{P_Q}(p)
\end{equation}

\noindent where the $\phi$'s are S-wave momentum 
space wavefunctions, $E_i=(p^2+m_i^2)^{1 \over 2}$, and 
$J_{V_qP_Q}=\int d^3p\phi^*_{V_q}(p)\phi_{P_Q}(p)$.  For a heavy 
quark transition in the 
heavy quark limit $C_f=1$, but for a light quark transition, the analog 
of eq. (26) for $n \rightarrow p$ would give roughly the required 
reduction of $g_A$.  
In ISGW2 we adopt this correction factor as being at the least a 
reasonable interpolation between these two extremes. The correction factors
resulting from eq.(26) using the masses and wavefunctions of Appendix A
are given in Table I.

     Two potential deficiencies of this approach should be noted.  There 
is in the first place no reason to suppose that there are not other 
more dynamical effects which renormalize the matrix elements of the 
light axial quark currents:  the effect taken into account by eq. (26)
should be only part of the story [34].  In addition, it is not 
clear that only the S-wave form factor $f$ will be affected by 
relativistic corrections.  We nevertheless take this as the simplest 
working hypothesis, and assume that the effective constituent quark 
mass subsumes other relativistic corrections as it does for quark 
model magnetic moments [27].

     A second class of relativistic corrections to the quark model appears 
in the wavefunctions themselves.  For simplicity, ISGW ignored the 
effect of relativistic corrections to the effective interquark
potential.  In particular, although quark model hyperfine interactions 
are responsible for the $\bar B^*-\bar B$, $D^*-D$, 
$\bar K^*- \bar K$, and $\rho - \pi$ splittings, their 
effects on the wavefunctions were not taken into account.  (In HQET
[6], this origin of the $\bar B^*- \bar B$ 
and $D^*-D$ splittings 
can be given a firm foundation $via$ the ${\sigma_{\mu \nu} G^{\mu \nu}}
/ {2m_Q}$ operator appearing at 
order $1/m_Q$ in the heavy quark expansion.  The quark model assumes 
the continuing relevance of this mechanism for light quarks as well). 
For our purposes, the net effect is that pseudoscalar and vector 
particles of a given flavor are no longer characterized by the same
wavefunction parameter $\beta_S$ (see Table II of ISGW).  An update of this 
Table which takes into account this splitting is given in Appendix A. 
Given that both hyperfine and spin-orbit effects in P-wave mesons are 
empirically very weak, we ignore such effects. 

\bigskip

\subsection {more realistic form factors}

\medskip

    As mentioned above, ISGW used the gaussian form factors 
generated by their highly truncated harmonic oscillator 
basis; moreover, they used them out to relativistic recoils.  
Here we attempt a more accurate parameterization of the form factors
which will
have a more realistic behaviour at large $(t_m-t)$ by
making the replacement

\begin{equation}
     \exp {\Bigl[-{1\over 6} r^2_{wf}(t_m-t) \Bigr]} 
               \rightarrow \Bigl[1+{1\over 6N}r^2(t_m-t) \Bigr]^{-N}
\end{equation}
where $r^2$ is given by eq. (24). In eq. (27), $N=2+n+n'$ where $n$ and $n'$ are
the harmonic oscillator quantum numbers of the initial and final wavefunctions ($i.e.$,
$N=2$ for $S$-wave to $S$-wave, $N=3$ for $S$-wave to $P$-wave, $N=4$ for $S$-wave
to $S'$-wave, $etc.$). These form factors all have the charge radii dictated by the quark model
in the nonrelativistic limit, approach the gaussian form factors of the harmonic oscillator model 
as $N \rightarrow \infty$, but provide a much better global fit to the pion form factor
(see Fig. 1). In fact, with eq. (24) we predict 
$<r^2_{\pi}>^{1 \over 2}=0.61$ fm, in satisfactory agreement with the observed [18]
value of $0.71 \pm 0.02$ fm. Since the $Q^2$ range covered by this figure corresponds to a 
$(t_m-t)$ range that covers the recoils available in the semileptonic
decays we treat here, we adopt the substitution of eq. (27) for all our
decays. We emphasize that these substitutions should be viewed as 
low energy parameterizations of the form factors and {\it not} as
appropriate descriptions of their analytic or high $(t_m-t)$ forms.

\vfill\eject

\section{\bf results and discussion}

\bigskip

     The formulas we require to predict semileptonic form factors and rates
may all be obtained from ISGW [1] 
(supplemented by formulas given in refs. [25] and [29] for decays in which the
lepton mass is not negligible)  by making the few simple 
modifications described in the text. The required 
changes are described explicitly in Appendix C. To calculate rates
we insert into these formulas the constituent quark masses and $\beta$-values
from Tables A1 and A2 of Appendix A.

     We now present our results, organized by the underlying quark decay
and arranged in order of increasing spectator quark mass. We will compare these
results to the predictions of Heavy Quark Symmetry
in Section V and to experiment in Section VI.

\subsection {$\underline{ b {\mbox{$\rightarrow$}} c {\mbox{$\ell \bar \nu_{\ell}$}}}$}

 These decays are generally the most stable predictions of our model,
and those that are underwritten by Heavy Quark Symmetry are the 
most reliable.  All states
contain a heavy quark and the available recoil is limited, reducing
the sensitivity to form factor slopes. Since the $b$ and $c$ quarks
are not only heavy but also have a modest mass difference, the
Shifman-Voloshin limit [15,16] is also relevant
to the decays with a light spectator and thus provides a simple
explanation for why the electron spectral shape is
very similar to that
of the
free quark decay model despite dominance by the ground state pseudoscalar and
vector final states.  

\bigskip
\noindent 1.{${{\mbox{$\bar B$}} {\mbox{$\rightarrow$}} X_{c\bar d} {\mbox{$\ell \bar \nu_{\ell}$}}}$}

Our results for ${\mbox{$\bar B$}} {\mbox{$\rightarrow$}} X_{c\bar d} {\mbox{$e \bar \nu_e$}}$ are shown in Figure 2;
the partial widths are given in Table II.
This decay is dominated by the pseudoscalar
and vector meson final states, which contribute 29\% and 61\% of the total
semileptonic rate respectively. Our absolute prediction for the inclusive
decay rate for ${\mbox{$\bar B$}} {\mbox{$\rightarrow$}} X_{c \bar d} {\mbox{$e \bar \nu_e$}}$ is $\Gamma = 4.06 \times 10^{13} |V_{cb}|^2
\sec^{-1}$, about the same as the ISGW result.  
The approximate validity of the SV
limit gives an electron spectral shape {\it very} similar to the free
quark model despite the dominance by the ground states of $X_{c}$. Our predicted
form factors are also close to those of the heavy quark limit; a detailed
discussion of this limit will be given in Section V.

\bigskip
\noindent 2.{${{\mbox{$\bar B_s$}} {\mbox{$\rightarrow$}} X_{c \bar s} {\mbox{$\ell \bar \nu_{\ell}$}}}$}

The small difference in the ${m_d}$ and ${m_s}$ constituent quark masses on the
${m_b}$ or ${m_c}$ scale results in ${\mbox{$\bar B_s$}} {\mbox{$\rightarrow$}} X_{c \bar s} {\mbox{$e \bar \nu_e$}}$
decays behaving in a very similar fashion to the previous case.  This may be seen in both
Figure 3 and Table II
where our results are displayed.
As expected, there is a small increase $(\sim 5\% )$ in the total fraction of
the 1P and 2S states compared to ${\mbox{$\bar B$}}$ decay since the SV limit 
holds here to a slightly reduced degree.  Our absolute prediction for the
inclusive decay ${\mbox{$\bar B_s$}} {\mbox{$\rightarrow$}} X_{cs} {\mbox{$e \bar \nu_e$}}$ is $\Gamma = 3.90 \times 10^{13}
| V_{cb}|^2 \sec^{-1}$, slightly smaller than the previous case.

\bigskip
\noindent 3.{${\bar B_c {\mbox{$\rightarrow$}} X_{c \bar c} {\mbox{$\ell \bar \nu_{\ell}$}}}$}

This decay is different from the preceding two cases in several ways. As the
spectator quark is no longer light, both the parent and daughter mesons are
approximately non-relativistic and are thus appropriately described 
by our model. The results, which  should
therefore be quite reliable, are shown in Figure 4 and
Table II.  They are still reminiscent of the previous results with
lighter spectators even though the spectator approximation
prediction that the inclusive semileptonic decay rates should be equal fails by
about 25\% in going from $\bar B_d$ to $\bar B_s$ to $\bar B_c$ deacys.  The
contributions from the pseudoscalar and vector final states are, however, reduced (to
29\% and 47\%, respectively) as expected from the spectator arguments given
in the ISGW papers and from the inapplicability of the SV limit.

The measurement of the slopes of the form factors for these decays would provide
an interesting test of the arguments made in Refs.
[1,2,17] that naive dispersion relations for these slopes
will fail.  These systems are predicted to have charge radii determined by their
Bohr radii $\sim \bigl[ {4(m_b+m_c) \over{3 
m_bm_c\alpha_s}} \bigr]^{-1}$ while
dispersion relations would lead one to believe (unless one were {\it very}
careful [17]) that the charge radii will be of order
$({m_b} + {m_c} )^{-1}$.  Discussing the possibility of studying these states
may not be completely far-fetched: there are suggestions [37,38]
for experiments to observe them.

\subsection {\underline {${c {\mbox{$\rightarrow$}} s {\mbox{$\ell^+ \nu_{\ell}$}}}$}}

     The decays induced by the quark level process $c {\mbox{$\rightarrow$}} s {\mbox{$\ell^+ \nu_{\ell}$}}$ are
dominated by the ground state pseudoscalar and vector daughter mesons.  This
is a consequence of the low available recoil momentum which has little probability
of producing excited states.
These decays are nevertheless not expected to be as accurately described as the $b {\mbox{$\rightarrow$}} c$ case
since the $s$ quark is too light for Heavy Quark Symmetry to apply.

\bigskip
\noindent 1.{${{\mbox{$D$}} {\mbox{$\rightarrow$}} X_{s\bar u} {\mbox{$\ell^+ \nu_{\ell}$}}}$}

Our results for ${\mbox{$D$}} {\mbox{$\rightarrow$}}  X_{s\bar u} {\mbox{$e^+ \nu_e$}}$ are shown in
Figure 5; the partial widths are given in Table III.
This decay is predicted to be almost totally dominated by the pseudoscalar and vector meson
final states, which contribute 63\% and 34\% of the total semileptonic rate
respectively. Our absolute prediction for the rate of the inclusive decay ${\mbox{$D$}}
{\mbox{$\rightarrow$}}  X_{s\bar u} {\mbox{$e^+ \nu_e$}}$ is $\Gamma = 0.17 \times 10^{12} |V_{sc}|^2
\sec^{-1}$, down about $10\%$ from ISGW. This decrease arises from an increase of the
$K$ rate of 18\% and a decrease in the $K^*$ rate of 41\% which dramatically alter
the ISGW $K^*/K$ ratio; for details, see Section VI below.
Our total predicted width is about one half that of a simple
free quark model using our constituent quark masses.

    It is amusing to note that while neither the heavy quark nor SV limits 
should be applicable here, they both seem to have strong residual influences
on this decay. The comparison of our predicted form factors with those of 
Heavy Quark Symmetry assuming that $s$ is a heavy quark will be given below.

\bigskip
\noindent 2.{${{\mbox{$D_s$}} {\mbox{$\rightarrow$}} X_{s \bar s} {\mbox{$\ell^+ \nu_{\ell}$}}}$}

While $m_u$ and ${m_s}$ are very similar on the scale of the charm
quark mass, they are noticeably
different on the scale of the daughter quark mass ${m_s}$.
As a result we do not expect as strong a similarity between ${\mbox{$D$}}$ and ${\mbox{$D_s$}}$
decays as that which existed between ${\mbox{$\bar B$}}$ and ${\mbox{$\bar B_s$}}$ decays.  
In addition,
non-ideal mixing in the pseudoscalar sector of ${\mbox{$D_s$}}$ decays leads to a very
different spectral shape due to the comparatively low mass of the $\eta$. This
low mass gives a much higher electron endpoint than the corresponding
free quark decay endpoint, as may be seen in Figure 6.  
The fractions of the semileptonic rate going to $\eta$, $\eta'$, and $\phi$ are 
$31\%$, $26\%$, and $40\%$, respectively, with the distribution of rate
to the $\eta$ and $\eta'$ sensitive to the assumed pseudoscalar mixing angle of
$-20^{\circ}$, but with the sum relatively insensitive (see Table III).
The 1P and 2S contributions 
are once again predicted to be small: only 3\% of the semileptonic width. The ratio
of rates for ${\mbox{$D$}} {\mbox{$\rightarrow$}} K^* {\mbox{$e^+ \nu_e$}}$ and ${\mbox{$D_s$}}
{\mbox{$\rightarrow$}} \phi {\mbox{$e^+ \nu_e$}}$ is

\begin{equation}
{{\Gamma({\mbox{$D_s$}} {\mbox{$\rightarrow$}} \phi {\mbox{$e^+ \nu_e$}} ) }\over { \Gamma(
{\mbox{$D$}} {\mbox{$\rightarrow$}} K^* {\mbox{$e^+ \nu_e$}} )}} = 0.84 ,
\end{equation}
down about 20\% from the ISGW  value [39] of 1.02. This decrease is mainly 
due to hyperfine interaction effects and the new $\bar m$ prescription of Section III.A.
We note before leaving these decays that the $D_s$ inclusive semileptonic 
decay rate is itself down
by more than 25\% from $D$ decay. This substantial failure of the 
spectator approximation will be discussed in Section VI.

\bigskip
\noindent 3.{${{\mbox{$B_c$}} {\mbox{$\rightarrow$}} X_{s \bar b} {\mbox{$\ell^+ \nu_{\ell}$}}}$}

Our results for ${\mbox{$B_c$}} {\mbox{$\rightarrow$}} X_{s \bar b} {\mbox{$e^+ \nu_e$}}$ are shown in
Figure 7.  The explicit partial widths are given in
Table III.  Not surprisingly, the extreme mass of the
spectator in this case
results in a spectrum that is very different from the naive free quark spectrum.
It is dominated by decays to
the pseudoscalar (43\%) and vector (55\%), as the available energy is small.
Recoil effects are very small due to the large daughter mass.  The softening of
the lepton spectra expected due to the high spectator  mass is
pronounced as is the reduction of the inclusive rate. Our
absolute rate $\Gamma({\mbox{$B_c$}} {\mbox{$\rightarrow$}} X_{s\bar b} {\mbox{$e^+ \nu_e$}}) = 0.50 \times 10^{11}
|V_{sc}|^2 \sec^{-1}$ is less than a third that of ${\mbox{$D$}} {\mbox{$\rightarrow$}} X_{s\bar u}
{\mbox{$e^+ \nu_e$}}$, corresponding to an even more dramatic failure of the spectator approximation.

The ratio of ${\mbox{$B_c$}}$ decay via $b {\mbox{$\rightarrow$}} c$ to $c {\mbox{$\rightarrow$}} s$ decay is
\begin{equation}
\frac{\Gamma({\mbox{$B_c$}} {\mbox{$\rightarrow$}} X_{s\bar b} {\mbox{$e^+ \nu_e$}}) }{\Gamma({\mbox{$B_c$}} {\mbox{$\rightarrow$}} X_{c\bar c}
   {\mbox{$e^+ \nu_e$}})} = 0.0014 \left| \frac{V_{sc}}{V_{cb}}\right|^2 \sim 1
\end{equation}
for $|V_{cb}| \simeq 0.04$ and $|V_{sc}| \simeq 1$. Thus,
amusingly,  ${\mbox{$B_c$}}$ semileptonic decays
will be roughly evenly split between the two very different quark level processes
$b {\mbox{$\rightarrow$}} c$ and $c {\mbox{$\rightarrow$}} s$.

\subsection {\underline {${c {\mbox{$\rightarrow$}} d {\mbox{$\ell^+ \nu_{\ell}$}}}$}}

We now consider the Cabibbo-suppressed decays involving the quark level process
$c {\mbox{$\rightarrow$}} d {\mbox{$\ell^+ \nu_{\ell}$}}$ which are also predicted to be
dominated by the
ground state pseudoscalar and vector final states.  These
decays have taken on a new importance since the realization that
their measured form factors can be related to
the form factors of $b {\mbox{$\rightarrow$}} u$ via Heavy Quark Symmetry. As indicated
in Section V below, these relations should
eventually lead to accurate model-independent determinations of $|V_{ub}|$.

\bigskip
\noindent 1.{${{\mbox{$D$}}^0 {\mbox{$\rightarrow$}} X_{d\bar u} {\mbox{$\ell^+ \nu_{\ell}$}}
\mbox{\rm ~~ and ~~} {\mbox{$D$}}^+ {\mbox{$\rightarrow$}} X_{d\bar d} {\mbox{$\ell^+ \nu_{\ell}$}}}$}

Our results for ${\mbox{$D$}}^0 {\mbox{$\rightarrow$}} X_{d\bar u} {\mbox{$e^+ \nu_e$}}$ and
${\mbox{$D$}}^+ {\mbox{$\rightarrow$}} X_{d\bar d} {\mbox{$e^+ \nu_e$}}$ are shown in Figures 8a) and
b) respectively.  The partial widths are given in Tables IV and V, respectively.
${\mbox{$D$}}^0 {\mbox{$\rightarrow$}} X_{d\bar u} {\mbox{$e^+ \nu_e$}}$ is dominated by $\pi$ and
$\rho$ final states which contribute 63\% and 31\% of the total respectively, compared to 43\%
and 52\% in ISGW.  This shift in relative probability comes mainly
from a substantial decrease in the $\rho$ rate. However,
there is also a sizeable $5\%$ rate predicted to the $J^P=1^+$
$P$-wave states. The longitudinal to transverse ratio for the $\rho$ is
$0.67$.
The ${\mbox{$D$}}^+$ decays look somewhat different as the final states now include
both the $I=0$ and $I=1$ neutral states.
Note that
$\Gamma({\mbox{$D$}}^+ {\mbox{$\rightarrow$}} X_{d\bar d} {\mbox{$e^+ \nu_e$}})  /
\Gamma({\mbox{$D$}}^0 {\mbox{$\rightarrow$}} X_{d\bar u} {\mbox{$e^+ \nu_e$}}) = 0.92 $, which is
mostly due to the effects of the $\eta$ and $\eta'$ channels. We also note in passing that
Cabibbo-forbidden decays are predicted to represent approximately 5\% of $D^0$ decays and 4\% of
$D^+$ decays.

\bigskip
\noindent 2.{${{\mbox{$D_s$}}^+ {\mbox{$\rightarrow$}} X_{d \bar s} {\mbox{$\ell^+ \nu_{\ell}$}}}$}

${\mbox{$D_s$}} {\mbox{$\rightarrow$}} X_{d \bar s} {\mbox{$e^+ \nu_e$}}$ decays of Figure 9 and
Table IV  are again dominated by the pseudoscalar
and vector ground states which contribute 60\% and 29\% of the total
resonant semileptonic rate with a $10\%$ contribution from the $J^P=1^+$
$P$-wave states.
The absolute rate is almost a factor of three times smaller than the free quark decay
rate, and the lepton spectrum is much softer. Note that these 
Cabibbo-forbidden decays are predicted to contribute approximately 6\% of the inclusive
$D_s$ decay rate.

\subsection { \underline {${b {\mbox{$\rightarrow$}} u {\mbox{$\ell \bar \nu_{\ell}$}}}$}}

    We now consider  the decays corresponding to the quark level
process $b {\mbox{$\rightarrow$}} u {\mbox{$\ell \bar \nu_{\ell}$}}$.  These decays are very
important in the determination of the $|V_{ub}|$ matrix element, which is itself important for CP
violation in the Standard Model. Here large recoils are available; as a result
we are not surprised to find large 
contributions from the
1P and 2S states in our lepton spectra. 
In ISGW the decays to the radially excited pseudoscalars $n^1S_0$ were
explicitly checked to confirm that the calculation would converge to
the inclusive rate in the appropriate limit. As in ISGW, however, ISGW2
only sums over the low-lying resonances and so for $b \rightarrow u$
decays it can be used as a model for the inclusive spectrum only in the
endpoint region. This point is discussed at greater length in Section VI.

\subsubsection
{$\bar B^0 \rightarrow X_{u\bar d} \ell \bar \nu_{\ell}$ {\rm and} 
$\bar B^- \rightarrow X_{u\bar u} \ell \bar \nu_{\ell}$}

We consider both the decays ${\mbox{$\bar B$}}^0 {\mbox{$\rightarrow$}} X_{u\bar d}{\mbox{$e \bar \nu_e$}}$ and
${\mbox{$B$}}^- {\mbox{$\rightarrow$}} X_{u\bar u} {\mbox{$e \bar \nu_e$}}$ which are shown in Figure
10a) and b), respectively.  Detailed partial widths are given
in Table VI.  As in ISGW, there are large contributions from the
$1P$ and $2S$ states.  On comparing  with the results of ISGW, 
one sees that our more realistic
form factors have increased the rate to 
the sum of the rates to the $1S$, $1P$, and $2S$ states by about $25\%$ and
transferred some of the rate from the heavier to the lighter states. ISGW2 therefore
predicts a somewhat hardened
endpoint spectrum relative to ISGW.
The change in individual exclusive rates is most
pronounced for the pion, which has increased by about a factor of four
over the ISGW result.
As discussed above, and as is apparent from  Fig.~1, ISGW was
designed to produce a conservative estimate of the endpoint rate.  
The effect on the $\pi$ rate is uncharacteristic
since it vanishes for kinematic reasons
at zero recoil where the ISGW form factor is nearly equal to ours,
and grows into the high recoil region where their form factor is far below ours and
the measured pion form factor.  The consequent large uncertainty in the $\pi$
rate is compounded by the potential effect (to be discussed below) of the nearby
$\bar B^*$ pole. In contrast, the ISGW2 rate to the $\rho$ is
only about 70\% larger than that of ISGW. A similar increase is obtained for the total
rate to the $1P$ states, 
while the rate to the 
radial excitations of the $\pi$ and the
$\rho$ decreased by almost a factor of three.  We also note that the ratio
$\Gamma_L/\Gamma_T$ for ${\mbox{$\bar B$}} {\mbox{$\rightarrow$}} \rho {\mbox{$e \bar \nu_e$}}$ has remained equal to the
ISGW value of 0.30, even though this value is sensitive to the method used to treat
large recoils and the axial current form factors.

    As in ISGW, we have included all
resonance states with masses   $\le 1.7$ GeV, which implies that our
lepton spectra are complete for lepton energies greater than about 2.4 GeV.  In this
region our spectra are considerably softer than that of the free quark decay.
Since our sum over exclusive channels is incomplete, we cannot quote a total
rate, although the treatment of the pion radial excitations described in ISGW
suggests that it will be within a factor of 
two of our free quark rate of  $1.28 \times 10^{14} |V_{ub}|^2 \sec^{-1}$.

\subsubsection
{${{\mbox{$\bar B_s$}} {\mbox{$\rightarrow$}} X_{u \bar s} {\mbox{$\ell \bar \nu_{\ell}$}}}$}

Our results for ${\mbox{$\bar B_s$}} {\mbox{$\rightarrow$}} X_{u \bar s} {\mbox{$e \bar \nu_e$}}$ are shown in
Figure 11, with the explicit partial widths in
Table VII.  As expected, this decay is very similar to that of the
$\bar B$-meson.  There is, however,
a noticeable softening of the spectrum due to the heavier spectator.

\subsubsection
{${\bar B_c \rightarrow X_{u \bar c} \ell \bar \nu_{\ell}}$}

Our results for $\bar B_c {\mbox{$\rightarrow$}} X_{u \bar c} {\mbox{$e \bar \nu_e$}}$ are shown in
Figure 12.  The explicit partial widths are given in
Table VII.  This decay is similar to the other
$b {\mbox{$\rightarrow$}} u {\mbox{$e \bar \nu_e$}}$ decays. However,
the softening of the spectrum due to increased spectator mass is much more
pronounced, as is the shifting of probability to states with masses above
those of our calculation.

\vfill\eject

\section
{\bf Comparison to Heavy Quark Symmetry}

   In this section we compare our results to those of Heavy Quark 
Symmetry, which provides  model-independent predictions 
for some aspects of
the weak hadronic
matrix elements presented here. While these model-independent
predictions are very interesting theoretically, the sizes of the
corrections to this limit may
restrict its validity to a limited number of processes, or to a small region of
phase space. One way to estimate the effects of such $\Lambda_{\rm QCD} /{m_q}$
corrections to these limiting predictions is to compute these corrections in a model such as 
ISGW2. ISWG2 is in fact most  reliable
precisely at the key zero-recoil point of Heavy Quark Symmetry,
and indeed our form factors reduce to those required when taken
to the symmetry limit. Away from this limit our results constitute model-dependent predictions
for the effects of the finite quark masses. 
Estimates of such corrections from other hadronic models and
from quenched lattice QCD have also been made [40].

   The predictions of Heavy Quark Symmetry 
for the decays ${P_{\! \scriptscriptstyle Q}} {\mbox{$\rightarrow$}} {P_q} \ell \bar \nu_{\ell}$ and ${P_{\! \scriptscriptstyle Q}} {\mbox{$\rightarrow$}} {V_q} \ell \bar \nu_{\ell}$
were first worked out in Refs. [4].
The relationship between our form
factors 
and those of Heavy Quark Symmetry described by eqs. (3) and (4) which are defined
in terms of four-velocity variables is
\begin{eqnarray}
   \tilde{f_+} & = & \frac{1}{2} \left( \sqrt{\frac{m_{{P_{\! \scriptscriptstyle Q}}}}{m_{{P_q}}}} +
             \sqrt{\frac{m_{{P_q}}}{m_{{P_{\! \scriptscriptstyle Q}}}}} \right) f_+ +
             \frac{1}{2} \left( \sqrt{\frac{m_{{P_{\! \scriptscriptstyle Q}}}}{m_{{P_q}}}} - \sqrt{
             \frac{m_{{P_q}}}{m_{{P_{\! \scriptscriptstyle Q}}}}} \right) f_- \\
   \tilde{f_-} & = & \frac{1}{2}\left( \sqrt{\frac{m_{{P_{\! \scriptscriptstyle Q}}}}{m_{{P_q}}}} -
             \sqrt{\frac{m_{{P_q}}}{m_{{P_{\! \scriptscriptstyle Q}}}}} \right) f_+ +
             \frac{1}{2} \left( \sqrt{\frac{m_{{P_{\! \scriptscriptstyle Q}}}}{m_{{P_q}}}} +
             \sqrt{\frac{m_{{P_q}}}{m_{{P_{\! \scriptscriptstyle Q}}}}} \right) f_-                \\
   \tilde{g} & = & 2\sqrt{m_{{P_{\! \scriptscriptstyle Q}}} m_{{V_q}}}\, g \\
   \tilde{f} & = & \frac{f(1+w)^{-1}}{\sqrt{m_{{P_{\! \scriptscriptstyle Q}}} m_{{V_q}}}} \\
   \tilde{a}_+ + \tilde{a}_- & = & -~\frac{m_{{P_{\! \scriptscriptstyle Q}}}^2}{\sqrt{m_{{P_{\! \scriptscriptstyle Q}}}
             m_{{V_q}}}} \left( a_+ + a_-\right) \\
   \tilde{a}_+ - \tilde{a}_- & = & -~\sqrt{m_{{P_{\! \scriptscriptstyle Q}}} m_{{V_q}}}\left( a_+ -
             a_-\right) .
\end{eqnarray}
Recall from Section III.A that since these relations involve the physical hadron masses
$m_H$, and not the hyperfine-averaged masses $\bar m_H$, these $\tilde f_i$ differ from the
$\tilde f_i^{qm}$ calculated directly at the quark model level by $1/m_Q^2$ terms.

   The $1/m_q$ and $1/m_Q$ corrections to these predictions have been
considered by various authors [41-45].  In particular, the first of 
Refs. [45] gives a general form for such
corrections which is, as we will see, particularly suited to our quark model, namely
\vskip 0.2in
\begin{eqnarray}
   \frac{\tilde{f}_+}{\xi(w)} & = &  1 + \frac{\rho_1(w)}{\mu_+} \\
   \frac{\tilde{f}_-}{\xi(w)} & = & \frac{1}{\mu_-} \left( -\frac{\bar \Lambda}{2} + \rho_4(w) \right)  \\
   \frac{\tilde{g}}{\xi(w)} & = & 1 +\frac{\bar \Lambda}{2\mu_+} + \frac{1}{{m_q}}\rho_2(w) + \frac{1}{{m_{\scriptscriptstyle Q}}} \left(
     \rho_1(w) - \rho_4(w) \right) \\
   \frac{\tilde{f}}{ \xi(w)}  & = & 1 + \frac{\bar \Lambda}{2\mu_+}
      \left(\frac{w-1}{w+1}\right)
      + \frac{1}{{m_q}} \rho_2(w) + \frac{1}{{m_{\scriptscriptstyle Q}}} \left( \rho_1(w) - \frac{w-1}{w+1} \rho_4(w)\right)
       \label{eq:hqetmf} \\
   \frac{(\tilde{a}_+ +\tilde{a}_- )}{\xi(w)} & = & -~\frac{1}{(w+1)}\frac{1}{{m_q}}
     \left( \bar \Lambda - (w+1)\rho_3(w) + \rho_4(w)\right) \\
    \frac{(\tilde{a}_+ - \tilde{a}_- )}{\xi(w)} & = & 1 +
    \frac{\bar \Lambda}{2}
     \left(\frac{w-1}{w+1} \frac{1}{{m_q}} + \frac{1}{{m_{\scriptscriptstyle Q}}} \right) + \frac{1}{{m_q}}
     \left(\rho_2(w) - \rho_3(w) - \frac{1}{w+1}\rho_4(w)\right)  \nonumber \\
   & & \mbox{} + \frac{1}{{m_{\scriptscriptstyle Q}}}\left(\rho_1(w) - \rho_4(w)\right)
\end{eqnarray}
\vskip 0.2in
\noindent where $\bar \Lambda $ is a constant and $1/\mu_{\pm} \equiv 1/m_q \pm 1/m_Q$.  
The inclusion of these effects thus
results in the appearance of four additional unknown functions $\rho_n(w)$
with unknown normalizations (although it can be shown that $\rho_1(1)=\rho_2(1)=0$).
Alternative parameterizations have also been given; we will comment on one of these below. 

   We can map onto our results by expanding them to leading order
in $1/m_q$ and $1/{m_{\scriptscriptstyle Q}}$; we find

\begin{eqnarray}
  \frac{\tilde{f}_+^{\rm qm}}{\xi} & = & 1-{{R_{P}(w)} \over \mu_+}  \\
  \frac{\tilde{f}_-^{\rm qm}}{\xi} & = & -\frac{m_{sp}}{2\mu_-}  \\
  \frac{\tilde{g}^{\rm qm}}{\xi} & = &  1 + \frac{m_{sp}}{2\mu_+} -{{R_{V}(w)} \over m_q} 
-{{R_{P}(w)} \over m_Q}    \\
  \frac{\tilde{f}^{\rm qm}}{\xi} & = & 1 -{{R_{V}(w)} \over m_q} 
-{{R_{P}(w)} \over m_Q} +\left( \frac{w-1}{w+1} \right) \frac{m_{sp}}{2\mu_+}\\
  \frac{\tilde{a}_+^{\rm qm} + \tilde{a}_-^{\rm qm}}{\xi} & = &
     -~\frac{m_{sp}}{m_q(1+w)}
\end{eqnarray}
and
\begin{eqnarray}
  \frac{\tilde{a}_+^{\rm qm} - \tilde{a}_-^{\rm qm}}{\xi}
     & = & 1 + \frac{m_{sp}}{2\mu_+} -\frac{m_{sp}}{m_q(1+w)} -{{R_{V}(w)} \over m_q} 
-{{R_{P}(w)} \over m_Q} ~~.
\end{eqnarray}
Here
\begin{equation}
   R_{P(V)}(w) \equiv \frac{(w-1)[\frac{1}{4} \delta_{sp}+ \frac{m_{sp}^2}{2\beta_{sp}^2}
\lambda_{Psp(Vsp)}]}{1+\frac{1}{2}\rho^2(w-1)}
\end{equation}
with $\delta_{sp} \equiv \bar m_{P_{Qsp}V_{Qsp}} -m_Q$ as $m_Q \rightarrow \infty$
and with the $\lambda$'s parameterizing
the approach of the $\beta$'s to the heavy quark limit {\it via}
\begin{equation}
    \beta_{P(V)_{Qsp}}^2 \equiv \beta_{sp}^2 \left(1-\frac{2\lambda_{Psp(Vsp)}}
{\tilde m_{Qsp}} \right) ~~,
\end{equation}
with $\lambda_{Psp}=k_{sp}-\frac{3}{4}h_{sp}$ and $\lambda_{Vsp}=k_{sp}+\frac{1}{4}h_{sp}$.
Here $k_{sp}$ describes the perturbation of $\beta^2_{Qsp}$ in a heavy quark meson with heavy quark $Q$ 
and a spectator $sp$ due to the heavy quark kinetic energy (which is spin-independent)
and $h_{sp}$ is the analogous perturbation due to the residual hyperfine interaction of $Q$
and $sp$.
From Table A2 one can see that
$k_d \simeq +0.14  $ GeV,
$h_d \simeq +0.36  $ GeV,
$k_s \simeq +0.26  $ GeV, and
$h_s \simeq +0.50  $ GeV. 
Using the measured masses
and the constituent quark masses of Table A1, and correcting for the residual
heavy quark kinetic energy, one can estimate that $\delta_d
\simeq 0.09$ GeV, and $\delta_s \simeq 0.17$ GeV.

     This decomposition allows us to identify
\begin{eqnarray}
   \bar \Lambda &=&m_{sp} \\
   \rho_1(w)    &=& -R_{P}(w) \\
  \rho_2(w)    &=& -R_{V}(w) \\
  \rho_3(w)    &=& 0 \\
  \rho_4(w)    &=& 0 ~~~,
\end{eqnarray}
from which one can easily see that the predicted corrections to the Heavy Quark
Symmetry limit are all of modest size. This assessment is made quantitative
by Table VIII
for $b {\mbox{$\rightarrow$}} c$ and $c {\mbox{$\rightarrow$}} s $ decays with an up or
down spectator. 
In terms of the common $\psi_1$, $\psi_2$, $\psi_3$, $\psi_+$ parameterization [44]
of $1/m_Q$ and $1/m_q$ corrections, these results are $\psi_1=-({R_P+3R_V}/{2 \bar \Lambda}) \xi$,
$\psi_2=0$, $\psi_3=-( {R_V-R_P}/{4 \bar \Lambda}) \xi$, and 
$\psi_+=-( {w-1}/{w+1}) \xi$. Note that, as required, $\psi_1$ respects heavy quark spin symmetry,
while $\psi_3$ is responsible for breaking it.
Conclusions similar to ours (couched in terms of the $\psi$-parameterization)
have been reached previously in the quark model [7].

    As an aside, let us note that if we focus on $\bar B \rightarrow D^* \ell \bar \nu_{\ell}$
transitions alone, and assume {\it only} that $\rho_3(w)=\rho_4(w)=0$, we may define 
a ``preasymptotic Isgur-Wise function" $\xi_{D^*\bar B}(w)$ such that

\begin{eqnarray}
   \frac{\tilde{g}}{\xi_{D^*\bar B}(w)} & = & 1 +\frac{\bar \Lambda}{2\mu_+} \\
   \frac{\tilde{f}}{ \xi_{D^*\bar B}(w)}  & = & 1 + \frac{\bar \Lambda}{2\mu_+}
      \left(\frac{w-1}{w+1}\right)\\  
   \frac{(\tilde{a}_+ +\tilde{a}_- )}{\xi_{D^*\bar B}(w)} & = &-~ \frac{\bar \Lambda}{m_q(w+1)}\\
   \frac{(\tilde{a}_+ - \tilde{a}_- )}{\xi_{D^*\bar B}(w)} & = &1 +
    \frac{\bar \Lambda}{2\mu_+} -\frac{\bar \Lambda}{m_q(w+1)}  ~~.
\end{eqnarray}
Under this assumption, therefore, the predictions of Heavy Quark Symmetry to order $1/m_Q$ and
$1/m_q$ can be described by {\it one} unknown parameter $\bar \Lambda$ known to be approximately
$m_{sp}$ and the unknown shape of the ``preasymptotic Isgur-Wise function" $\xi_{D^*\bar B}(w)$
which retains its normalization to unity at $w=1$. We also note that the measured slope of
this function is predicted by our model to be
\begin{equation}
\rho^2_{D^*\bar B} \simeq 0.74
\end{equation}
which value includes a contribution
\begin{equation}
\Delta \rho^2_{pert} \simeq \frac{16}{81} {\it ln} \left[ \frac{\alpha_s(\mu_{qm})}{\alpha_s(m_c)}\right]
\simeq 0.13~.
\end{equation}
(The numerical value of $\Delta \rho^2_{pert}$ is quite uncertain: it depends on
the leading logarithmic expansion in $m_c/\mu_{qm}$ and 
on the assumption that $\alpha_s(\mu_{qm})$, 
where $\mu_{qm}$ is the ``quark model scale", is the ``frozen-out" value
$\alpha_s=0.6$ from Appendix A.
We accordingly assign a theoretical error of $\pm 0.05$ to it.)

Another important set of predictions [10] of Heavy Quark Symmetry are those
which relate, e.g., the form factors of $\bar B\rightarrow \rho \ell \bar \nu_{\ell}$ 
to those of $D \rightarrow
\rho \ell^+ \nu_{\ell}$.  These predictions could play a vital role in determining
$V_{ub}$ if corrections to the symmetry limit are not too severe.  With ISGW2 we
can check these relations.  For example, in the ideal symmetry limit one should
have 
\begin {equation}
\alpha_s(m_b)^{-a_I(m_b)}
{f^{\bar B\rightarrow \rho}(p_{\rho} \cdot v_B) \over 2 \sqrt{m_{\rho} m_B}}=
\alpha_s(m_c)^{-a_I(m_c)}
{f^{D\rightarrow \rho}(p_{\rho} \cdot v_D) \over 2 \sqrt{m_{\rho} m_D}}
\end {equation}
where $p_\rho$ and $v_P$  are the four momentum of the $\rho$ and the four
velocity of the decaying meson $P$ and $a_I(m_Q)$ is given by eq. (7) 
with $N_f$ appropriate to $m_Q$. (Note that $m_\rho$ has no 
special significance in these formulas:  we are
simply using it to create dimensionless quantities. Also note that
we have removed the known quark mass dependence of the leading logarithmic matching
condition, but not attempted to remove the mass dependence contained in the
$\alpha_s/ \pi$ corrections
since, while relatively weak given that $\alpha_s(\mu_{ub}) \simeq \alpha_s(\mu_{dc})$,
it is model-dependent.)  
We find, {\it e.g.}, that at zero recoil
\begin {eqnarray}
\alpha_s(m_b)^{-a_I(m_b)}{f^{B\rightarrow \rho}(m_\rho) \over 2 \sqrt{m_{\rho} m_B}}= 0.49 \\ \cr
\alpha_s(m_c)^{-a_I(m_c)}{f^{D\rightarrow \rho}(m_\rho) \over 2 \sqrt{m_{\rho} m_D}}= 0.45.
\end {eqnarray}
  One also expects
\begin {equation}
2\alpha_s(m_b)^{-a_I(m_b)}\sqrt{m_\rho m_B} g^{B\rightarrow \rho}= 
2\alpha_s(m_c)^{-a_I(m_c)}\sqrt{m_\rho m_D} g^{D
\rightarrow \rho}
\end{equation} 
while our model predicts (once again at zero recoil)
\begin{eqnarray}
2\alpha_s(m_b)^{-a_I(m_b)}\sqrt{m_\rho m_B} g^{B\rightarrow \rho}(m_{\rho})&=& 1.16 \\ \cr
2\alpha_s(m_c)^{-a_I(m_c)}\sqrt{m_\rho m_D} g^{D \rightarrow \rho}(m_{\rho})&= &1.15.
\end {eqnarray}
The form factors $f_{+}$ and $a_{+}$ are more complex since it is the
combinations $f_{+}\pm f_{-}$ and $a_{+}\pm a_{-}$ which obey simple scaling
relations.  However, since the objects which scale are $\sqrt{{m_B\over m_\rho}}
(f_{+}+ f_{-}), \ \sqrt{{m_\rho \over m_B}}(f_{+}-f_{-}), \ m_B \sqrt{{m_B\over
m_\rho}}(a_{+}+a_{-}), \ {\rm and}\ \sqrt{m_Bm_\rho}(a_{+}-a_{-}),$ in the
heavy quark limit $f_{-}=-f_{+}$ and $a_{-}=-a_{+}$ so that in fact the simple
scaling laws
\begin{equation}
\alpha_s(m_b)^{-a_I(m_b)}\sqrt{m_\rho \over m_B} f^{B\rightarrow \pi}_{+}=
\alpha_s(m_c)^{-a_I(m_c)}\sqrt{m_\rho \over m_D}
f^{D\rightarrow \pi}_{+}
\end{equation} 
and
\begin{equation}
2\alpha_s(m_b)^{-a_I(m_b)}\sqrt{m_\rho m_B} a^{B\rightarrow \rho}_{+}=  
2\alpha_s(m_c)^{-a_I(m_c)}\sqrt{m_\rho m_D} a^{D\rightarrow
\rho}_{+}
\end{equation}
emerge.  Our model in fact gives 
\begin{eqnarray}
\alpha_s(m_b)^{-a_I(m_b)}\sqrt{m_\rho \over m_B} f^{B\rightarrow \pi}_{+}(m_{\pi})=0.68 \\ \cr
\alpha_s(m_c)^{-a_I(m_c)}\sqrt{m_\rho \over m_D} f^{D\rightarrow \pi}_{+}(m_{\pi})=0.66 
\end{eqnarray}
and
\begin{eqnarray}
2\alpha_s(m_b)^{-a_I(m_b)}\sqrt{m_\rho m_B} a^{B\rightarrow \rho}_{+}(m_{\rho})=-0.66 \\ \cr
2\alpha_s(m_c)^{-a_I(m_c)}\sqrt{m_\rho m_D} a^{D\rightarrow \rho}_{+}(m_{\rho})=-0.60. 
\end {eqnarray}
We conclude that our model strongly supports the conclusion that $1/ m_Q$
effects will not obscure the extraction of $V_{ub}$  for exclusive $B$ decays
via the scaling relations of Heavy Quark Symmetry so that the proposal [10] to
do so appears to be sound. 
(It should be noted that in the case of $f^{B\rightarrow \pi}_{+}$
and $f^{D\rightarrow \pi}_{+}$, our quark model contributions at zero recoil must
be supplemented by the $B^*$ and $D^*$ pole terms [46], respectively, before they may
be compared to experiment. These pole terms carry with them large but known $1/m_Q$ effects
related to the smallness of $m_{\pi}$ relative to the $B^*-B$ and $D^*-D$ hyperfine
splittings.)

Similar conclusions follow for the validity of relations between
$c \rightarrow s$ and $b \rightarrow s$ matrix elements which enter into
the prediction of exclusive $b \rightarrow s \gamma$ decays.

\vfill\eject

\section{\bf Comparison to Experiment}

\subsection{Magnetic Dipole Decays}

    Magnetic dipole decays of
mesons like $\omega \rightarrow \pi \gamma$, $K^* \rightarrow K \gamma$, and $\psi \rightarrow
\eta_c \gamma$ proceed through a transition magnetic dipole moment form factor which is
precisely analogous to the vector current form factor $g$ in weak decays of ground state 
pseudoscalar mesons to ground state vector mesons. The ability of our model to describe such decays is
therefore relevant to the reliability of the model for the weak decays which are the focus of this paper.
For the transition magnetic dipole moment $\mu_{PV}=\mu_{VP}$ underlying the transition 
$V \rightarrow P \gamma$ (or $P \rightarrow V \gamma$ when {\it it} is energetically
allowed), theory (experiment [33]) gives, in units of the nucleon magneton,
$\mu_{\pi \rho}=0.52~~( 0.69\pm 0.04 )$,
$\mu_{\pi \omega}=1.56~~ (2.19 \pm 0.09)$,
$\mu_{\pi \phi}=0.07~~ (0.13 \pm 0.01)$,
$\mu_{\eta \rho}=2.16~~ (1.77 \pm 0.17)$,
$\mu_{\eta \omega}=0.68~~ (0.57 \pm 0.07)$,
$\mu_{\eta \phi}=0.61~~ (0.66 \pm 0.02)$,
$\mu_{\rho \eta '}=1.53~~ (1.20 \pm 0.08)$,
$\mu_{\omega \eta '}=0.58~~ (0.42 \pm 0.04)$,
$\mu_{\eta ' \phi}=-0.94~~ (\vert \mu_{\eta ' \phi} \vert <1.8 )$,
$\mu_{K^+K^{*+}}=0.95~~ (0.79 \pm 0.03)$,
$\mu_{K^0K^{*0}}= -1.27~~(-0.98 \pm 0.26)$, and
$\mu_{\eta_c \psi}= 0.76~~(0.55 \pm 0.12)$.
As in the main calculations we have taken the pseudoscalar mixing angle here to be $-20^{\circ}$;
we have also assumed that the vector mixing angle is $39^{\circ}$.

    We conclude from this comparison that the quark model will probably be able to predict 
the  form factor $g$ with the typical quark model accuracy of $\pm 25\%$ for transitions involving
light quarks. Since Heavy Quark Symmetry guarantees that our formulas for $g$ 
will be correct in the heavy quark limit, this should be an upper bound to the probable error in 
such predictions.

\subsection{$K {\mbox{$\rightarrow$}} \pi {\mbox{$\ell \bar \nu_{\ell}$}}$}

  Although the form factors for these decays are usually referred to the $SU(3)$
symmetry normalization point $t=0$, we prefer to refer them to the point $t=t_m$ where 
Heavy Quark Symmetry will develop. 
We find that $f_+(t_m) = 1.04$ and 
$f_-/f_+ = -0.28$. The latter  is in reasonable agreement with the measured value 
[33] though there is a substantial uncertainty
since $K^{\pm}$ decay gives $f_-/f_+ = -0.35 \pm 0.15$ while $K^0_L$ decay 
gives $-0.11 \pm  0.09$.  Our equation for $f_+$ is
consistent with the Ademollo-Gatto theorem [47] which protects
$f_+$ from substantial deviations from unity. Our prediction for $f_+(t)$ can be compared
with the ``standard" [48,49] used to extract $V_{us}$ from these decays. If we convert to the linearized form
$f_+(t)=f_+(0) \bigl[1+\frac{1}{6} r^2_{\pi K} t \bigl]$, then we predict
$f_+(0)=0.93$ and $r_{\pi K}=0.48$ $fm$ versus the ``standard" [49] 
$f_+(0)=0.97 \pm 0.01$ and $r_{\pi K}=0.53$ $fm$ corresponding to $K^*$ pole dominance.
The best current fit value to this transition radius gives [33] $r_{\pi K}=0.59 \pm 0.02$ $fm$.

\subsection{Meson Decays through $b {\mbox{$\rightarrow$}} c {\mbox{$\ell \bar \nu_{\ell}$}}$} 

    Our results for semileptonic meson decays involving the quark level
decay $b {\mbox{$\rightarrow$}} c {\mbox{$\ell \bar \nu_{\ell}$}}$ were given in Section IV.A. Their
relatively low recoil and  heavy quark masses provide a theoretical stability that makes them our 
most reliable predictions. 

    Reviews of the experimental status of
semileptonic $B$ meson decays can be found in   Refs. [19].
From the measured rate (here we use the latest  CLEO result [21])
    
\begin{equation}
   {\Gamma({\mbox{$B$}} {\mbox{$\rightarrow$}} {\mbox{$D$}}^* {\mbox{$\ell \bar \nu_{\ell}$}}) } = 
    2.99 \pm 0.39 \times 10^{10} sec^{-1},
\end{equation}
and our prediction $\Gamma({\mbox{$B$}} {\mbox{$\rightarrow$}} {\mbox{$D$}}^* {\mbox{$\ell \bar
\nu_{\ell}$}})=2.48
\times 10^{13} 
\vert V_{cb} \vert^2 sec^{-1}$ we  obtain 

\begin{equation}
        \vert V_{cb} \vert = 0.035  \pm 0.002 .
\end{equation}
The measured rate for ${\mbox{$\bar B$}} {\mbox{$\rightarrow$}} {\mbox{$D$}} {\mbox{$\ell \bar
\nu_{\ell}$}}$ is 

\begin{equation}
   \Gamma({\mbox{$\bar B$}} {\mbox{$\rightarrow$}} {\mbox{$D$}} {\mbox{$\ell \bar \nu_{\ell}$}}) = 
1.3 \pm 0.3 \times 10^{10} \sec^{-1}~.
\end{equation}
Using our predicted rate of $\Gamma({\mbox{$\bar B$}} {\mbox{$\rightarrow$}} {\mbox{$D$}}
{\mbox{$\ell \bar \nu_{\ell}$}}) = 1.19 \times 10^{13} |V_{cb}|^2 \sec^{-1}$ implies that

\begin{equation}
   |V_{cb}| = 0.033 \pm 0.004 .
\end{equation}
The consistency between eqs. (74) and (76) of course means that the model correctly
predicts the ratio of the rates to $D$ and $D^*$. However, these determinations of 
$|V_{cb}|$ depend on the prediction of the recoil dependence of 
the relevant form factors and so have a theoretical error of
order $10\%$. A comparison with data near zero recoil using Heavy Quark 
Symmetry, as has become standard, remains the reliable way to determine $|V_{cb}|$.

   The measurements of $\Gamma_L/\Gamma_T$ for ${\mbox{$B$}} {\mbox{$\rightarrow$}}
{\mbox{$D$}}^*{\mbox{$e
\bar \nu_e$}}$ are quite sensitive to the relative  importance of the $f$, $g$, and $a_+$ form
factors. Experiment gives
\begin{equation}
   \frac{\Gamma_L}{\Gamma_T} = \left\{ \begin{array}{l}
      0.85 \pm 0.45 \mbox{ (ARGUS [21])} \\
      0.83 \pm 0.33 \pm 0.13 \mbox{ (CLEO [21])} \\
      1.24 \pm 0.16          \mbox{ (CLEO [50])}  \end{array} \right.
\end{equation}
consistent with our prediction of 1.04 (versus 0.97 for ISGW).
Furthermore, the predictions of both ISGW and ISGW2 for the 
$q^2$ dependence of ${\mbox{$\bar B$}}^0 {\mbox{$\rightarrow$}} {\mbox{$D$}}^{*\,+} {\mbox{$\ell \bar
\nu_{\ell}$}}$ agrees reasonably well with the measured results: see  Refs.~[19].
In particular, we predict that the slope of the preasymptotic Isgur-Wise function
$\xi_{D^*\bar B}(w)$ will be $\rho_{D^*\bar B}^2=0.74 \pm 0.05$ (see the text below eq. (60) for
an explanation of the theoretical error), while the latest fits to the data
(see the last of Refs. [21]) 
give for the closely related quantity $\hat \rho^2$ the value
$0.84 \pm 0.14$. (The ISGW prediction was 0.69.) 
Table IX shows our predictions for the individual form factors in terms of
the HQS form factors defined in Section V. It also compares them with the
predictions of
ISGW,  Heavy Quark 
Symmetry, and HQET (with matching but no $1/m_Q$ corrections). This comparison
illustrates the relatively model-independent nature of these predictions. A recent
measurement [50] gives for $\bar B \rightarrow D^* \ell \bar \nu_{\ell}$ decay
\begin{eqnarray}
\frac{g}{f} = 0.031 \pm 0.009(stat) \pm 0.004(syst) ~GeV^{-2} \\ \cr
\frac{a_+}{f} = - 0.015 \pm 0.006(stat) \pm 0.003(syst) ~GeV^{-2}
\end{eqnarray}
in reasonable agreement with our predictions of $0.030~GeV^{-2}$ and $-0.024~GeV^{-2}$,
respectively.

   Both  CLEO and ARGUS currently find indications that the $D$ and
$D^*$ final states account for much less than 
all of the  semileptonic decay width of the $\bar B$ meson.  We
predict that these final states account for $\simeq 90$\% of the total rate
to the states included in our calculation.  If confirmed,
these  experimental results may
indicate that non-resonant processes are important, or, perhaps, that we have
underestimated the effects of the 1P and 2S states. Note that the Bjorken sum
rule [9] requires that the rate missing from the $D$ and $D^*$ channels
be approximately proportional to $\rho^2$. Thus doubling the missing rate
would require doubling $\rho^2$, in apparent contradiction to the existing agreement
between theory and experiment described above.

    Finally we note that the decays of the $\bar B_d$, $\bar B_s$, and $\bar B_c$
sequence show a marked departure from the spectator approximation in which
their inclusive semileptonic decay rates would all be equal. This phenomenon, which is
more pronounced in the $c \rightarrow s$ decays, is addressed in the next subsection.

\subsection{Meson Decays through $c {\mbox{$\rightarrow$}} s {\mbox{$\ell^+ \nu_{\ell}$}}$} 

    The quark level decays $c {\mbox{$\rightarrow$}} s {\mbox{$\ell^+ \nu_{\ell}$}}$ are at this time
better measured than the $b {\mbox{$\rightarrow$}} c {\mbox{$\ell \bar \nu_{\ell}$}}$ decays.  They
also provide a greater challenge for our model since in these decays Heavy Quark
Symmetry does not guarantee the success of the leading approximation
to their form factors: strange quarks do not qualify as heavy
quarks! Note that since the CKM matrix element $|V_{sc}|$
may be related to $|V_{ud}|$ {\it via} the unitarity of the CKM matrix,  direct
measurements of the form factors can be made.  
The experimental status of weak charmed meson decays was recently reviewed in Refs. [23].

    Averaging over measurements [24]  and using isospin gives [23]

\begin{equation}
  \Gamma(D \rightarrow \bar K \ell \nu_{\ell})=  9.0 \pm 0.5  \times 10^{10}
             \sec^{-1}
\end{equation}
which compares reasonably well to our prediction of

\begin{equation}
  \Gamma ({\mbox{$D$}} {\mbox{$\rightarrow$}} K {\mbox{$\ell^+ \nu_{\ell}$}}) = 10.0 \times 10^{10}
\sec^{-1} .
\end{equation}
In this decay one can also measure the pole mass for the $f_+$
form factor assuming a monopole shape.  CLEO obtains
$M_{\rm pole}^{f_+} = 2.00 \pm 0.12 \pm 0.18$ GeV, consistent with earlier experiments
but with smaller errors.  This mass corresponds to a transition radius
$r_{KD}=\frac{\sqrt{6}}{M_{\rm pole}^{f_+}}=0.24\pm0.03$ $fm$ compared to our prediction
of $0.22$ $fm$. The data cannot currently
distinguish between the common choices (monopole, dipole, exponential)
for the shape of this form factor as the available range of $t_m-t$ is limited
and all these shapes give an
approximately linear dependence over this  range.

    Assuming the measured form factor, the
rate may be transformed [23] into a measurement of

\begin{equation} 
  f_+(t_m) = 1.42 \pm 0.25
\end{equation}
(or equivalently $f_+(0) = 0.75 \pm 0.03$).
We predict $f_+(t_m) = 1.23$  (or equivalently
$f_+ (0) = 0.85$ using our predicted $t$
dependence and 0.80 using our form factor with the central experimental value of $r_{KD}$. 

    As an aside, we would like to explain why such form factor measurements should 
be referred to $t=t_m$ and not $t=0$. Heavy Quark Symmetry establishes that
heavy to light transition form factors are all related in the region of $t_m$ [4,10], 
{\it i.e.}, are independent of the heavy quark mass $m_Q$ as $m_Q \rightarrow \infty$
when scaled by an appropriate power of $m_Q$. Measurements near $t_m$ are therefore determinations
of universal transition form factors (up to $1/m_Q$ corrections). Form factors at $t=0$ are, in 
contrast, ``random numbers" since they are the product of the universal 
transition amplitudes relevant at 
$t_m$ and a complicated dynamical function which depends on the microscopic details
of the high momentum tails of the initial and final state wavefunctions. This is because
$t=0$ corresponds to a final state $X$ recoiling with {\it maximum} three momentum
$\vert \vec p_X \vert = {{m_{P_Q}^2-m_X^2} \over {2 m_{P_Q}}}$ in the rest frame of
$P_Q$. This momentum increases with $m_Q$ so that $t=0$ form factors are ever-decreasing
functions of $m_Q$.

   The D meson semileptonic decay to the $K^*$ final state
has been the subject of much interest. An early measurement
found a value for $\Gamma_L/\Gamma_T$
approximately two times larger than expected while the ratio of vector to
pseudoscalar branching ratios was about one half what was expected from many
models.  Attempts were made [25] to
accomodate these results within the ISGW model by allowing
for the theoretical uncertainties inherent to the quark model ($\pm$20\%).
It was found that the model
could accommodate the vector to pseudoscalar ratio but not the
$\Gamma_L/\Gamma_T$ ratio with such variations.  The current
experimental situation is more precise with at least two independent
measurements of each quantity.  In addition to the above quantities, 
measurements of the form factors themselves 
have now been made.

    The averaged  experimental measurements of Mark III, CLEO,
E691, ARGUS, E653, and WA82 are [23] 

\begin{equation}
  \frac{\Gamma(D{\mbox{$\rightarrow$}} K^*{\mbox{$e^+ \nu_e$}} )}{\Gamma(D{\mbox{$\rightarrow$}} K{\mbox{$e^+ \nu_e$}})}  =  0.57 \pm 0.08 ~, 
\end{equation}
\begin{equation}
     \Gamma(D{\mbox{$\rightarrow$}} K^*{\mbox{$e^+ \nu_e$}} )=5.1 \pm 0.5 \times 10^{10} \sec^{-1}
\end{equation}
and
\begin{equation}  
\frac{\Gamma_L}{\Gamma_T}  =  1.15 \pm 0.17 ~~~ .
\end{equation}
This compares reasonably well with our model values of

\begin{equation}
  \frac{\Gamma(D{\mbox{$\rightarrow$}} K^*{\mbox{$e^+ \nu_e$}} )}{\Gamma(D{\mbox{$\rightarrow$}} K{\mbox{$e^+ \nu_e$}})}  =  0.54
\end{equation}
\begin{equation}
     \Gamma(D{\mbox{$\rightarrow$}} K^*{\mbox{$e^+ \nu_e$}} )=5.4 \times 10^{10} \sec^{-1}
\end{equation}
and
\begin{equation} 
  \frac{\Gamma_L}{\Gamma_T}  =  0.94 ~~~ .
\end{equation}

   As anticipated in Ref. [25], agreement with the data relative to ISGW has come about
{\it via} a modest shift in the form factor $f$. In fact, {\it four} different effects contribute:
the matching conditions lower $f(t_m)$ by 11\%, $C_f$ from Table I lowers it by about 
7\%, the wavefunction mismatch induced by hyperfine effects (see Table A2) decreases
$f(t_m)$ by another 7\%, while the new factor of $\frac{1}{2} (1+w)$ raises the average
of the amplitude over the Dalitz plot by about 7\%.
For this decay the form factors themselves  have
been determined. 
The comparison of our predictions to the measured results [23] are given in
Table X. Before leaving these decays, we note that (if we subtract our predicted Cabibbo-suppressed
rate) the inclusive Cabibbo-allowed $D$ semileptonic decay rate is measured [33] to be
$(16.2 \pm 1.5) \times 10^{10} sec^{-1}$. This compares favorably with our prediction
of $15.8 \times 10^{10} sec^{-1}$.

    Our predictions for the analogous form factors for ${\mbox{$D_s$}} {\mbox{$\rightarrow$}} \phi
{\mbox{$e^+ \nu_e$}}$ are

\begin{equation}
  \begin{array}{rcl}
   f(t_m)  & = &+2.03~GeV \\
   g(t_m)  & = & +0.52~GeV^{-1} \\
   a_+(t_m) & = &-0.29~GeV^{-1}.
  \end{array}
\end{equation}
These results are in reasonably good agreement with recent measurements [51]
which give (see Ref. [50]) $g(t_m)/ f(t_m) =(+0.20  \pm 0.07)~GeV^{-2}$ 
and  $a_+(t_m)/f(t_m)=(-0.21 \pm 0.05 )~GeV^{-2}$
to be compared with our predictions for these ratios of $+0.26~GeV^{-2}$ 
and $-0.14~GeV^{-2}$, respectively. CLEO [51] also quotes 
$\Gamma (D_s \rightarrow \eta e^+ \nu_e)/\Gamma (D_s \rightarrow \phi e^+ \nu_e)
=1.7 \pm 0.4$
and
$\Gamma (D_s \rightarrow \eta' e^+ \nu_e)/\Gamma (D_s \rightarrow \phi e^+ \nu_e)
=0.7 \pm 0.2$
to be compared with our predictions of $0.8$ and $0.7$, respectively, for a 
pseudoscalar mixing angle of $-20^{\circ}$, and $1.2$ and $0.5$, respectively, for
$-10^{\circ}$. 

    Ref. [51] also presents an extraction of the rate 
$\Gamma (D_s \rightarrow \phi e^+ \nu_e)$ based, among other things,
on the assumption that the {\it inclusive}
semileptonic decay rates of the $D$ and $D_s$ are equal. This 
assumption would appear to be justified on the basis of recent work on
the $1/m_Q$ expansion of inclusive heavy quark semileptonic decays [52-55].
However, it is inconsistent with the results quoted here, which predict that
$\Gamma (D_s \rightarrow X e^+ \nu_e)$ is 27\% smaller than
$\Gamma (D \rightarrow X e^+ \nu_e)$, largely as a consequence of 
the restricted phase space in the $\eta '$ decay of the $D_s$. We speculate that
the unexpectedly [55] large corrections we predict arise from an inapplicability
of the assumptions under which the strong version of the
results of Ref. [52] were derived: since these decays (along with those induced by
the $b \rightarrow c$ transition) are dominated by the lowest few resonances, the
spectral decomposition of the decay is imperfectly described by the smooth partonic
spectral function. (As explained by the authors of Ref. [52], this is analogous
to $R$ in $e^+e^-$ annihilation being smooth and well-approximated by its partonic
value only well above a threshold. We note that 
heavy quark semileptonic decays can be deceptive
in regard to when they are ``well above a threshold" because, while the recoil mass
can kinematically run up to the mass of the decaying quark, the hadronic spectrum
in $Q \rightarrow q \ell \bar \nu_{\ell}$ in fact cuts off at a recoil mass-squared
only of order $\Lambda_{QCD}(m_Q-m_q)^2/m_Q << m_Q^2$ above threshold in the decay
of a $Q \bar d$ meson.) With the assumptions made, Ref. [51] obtains 
$\Gamma (D_s \rightarrow \phi e^+ \nu_e)=(4.4 \pm 0.7) \times 10^{10}$ sec$^{-1}$;
if their assumptions are modified to correspond 
to our predictions from Table III for the ratio of inclusive rates and
for the degree to which $\Gamma (D_s \rightarrow (\eta+\eta '+\phi) e^+ \nu_e)$
saturates $\Gamma (D_s \rightarrow X e^+ \nu_e)$, this extracted rate would be changed to
$(3.5 \pm 0.5) \times 10^{10}$ sec$^{-1}$. These results are both roughly
consistent with our prediction that 
$\Gamma (D_s \rightarrow \phi e^+ \nu_e)= 4.6 \times 10^{10}$ sec$^{-1}$.

\subsection{Meson Decays through $c {\mbox{$\rightarrow$}} d {\mbox{$\ell^+ \nu_{\ell}$}}$}

\bigskip
   The Cabibbo-suppressed charmed meson decays via the quark-level
process $c \rightarrow d  \ell^+ \nu_{\ell}$ have taken on
an enhanced importance recently. 
As described in Section V, Heavy Quark Symmetry relates the form
factors for such decays near $t=t_m$ to their analogues induced by the crucial
$b \rightarrow u \ell \bar \nu_{\ell}$ processes. In the short term the
better measured $c \rightarrow s  \ell^+ \nu_{\ell}$ processes, combined
with $SU(3)$ flavor symmetry, can substitute for these decays, but
precision determinations of $\vert V_{ub} \vert$ will probably require
accurate determinations of the Cabibbo-suppressed form factors.

    Experimental studies of such decays have begun. The decays 
${\mbox{$D$}}^0 {\mbox{$\rightarrow$}} \pi^- {\mbox{$e^+ \nu_e$}}$ and $D^+ \rightarrow \pi^0 e^+
\nu_e$ have  been measured by Mark III and CLEO II, respectively [24]. 
Using the value of $\vert V_{cd}/V_{sc}| \simeq 0.227 \pm 0.003$ which follows from
CKM unitarity [33], their quoted results can be translated into the form
\begin{equation}
    \frac{f_+^{D \rightarrow \pi}(0)}
{f_+^{D \rightarrow \bar K}(0)}
           = 1.17 \pm 0.19
\end{equation}
where a pole model for the $t$-dependence of the form factors has been 
assumed in the determination of the numerical factor, and we have averaged the results of
the two experiments. Our model predicts the value $0.71$ for this ratio.

\subsection{Meson Decays through $b {\mbox{$\rightarrow$}} u {\mbox{$\ell \bar \nu_{\ell}$}}$} 

   Our results for semileptonic meson decays involving the quark level decay
process $b {\mbox{$\rightarrow$}} u {\mbox{$\ell \bar \nu_{\ell}$}}$ were presented above.  Only the
decays with a light spectator have been observed.  Both CLEO [20]
and ARGUS [20] have observed leptons in the 2.4--2.6 GeV energy
range that can be populated only by leptons from a $b {\mbox{$\rightarrow$}} u {\mbox{$\ell \bar
\nu_{\ell}$}}$ process.  In addition, searches for exclusive modes like
$\bar B \rightarrow \pi  {\mbox{$\ell \bar \nu_{\ell}$}}$,
$\bar B \rightarrow \omega  {\mbox{$\ell \bar \nu_{\ell}$}}$, and
$\bar B \rightarrow \rho  {\mbox{$\ell \bar \nu_{\ell}$}}$ have begun.

The extraction of $|V_{ub}|$ from these data is based on kinematics. 
For ${\mbox{$\bar B$}}$ mesons produced at the $\Upsilon(4S)$ resonance, decays via the 
quark level process $b {\mbox{$\rightarrow$}} c {\mbox{$\ell \bar \nu_{\ell}$}}$ have a maximum
lepton energy of  2.4 GeV/c, while the leptons
from a $b {\mbox{$\rightarrow$}} u {\mbox{$\ell \bar \nu_{\ell}$}}$ process may have energies up to
2.6 GeV/c.  Consequently, these inclusive decay
processes can be unravelled  
in the endpoint region of the lepton spectrum. 

    As previously mentioned, the physics of
this endpoint region has been the subject of intense discussion [56]. 
The ISGW papers met with strong criticism by many who argued that its
treatment of the endpoint region was inconsistent with the parton model.
This issue has  recently
been clarified in favor of ISGW by rigorous $1 / m_Q$ expansions of the inclusive rate.
The zeroth order argument was given in Ref. [57]. It is shown there
how, in a $b \rightarrow u \ell \bar \nu_{\ell}$ transition, the zeroth
order lepton spectrum is controlled by quark level kinematics. The key
observation is that decays to low mass {\it hadronic} final states (in this
approximation, the  hadronic mass is just
the invariant mass of the recoiling u quark and the noninteracting
spectator quark) only populate the high $t$ (low recoil) region of their
Dalitz plot which therefore cuts off their electron
spectrum at the quark level endpoint energy. 
Thus while from kinematics alone such decays might have 
produced electrons with energies out to the physical ({\it i.e.} 
hadronic) endpoint, they
do not for dynamical reasons. Conversely, high mass hadronic final states
produce electrons out to their kinematic endpoint, but the highest mass
hadrons have an endpoint which exactly coincides with the quark level
endpoint in zeroth order. Recent work [52,53] 
has demonstrated that this picture
is the beginning of a rigorous $1 / m_Q$ expansion of inclusive
decays, and that the $1 / m_Q$ corrections have exactly the character
anticipated by ISGW and Ref. [57] and continued in this work.

    It has very recently been speculated that, within the $1/m_Q$ expansion,
even the endpoint region is amenable to treatment {\it via} an operator product
analysis [54]. If, as indicated by Fig. 10(a), this region is really
dominated by a few resonances (mainly the $\rho$, $a_1$, and $b_1$), then this
analysis may not apply. Thus while for $b \rightarrow u$ decays,
in contrast to $c \rightarrow s$ and $b \rightarrow c$ decays,
the hadronic spectral function over most of the Dalitz plot will be well-approximated 
by the partonic spectral function (corresponding to strong applicability 
of the results of Ref. [52]), the endpoint region is only dual to the partonic
spectral function in an average sense.

  The experimental analysis of this data requires simultaneous fits to the $b
{\mbox{$\rightarrow$}} c {\mbox{$\ell \bar \nu_{\ell}$}}$ and $b {\mbox{$\rightarrow$}} u
{\mbox{$\ell \bar \nu_{\ell}$}}$ inclusive spectra combined with the measured continuum backgrounds. 
Consequently, it is not possible to simply convert the values determined by various experiments which
have used the ISGW model  to the modified version of the model
presented here, since the results are dependent upon the shape as well
as the integrated inclusive rate. Since our spectrum is considerably harder than that of ISGW, 
it seems clear that the ISGW value of $\vert V_{ub}/V_{cb} \vert$ will decrease when reanalyzed.
However, the change seems unlikely to be very large. 
While the rate to the $\rho$, which is most important in the extreme
endpoint, has increased by $70\%$, the total rate
to the states we consider has only increased by 23\%. Given that these rates are
proportional to $\vert V_{ub} \vert ^2$,
the decrease in $\vert V_{ub}/V_{cb} \vert$ itself seems likely to be less than 25\%.

    We would like to caution against interpreting this decrease, which 
brings ISGW into better agreement with other models of the endpoint region [58], 
as leading to a more reliable value for $\vert V_{ub} \vert $ from the inclusive spectrum.
In the first place, it is a mistake to use models for this region which consider only
the $\pi$ and $\rho$ final states: the endpoint region is clearly going to be
populated by many more states than these. This exclusion leaves only ACCMM [11] and
ISGW as potentially realistic models for this region. However, we would continue to stress,
in spite of the real improvements of ISGW2, that our theoretical errors here
are of order $\pm 50\%$.
The recent clarification [53-55] of the status of 
the ACCMM calculation [11] in this region suggests that it should be assigned a very
substantial theoretical error as well.

\vfill\eject

\section{\bf CONCLUSIONS}

   We would argue that ISGW was already a good model  
for heavy meson semileptonic decay, and that with the improvements
added here ISGW2 is an even better model for this sector.  ISGW2 
behaves correctly in the Heavy Quark
Symmetry and Shifman-Voloshin
limits, including lowest
order corrections to these limits.  In taking into account the leading
corrections to the Heavy Quark Symmetry limit, ISGW2
adds physics to ISGW which corresponds to that demanded by 
Heavy Quark Effective Theory. These corrections are implemented
with a well-known, and well-tuned, model of quark dynamics.  

   In order to extend the range of validity of the model and to include
all relevant physics, ISGW2 also adds to ISGW other
effects. These extensions
have improved our agreement with the experimental data.
For example,
the mesonic decay rate for $P {\mbox{$\rightarrow$}} V \ell \bar \nu_{\ell}$, 
where $P$ and $V$ are pseudoscalar
and vector mesons, respectively, is sensitive to the S-wave axial current form
factor.  This form factor probably receives sizable relativistic corrections
(of order 10\%) which we have attempted to take into account.
The motivation for such extensions comes not only from first principles:
in this case, such a correction is needed in the quark model to understand  $g_A$ in
neutron beta decay.

    Given these points, the
extraction of $|V_{cb}|$ from the measurements of
$\Gamma({\mbox{$\bar B$}} {\mbox{$\rightarrow$}} D {\mbox{$e \bar \nu_e$}})$, and
$\Gamma({\mbox{$\bar B$}} {\mbox{$\rightarrow$}} D^* {\mbox{$e \bar \nu_e$}})$ should be reliable. Ultimately, a
precise determination of $|V_{cb}|$ will come from a careful
consideration of the heavy quark limit in which models like this one
have been used to estimate $1/{m_q}$ and $1/{m_{\scriptscriptstyle Q}}$ 
corrections.

   The determination of $|V_{ub}|$
is important for understanding  CP violation in the standard model, since
it is vital for determining the area of the``unitarity
triangle'' to which standard model CP violation is proportional.
In most decays considered in this paper the model dependence of our results is modest.  The $b
{\mbox{$\rightarrow$}} u {\mbox{$e \bar \nu_e$}}$ decays are, however, an exception.  The large available
recoil, the
relativistic nature of the $\pi$ and $\rho$, and 
the fact that such decays are far from any symmetry limits leave these 
predictions very exposed to uncertainties.
We estimate that the theoretical uncertainties within our model for extracting
$|V_{ub}/V_{cb}|$ from the inclusive endpoint spectrum are at the
50\% level. The uncertainties associated with individual exclusive channels
are even larger: we would estimate them to be almost a factor of two
for $\bar B \rightarrow \pi \ell \bar \nu_{\ell}$ (where our model uncertainties are compounded
by the uncertain effects of the nearby $\bar B^*$ pole [46]) and 50\% for 
$\bar B \rightarrow \rho \ell \bar \nu_{\ell}$ and
$\bar B \rightarrow \omega \ell \bar \nu_{\ell}$.
Fortunately, the
determination of $|V_{ub}|$ can be greatly 
improved by combining the observation of exclusive $\bar B$
decays with their analogous $D$ decays since such measurements 
can be related by Heavy Quark Symmetry. Here, once again, a model like ISGW2
has an important role to play, since it can assess the size of symmetry-breaking
effects in this procedure. As we showed in Section V, our model predicts
that this technique should allow the extraction of $|V_{ub}|$ with a theoretical
error of about 10\%.

\newpage

\bigskip
\leftline{\bf ACKNOWLEDGEMENTS}
 
    In the first instance we would like to thank our coauthors in
ISGW, Benjamin Grinstein and Mark Wise,  for their encouragement
in producing this update to ISGW, as well as for innumerable
discussions along the way.  We are also grateful to many of our 
experimental colleagues for their interest in this work and for their
many helpful comments; we would especially like to acknowledge conversations
with Marina Artuso, Arne Freyberger, Douglas Potter, and Sheldon Stone, who significantly
influenced the shape of this work.

\bigskip
\newpage

\noindent{\bf REFERENCES}

\begin{enumerate}

\item 
N.~Isgur, D.~Scora, B.~Grinstein,
and M.~B. Wise,  Phys. Rev. D
{\bf 39}, 799 (1989).

\item
B. Grinstein, M.B. Wise, and N. Isgur,
Caltech Report No. CALT-68-1311, and
University of Toronto Report No.
UTPT-85-37, 1985 (unpublished).

\item
B.~Grinstein, M.~B. Wise, and N.~Isgur,
 Phys. Rev. Lett. {\bf 56}, 298 (1986).

\item
N. Isgur and M.B. Wise, Phys.
Lett. {\bf B232} (1989) 113; Phys. Lett.
{\bf B237} (1990) 527. For an overview
of Heavy Quark Symmetry see N. Isgur
and M.B. Wise, ``Heavy Quark Symmetry"
in {\it B Decays}, ed. S. Stone (World
Scientific, Singapore, 1991), p. 158,
and in {\it ``Heavy Flavors"}, ed. A.J.
Buras and M. Lindner (World Scientific,
Singapore, 1992), p. 234.

\item
For some of the precursors to Heavy
Quark Symmetry see, in addition to
Refs. [1]-[3],
M.~B.~Voloshin and M.~A.~Shifman, Yad. Fiz. {\bf 47}, 801 (1988); Sov. J. Nucl.
  Phys. {\bf 47}, 511 (1988); M.~A.~Shifman in {\em Proceedings of the 1987
  International Symposium on Lepton and Photon Interactions at High Energies},
  Hamburg, West Germany, 1987, edited by W.~Bartel and R.~R{\"u}ckl, Nucl.
  Phys. B (Proc. Suppl.) {\bf 3}, 289
(1988); S. Nussinov and W. Wetzel,
Phys. Rev. {\bf D36}, 130 (1987);
 G.P. Lepage and B.A. Thacker, in {\it
Field Theory on the Lattice}, edited by
A.~Billoire, Nucl. Phys. B (Proc.
Suppl.) 4 (1988) 199;
E. Eichten, in {\it Field Theory on the
Lattice}, edited by A.~Billoire, Nucl.
Phys. B (Proc. Suppl.) 4, (1988) 170;
E. Shuryak, Phys. Lett. {\bf B93}, 134 (1980); Nucl. Phys. {\bf B198}, 83
(1982).

\item 
For an overview of Heavy Quark Effective Theory see the review cited in
Ref. [4] and the references therein,
especially  H. Georgi, Phys. Lett. {\bf
B240} (1990) 447; E. Eichten and B.
Hill, Phys. Lett. {\bf B234} (1990)
511; M.B. Voloshin and M.A. Shifman, Sov.
J. Nucl. Phys. {\bf 45} (1987) 463; H.D.
Politzer and M.B. Wise, Phys. Lett. {\bf
B206} (1988) 681; Phys. Lett. {\bf B208}
(1988) 504; A.F. Falk, H. Georgi, B.
Grinstein and M.B. Wise,  Nucl.
Phys. {\bf B343}, 1 (1990);
B. Grinstein,  Nucl. Phys. {\bf B339},
253 (1990); M.B. Wise, ``CP Violation"
in {\it Particles and Fields 3:
Proceedings of the Banff Summer
Institute (CAP)} 1988, p.~124, edited by
N.~Kamal and F.~Khanna, World Scientific
(1989). 

\item 
The phenomenological literature based on Heavy Quark Symmetry is too extensive for us to do 
more than quote 
some of the papers that have discussed issues related to those being addressed here. 
In particular, there have been a number of papers
which have shared with us as one of their main goals the study of $1/m_Q$ effects
using the quark model. C.O. Dib and F. Vera, Phys. Rev. {\bf D47}, 3938 (1993)
considered heavy-to-light transitions; 
J.F. Amundson, Phys. Rev. {\bf D49}, 373 (1994) has considered the $\bar B \rightarrow D$
and $\bar B \rightarrow D^*$ decays. See also J.F. Amundson and 
J.L. Rosner, Phys. Rev. {\bf D47}, 1951 (1993) for a model-independent discussion of the use
of the $c \rightarrow s$ transition as a gauge of $1/m_Q$ effects. For related work on $1/m_Q$ effects in the QCD sum rule
approach, see E. Bagan, P. Ball, V.M. Braun, and H.G. Dosch, Phys. Lett.
{\bf B278}, 457 (1992); M. Neubert, Phys. Rev. {\bf D46}, 3914 (1993); M. Neubert, Z. Ligeti, and Y. Nir,
Phys. Lett. {\bf B301}, 101 (1993); Phys. Rev. {\bf D47}, 5060 (1993). For a model
based on the Schwinger-Dyson equation, see B. Holdom and M. Sutherland, Phys. Rev. {\bf D47},
5067 (1993).

\item 
We accordingly
suggest that, when quoting the results
presented here, reference be made to
both this work {\it and} Ref. [1]. Some
very preliminary results of ISGW2 have
already been reported in
D.~Scora,
 ``${D}_s$ {S}emileptonic {D}ecay in the {Q}uark {M}odel",
 in {\em Particles and Nuclei, Proceedings of the Twelfth
  International Conference on Particles and Nuclei}, M.I.T., Cambridge, MA,
  USA, 1990,
 Nucl. Phys. A{\bf 527}, 743c (1991).

\item
 J.D. Bjorken, in Proceedings of
the $4^{th}$  Rencontre de
Physique de la Vallee d'Aoste, La
Thuile, Italy, 1990, ed. M. Greco
(Editions Frontieres, Gif-sur-Yvette,
France, 1990);
J.~D. Bjorken,
 {\em Recent Developments in Heavy Flavor Theory},
 in Proceedings of the XXVth
International Conference on High Energy
  Physics, Singapore, (World Scientific, Singapore, 1992);
 N.~Isgur and M.~B. Wise,
 Phys. Rev. D {\bf 43}, 819 (1991).

\item
N. Isgur and M.B. Wise, Phys. Rev. {\bf
D42}, 2388 (1990).

\item
G.~Altarelli, N.~Cabibbo, G.~Corb{\`o}, L.~Maiani, and G.~Martinelli,
 Nucl. Phys. B {\bf 208}, 365
(1982); N.~Cabibbo, G.~Corb{\`o}, and
L.~Maiani,  Nucl. Phys. B {\bf
155}, 93 (1979).

\item
M.K. Gaillard, B.W. Lee, and J.L.
Rosner, Rev. Mod. Phys. {\bf 47}, 277
(1975); J. Ellis, M.K. Gaillard, and
D.V. Nanopolous, Nucl. Phys. {\bf
B100}, 313 (1975);
A.~Ali and E.~Pietarinen,
 Nucl. Phys. B {\bf 154}, 519 (1979).

\item
N.~Cabibbo,
 Phys. Rev. Lett. {\bf 10}, 531
(1963); M.~Kobayashi and K.~Maskawa,
 Prog. Theor. Phys. {\bf 49}, 652 (1973).

\item 
See, e.g., E.H. Thorndike, in
Proceedings of the 1985 International
Symposium at High Energies, Kyoto,
Japan, 1985, ed. M. Konuma and K.
Takahashi (Research Institute for
Fundamental Physics, Kyoto University,
1986) p. 406.

\item
M.B. Voloshin and M.A. Shifman in Ref.
[5].

\item
N.~Isgur,
 Phys. Rev. D {\bf 40}, 101 (1989).

\item
N.~Isgur and M.~B. Wise,
 Phys. Rev. D {\bf 41}, 151
(1990); R.~Jaffe,
 Phys. Lett. B {\bf 245}, 221
(1990); R.~Jaffe and P.~F. Mende,
 Nucl. Phys. B {\bf 369}, 189
(1992); B. Grinstein and P.F. Mende, Phys. Rev. Lett. {\bf 69}, 1018 (1992).

\item
C.~Bebek et~al.,
 Phys. Rev. D {\bf 17}, 1693 (1978).

\item
The status of semileptonic B decays has recently been summarized by R.J. Morrison and J.D. Richman
for the Particle Data Group in p. 1602-1609 of the {\it Review of Particle Properties},
Phys. Rev. {\bf D50}, 1173 (1994).
For other recent reviews, see D.Z. Besson in {\it Lepton and Photon Interactions}, proceedings
of the XVI International Symposium, Ithaca, N.Y., ed. P. Drell and D. Rubin (AIP Conf. Proc.
No. 302)(AIP, New York, 1994), p. 221;
S.~Stone,
 ``Semileptonic {B} {D}ecays --
{E}xperimental", in {\em B
  Decays}, edited by S.~Stone (World
Scientific, Singapore, 1991), p. 210. See Refs. [20-22] for many of the individual
measurements of $\bar B$ decay.

\item
{CLEO Collaboration, R.~Fulton} et~al.,
 Phys. Rev. Lett. {\bf 64}, 16
(1990) and J. Bartelt {\it et al.}, Phys. Rev. Lett. {\bf 71}, 4111 (1993); 
{ARGUS Collaboration,
H.~Albrecht} et~al.,  Phys.
Lett. B {\bf 234}, 409 (1990).

\item
In addition to Ref. [19], see
{CLEO Collaboration, R.~Fulton} et~al.,
 Phys. Rev. D {\bf 43}, 651
(1991); 
{ARGUS Collaboration, H.~Albrecht} et~al.,
 Phys. Lett. B {\bf 219}, 121
(1989);
Phys. Lett. {\bf B229}, 175 (1985);
Z. Phys. {\bf C57}, 523 (1993);
Phys. Lett. {\bf B324}, 249 (1994);
Phys. Lett. {\bf B275}, 175 (1992);
 {CLEO Collaboration,
D.~Bortoletto} et~al.,  Phys.
Rev. Lett. {\bf 63}, 1667 (1989). For  a recent CLEO measurement of $\bar B \rightarrow 
D^* \ell \nu_{\ell}$ decay see preprint CLNS94/1285 of June 1994, to appear 
in Physical Review {\bf D}.

\item
{ARGUS Collaboration, H.~Albrecht} et~al.,
 Phys. Lett. B {\bf 255}, 297
(1991); M.~Danilov,
 {\em Heavy Flavour Physics (Non-LEP)}, presented at the 15th {I}nt.
  {C}onf. on {L}epton and {P}hoton {I}nteractions at {H}igh {E}nergy, Geneva
  (1991); The CLEO Collaboration, A. Bean {\it et al.}, Phys. Rev. Lett. {\bf 70}, 2681 (1993)
and as reported by D. Besson in Ref. [19].

\item
The status of semileptonic D decays has recently been summarized by R.J. Morrison and J.D. Richman
for the Particle Data Group in p. 1565-1572 of the {\it Review of Particle Properties},
Phys. Rev. {\bf D50}, 1173 (1994).
For  other recent reviews, see
S.~Stone,
 ``Charmed {M}eson {D}ecays",
in {\em Heavy
  Flavors}, edited by A.J.~Buras 
and H.~Lindner  (World Scientific,
Singapore, 1992), p. 334; M. Witherell in {\it Lepton and Photon Interactions}, proceedings
of the XVI International Symposium, Ithaca, N.Y., ed. P. Drell and D. Rubin (AIP Conf. Proc.
No. 302)(AIP, New York, 1994), p. 198.  See Refs. [24] for many of the individual
measurements of $D$ decay.

\item
{The Mark III Collaboration, J.~Adler} et~al.,
 Phys. Rev. Lett. {\bf 62}, 1821
(1989); {CLEO Collaboration,
G.~Crawford} et~al.,  Phys.
Rev. D {\bf 44}, 3394 (1991);
{Tagged Photon Collaboration, J.C.~Anjos} et~al.,
 Phys. Rev. Lett. {\bf 62}, 1587
(1989); {The Mark III Collaboration,
Z.~Bai} et~al.,  Phys. Rev.
Lett. {\bf 66}, 1011 (1991);
R.~Wanke, diploma thesis and private communication from H.~Schroder to
  S.~Stone, see note 51 of the review by Stone quoted in Ref. [23]; {E653 Collaboration,
K.~Kodama} et~al.,  Phys. Rev.
Lett. {\bf 66}, 1819 (1991);
{Tagged Photon Collaboration, J.C.~Anjos} et~al.,
 Phys. Rev. Lett. {\bf 67}, 1507
(1991); {ARGUS Collaboration,
H.~Albrecht} et~al.,  Phys.
Lett. B {\bf 255}, 634 (1991);
{E653 Collaboration, K.~Kodama} et~al.,
 Phys. Lett. B {\bf 274}, 246
(1992); {WA82 Collaboration,
M.~Adamovich} et~al.,  Phys.
Lett. B {\bf 268}, 142 (1991);
{Tagged Photon Collaboration, J.C.~Anjos} et~al.,
 Phys. Rev. Lett. {\bf 65}, 2630 (1990);
CLEO Collaboration, M.S. Alam {\it et al.}, Phys. Rev. Lett. {\bf 71}, 1311 (1993).

\item
D. Scora and N. Isgur,
 Phys. Rev. D {\bf 40}, 1491 (1989).

\item
N.N. Bogoliubov, Ann. Inst. Henri
Poincar\'e {\bf 8}, 163 (1963); A. Le
Yaouanc {\it et al.}, Phys. Rev. {\bf
D9}, 2636 (1974); {\bf D15}, 844 (1977)
and references therein; Michael J.
Ruiz, {\it ibid.} {\bf D12}, 2922
(1975); A. Chodos, R.L. Jaffe, K.
Johnson, and C.B. Thorn, {\it ibid.}
{\bf D10}, 2599 (1974); P. Ditsas, N.A.
McDougall, and R.G. Moorhouse, Nucl.
Phys. {\bf B146}, 191 (1978); J.S. Kang
and J. Sucher, Phys. Rev. {\bf D18},
2698 (1978); R.K. Bhaduri, L.E. Cohler,
and Y. Nogami, Phys. Rev. Lett. {\bf
44}, 1369 (1980).

\item
C.~Hayne and N.~Isgur,
 Phys. Rev. D {\bf 25}, 1944 (1982);
for some more recent related work, see
S.~Godfrey and N.~Isgur,
 Phys. Rev. D {\bf 32}, 189
(1985); P.~J. O'Donnell and H.~K.~K.
Tung,  Phys. Rev. D {\bf 44},
741 (1991).

\item 
G. Karl, private communication.

\item 
See N. Isgur and M.B. Wise in Ref. [9].

\item
N. Isgur and M.B. Wise, Phys. Rev.
Lett. {\bf 66}, 1130 (1991).

\item
The matching conditions first described in Refs. [6]
were performed to leading logarithmic order, valid for $m_i << m_j$. A.F. Falk
and B. Grinstein, Phys. Lett. {\bf B247}, 406 (1990) 
and Phys. Lett. {\bf B249}, 314 (1990), extended these results
to matching corrections of order $m_j/m_i$ and to the case of matching for
$m_i \simeq m_j$. The matching problem has recently been addressed in
a series of papers by Neubert and collaborators, and we adopt the 
general approach of
these papers here. See M. Neubert, Phys. Rev. {\bf D46}, 2212 (1992); Nucl. Phys. {\bf B371}, 149 (1992);
and references therein. In the first of these papers, matching is done up to
corrections of order $\alpha_s ^2 (\frac{m_c}{m_b} ln \frac{m_b}{m_c})^2$. Here we
simplify and retain only the leading corrections as in the second paper quoted.
The error inherent in this simplification is of order $5\%$, which is adequate
for our purposes and readily improved upon when necessary.

\item
A.F. Falk, Nucl. Phys. {\bf B378}, 79
(1992).

\item
Particle Data Group, 
 Phys. Rev.  {\bf D50}, 1173 (1994).

\item
{\it E.g.}, the axial current can be renormalized in the ``classical" way by hadronic
loop diagrams. For a very interesting recent approach to such effects, see S. Weinberg, 
Phys. Rev. Lett. {\bf 65}, 1181 (1990); Phys. Rev. Lett. {\bf 67}, 3473 (1991).

\item 
See S. Godfrey and N. Isgur in Ref.
[27].

\item
The CLEO Collaboration, preprint
CLNS94/1266 and CLEO 94-1 (1994).

\item
C.-H. Chang and Y.-Q. Chen,
 The production of ${B}_c$ or $\overline {B}_c$ meson associated with
  two heavy quark jets in ${Z}^0$ boson decay,
 Institute of Theoretical Physics, Academia Sinica, Beijing preprint
  {AS-ITP-91-64}. 

\item
E. Eichten and C.
Quigg, Phys. Rev. {\bf D49}, 5845
(1994).

\item
There has been some confusion over the
value of this ratio in our model. Phase
space alone gives 0.80, which value has
been quoted by CLEO in Phys. Rev. Lett.
{\bf 65}, 1531 (1990). A typographical
error in the manuscript copy of the
contribution by D. Scora quoted in Ref.
[8] gave a value of 1.08, which was
corrected to 1.02 before publication,
but still added to the confusion. This ratio is important because it is used to set the scale
for all $D_s$ decay rates using $\Gamma(D_s \rightarrow \phi e^+ \nu_e)/
\Gamma(D_s \rightarrow \phi \pi)$, 
$\Gamma(D \rightarrow \bar K^* e^+ \nu_e)/
\Gamma(D \rightarrow \bar K \pi \pi)$, and $\tau_{D_s}/\tau_{D^+}$ to determine
$\Gamma(D_s \rightarrow \phi \pi)$. For example, using our new prediction, the
branching ratio for $\Gamma(D_s \rightarrow \phi \pi)$ presented in F. Butler {\it
et al.}, Phys. Lett. {\bf B324}, 255 (1994) becomes $4.3 \pm0.3 \pm 0.3 \pm 0.6$
with a theoretical error which we estimate to be $\pm 0.4$.

\item
See the work quoted in Ref. [7]; for related estimates of $1/m_Q$ effects
using QCD Sum Rules see E. Bagan, P. Ball, V.M. Braun, and H.G. Dosch,
Phys. Lett. {\bf B278}, 457 (1992); M. Neubert, Phys. Rev. {\bf D46}, 3914 (1993);
M. Neubert, Z. Ligeti, and Y. Nir, Phys. Lett. {\bf B301}, 101 (1993);
Phys. Rev. {\bf D47}, 5060 (1993); for a Nambu-Jona-Lasinio-like approach see
B. Holdom and M. Sutherland, Phys. Rev. {\bf D47}, 5067 (1993). For related
work using lattice QCD, see the review by P.B. Mackenzie in {\it Lepton and Photon Interactions}, proceedings
of the XVI International Symposium, Ithaca, N.Y., ed. P. Drell and D. Rubin (AIP Conf. Proc.
No. 302)(AIP, New York, 1994),
p.634.

\item
The proof in Ref. [42] that certain predictions of Heavy Quark Symmetry are protected
from first order corrections in $1/m_q$ and $1/m_Q$ (``Luke's Theorem") plays an important
role in increasing the reliability of theory for the extraction of $V_{cb}$ from 
experiment. It is very similar in physics content to the analogous theorem for
extracting $V_{us}$ from $\bar K \rightarrow \pi \ell \bar \nu_{\ell}$ from Ref. [47].

\item
M.~E. Luke,
 Phys. Lett. B {\bf 252}, 447 (1990).

\item
H.~Georgi, B.~Grinstein,
and M.~B. Wise,  Phys. Lett. B
{\bf 252}, 456 (1990).

\item
C.~Boyd and D.~Brahm,
 Phys. Lett.  {\bf B257}, 393 (1991).

\item
M.~Neubert and V.~Rieckert, Nucl. Phys.
{\bf B382}, 97 (1992); A. Falk, M.E. Luke, and M. Neubert, Nucl. Phys. {\bf B388}, 363 (1992).

\item
N.~Isgur and M.B. Wise,
Phys. Rev. {\bf D41}, 151 (1990); M.B. Wise, Phys. Rev. {\bf D45}, 2188 (1992);
G. Burdman and J.F. Donoghue, Phys. Lett. {\bf B280}, 287 (1992); L. Wolfenstein, Phys. Lett. 
{\bf B291}, 177 (1992); G. Burdman, Z. Ligeti, M. Neubert, and Y. Nir, 
Phys. Rev. {\bf D49}, 2331 (1994); B. Grinstein and P.F. Mende, 
Phys. Rev. Lett. {\bf 69}, 1018 (1992) and preprint SMU-HEP/94-11 and UCSD/PTH 94-09.

\item
M.~Ademollo and R.~Gatto,
 Phys. Rev. Lett. {\bf 13}, 264 (1964).

\item 
For an excellent review of the status of the determinatiuon of the CKM 
parameters, see J.L. Rosner, in {\it B Decays}, ed. S. Stone (World
Scientific, Singapore, 1991), p. 312.

\item 
H. Leutwyler and M. Roos, Z. Phys. {\bf C25}, 91 (1984).

\item
The CLEO Collaboration, A. Ryd {\it et al.}, Proceedings of the APS/DPF Meeting,
Albuquerque, August 1994; for an earlier result see
S. Sanghera {\it et al.}, Phys. Rev. {\bf D47}, 791 (1993). Note that 
these papers quote their results in terms of the form factors $A_1$, $V$, and $A_2$ defined
in
K.~Hagiwara, A.~Martin, and M.~Wade,
 Phys. Lett. B {\bf 228}, 144
(1989); M.~Bauer and W.~Wirbel,
 Z. Phys. C {\bf 42}, 671
(1989); F.~J. Gilman and R.~L.
Singleton, Jr.,  Phys. Rev. D
{\bf 41}, 142 (1990). These form factors are related to ours {\it via} $f=(m_{P_Q}+m_{V_q})A_1$,
$g=(m_{P_Q}+m_{V_q})^{-1}V$, and $a_+= - (m_{P_Q}+m_{V_q})^{-1}A_2$.

\item
E653 Collaboration, K. Kodama {\it et al.}, Phys. Lett. {\bf B316}, 455 (1993);
E687 Collaboration, P.L. Frabetti {\it et al.}, Phys. Lett. {\bf B313}, 253 (1993);
CLEO Collaboration,
P. Avery {\it et al.}, CNLS 94/1290, to appear in Phys. Lett. {\bf B}. 
The quoted form factor ratios are based on our rough averaging of these
three measurements.

\item 
J. Chay, H. Georgi, and B. Grinstein,
Phys. Lett. {\bf B247}, 399 (1990).

\item
A.V. Manohar and M.B. Wise, Phys. Rev. {\bf D49}, 1310 (1994); I.I. Bigi, M. Shifman, N.G. Uraltsev, and A.I. 
Vainshtein, Phys. Rev. Lett. {\bf 71}, 496 (1993); B. Blok, L. Koyrakh, M. Shifman, and A.I. Vainshtein, ITP Report No.
NSF-ITP-93-68 (1993); T. Mannel, Nucl. Phys. {\bf B413}, 396 (1994); I.I. Bigi, N.G. Uraltsev, and A.I. 
Vainshtein, Phys.  Lett. {\bf B293}, 430 (1992); A.F. Falk, M. Luke, and M.J. Savage,
Phys. Rev. {\bf D49}, 3367 (1994).

\item
M. Neubert {\it et al.}, Phys. Rev. {\bf D49}, 3392 (1994);
Phys. Rev. {\bf D49}, 4623 (1994).

\item
M. Savage in Proceedings of ``Intersections of Particle and Nuclear Physics",
St. Petersburg, Florida (1994). 

\item
See for examples of the discussion of the
$b \rightarrow u$ endpoint problem
C.~Ramirez, J.~F. Donoghue, and 
G.~Burdman, Phys. Rev. {\bf D41}, 1496
(1990); G.~Altarelli and P.~Franzini,
 in {\em {\em Conf. Proc. Vol. 15}, {P}resent {T}rends, {C}oncepts and
  {I}nstruments of {P}article {P}hysics}, (Italian Phys. Soc., Bologna, 1988).

\item N. Isgur, Phys. Rev. {\bf D47},
2782 (1993).

\item
M.~Wirbel, B.~Stech, and M.~Bauer,
 Z. Phys. C {\bf 29}, 637
(1985); T.~Altomari and L.~Wolfenstein,
 Phys. Rev. Lett. {\bf 58}, 1583
(1987); T.~Altomari and L.~Wolfenstein,
 Phys. Rev. D {\bf 37}, 681
(1988); E.~Golowich et~al.,
 Phys. Lett. B {\bf 213}, 521
(1988); J.~G. K{\"o}rner and G.~A.
Schuler,  Z. Phys. C {\bf 38},
511 (1988);
J.~G. K{\"o}rner and G.~A. Schuler,
  {\bf 226}, 185 (1989);
Xin-heng Guo and Tao Huang,
 Phys. 
Rev. D {\bf 43}, 2931 (1991); see also
the three papers of Ref. [52].

\end{enumerate}

\newpage

\noindent  APPENDIX A: {\bf hyperfine-corrected wavefunctions}

\bigskip

   Of the two leading order effects which break Heavy Quark Symmetry, the heavy 
quark kinetic energy and its hyperfine interaction, only the first was included 
in the constituent quark model which formed the basis of the ISGW prediction 
for form factors.  In this Appendix we present a simple extension 
of their spectroscopic
model which qualitatively takes hyperfine interactions into account.

   The extended spectroscopic model remains a nonrelativistic 
constituent quark model
with essentially the same Coulomb-plus-linear central potential:

\begin{equation}
    V(r)= - {{4 \alpha_s} \over {3r}} +c +br~~.
\end{equation}

\noindent  One 
cannot simply extend such a model
by adding the non-relativistic hyperfine interaction since 
the Fermi spin-spin contact term, which is proportional to 
$\vec S_i \cdot \vec S_j~~\delta ^3(\vec r)$ is an illegal operator
in the Schrodinger equation.  The problem is that, in channels where it 
is attractive, 
this interaction is more singular than the kinetic energy so that the solutions of 
the Schrodinger equation collapse into $r_{ij}=0$ and 
to infinitely negative energies.  
This problem is solved by relativistic corrections which turn this operator into 
an extended and nonlocal one.  Here we model this behavior by taking 

\begin{equation}
    \bar H^{ij}_{hyp}= [{{m_im_j}\over {E_iE_j}}]^{1 \over 2}
                        \bigl( {{32 \pi a \alpha_s \vec S_i 
                  \cdot \vec S_j~~\delta ^3(\vec r) } \over {9m_im_j}} \bigr)
                          [{{m_im_j}\over {E_iE_j}}]^{1 \over 2}
\end{equation}

\noindent where the term in parentheses would be the ordinary 
Fermi contact term if the anomalous coupling coefficient $a$ 
were unity, and where $E_i=(m_i^2+p^2)^{1 \over 2}$.  
We have examined the effects of smearing out $\delta ^3(\vec r)$ and found 
it to be small compared to the very strong nonlocality created by the pre- 
and postfactors of $[{{m_im_j}\over {E_iE_j}}]^{1 \over 2}$. 
We have also ignored the tensor part of the hyperfine 
interaction as well as spin-orbit interactions.  Neither can play a leading-order 
role in the S-waves which dominate our discussions, and both are also observed to 
be relatively weak even in excited states. Finally, we have made the running of $\alpha_s$ in ISGW slightly
less crude by taking $\alpha_s$ for the $\rho$, $K^*$, $\phi$, $D^*$, $\bar B$, $D_s^*$,
$\bar B_s^*$, $\psi$, $\bar B_c$, and $\Upsilon$ families to be 0.60, 0.55, 0.55, 0.50, 0.50, 0.45, 
0.40, 0.40, 0.35, and 0.30, respectively on the basis of their reduced masses. Note that,
following Ref. [35] we assume $\alpha_s$ ``freezes out" at 0.60 at low mass scales. 

    Also as in ISGW, we solve the Hamiltonian variationally in a basis of harmonic 
oscillator states truncated to include only the $1S$, $1P$ and $2S$ states.  
An interesting 
feature of this procedure, realized numerically in ISGW but thought to be an 
accident, is that the solution will exhibit zero $1S-2S$ mixing.  
This is proved in
Appendix B.  As a result, we need only the diagonal matrix elements of 
$p^2$, $1/r$, $1$, $r$, and $(A1)$ to solve for the wavefunctions.  
All but the last are trivial; 
for it one easily finds that the diagonal matrix elements all vanish except in 
the S-waves where

\begin{equation}
   \langle n^{2s+1}S_{2s+1} \vert \bar H_{hyp} \vert n^{2s+1}S_{2s+1} \rangle=
          [{{2s(s+1)-3} \over 4}]  \bigl( {{32 \pi a \alpha_s} 
        \over {9m_im_j}} \bigr) \vert \bar \psi_{nS}(0) \vert^2
\end{equation}

\noindent with $s=0$ or $s=1$ the total quark-plus-antiquark spin,

\begin{equation}
    \bar \psi_{nS}(0)={1 \over {(2 \pi)^{3/2}}} 
       \int d^3p[{{m_im_j}\over {E_iE_j}}]^{1 \over 2} \phi_{nS}(p)
\end{equation}

\noindent with $m_1$ 
and $m_2$ the constituent quark and antiquark masses and $\phi_{nS}(p)$ 
the momentum space 
wavefunction.  (Note that $\bar \psi_{nS}(0)$ 
reduces to the nonrelativistic spatial wavefunction at $\vec r=0$ in 
the nonrelativistic limit).

   On minimizing energies with respect to the gaussian wavefunction 
parameters $\beta_i$
previously defined in ISGW and searching for a fit to the observed 
meson spectra, we found the results listed in Tables A1 and A2. 
We assume that defects like the $\rho-\pi$ splitting would 
improve with a larger basis 
space, but given the crude nature of this quasirelativistic model and our goal 
of a qualitative description of hyperfine effects, we do not attempt a better fit 
via a more complicated variant of the model.
We also emphasize that full consistency would require a parallel 
relativistic treatment of both the spectrum and the weak matrix elements; this 
far more ambitious program would be very worthwhile, but it is well beyond the 
scope of this work.

\newpage

\noindent  APPENDIX B: {\bf variational solution in a $1S$-$2S$ basis}

   Let $\beta_S^{(0)}$ be the value of $\beta_S$ 
which minimizes the full Hamiltonian in the 
harmonic oscillator ground state 

\begin{equation}
      \psi_{1S}^{(\beta_S)}={{{\beta_S^{3/2}} \over {\pi^{3/4}}}e^{-{1\over 2}\beta_S^2r^2}}.
\end{equation}

\noindent Thus

\begin{equation}
    {d \over {d \beta_S}} \int d^3r \psi_{1S}^{(\beta_S)}
                       H(p,r) \psi_{1S}^{(\beta_S)}=0
\end{equation}

\noindent for $\beta_S=\beta_S^{(0)}$. But 

\begin{equation}
     {d \over {d \beta_S}}  \psi_{1S}^{(\beta_S)}=({3 \over 2})^{1/2}
             \beta_S^{-1} \psi_{2S}^{(\beta_S)}
\end{equation}

\noindent since 

\begin{equation}
    \psi_{2S}^{(\beta_S)}=({2 \over 3})^{1/2}{{\beta_S^{7/2}} \over {\pi^{3/4}}}
                 (r^2-{3 \over 2} \beta_S^{-2})e^{-{1\over 2}
                 \beta_S^2r^2}~~.
\end{equation}

\noindent Thus, if $\beta_S^{(0)}$ minimizes H, 

\begin{equation}
      \int d^3r \psi_{2S}^{(\beta_S^{(0)})}
                       H \psi_{1S}^{(\beta_S^{(0)})}+
        \int d^3r \psi_{1S}^{(\beta_S^{(0)})}
                       H \psi_{2S}^{(\beta_S^{(0)})}=0~~.
\end{equation}

\noindent  But  $\langle \psi_{1S}^{(\beta_S)} \vert H(p,r) 
\vert \psi_{2S}^{(\beta_S)}\rangle=\langle \psi_{2S}^{(\beta_S)} \vert H(p,r) 
\vert \psi_{1S}^{(\beta_S)}\rangle ^*$
and both are real so 

\begin{equation}
     \langle \psi_{2S}^{(\beta_S^{(0)})} \vert H 
\vert \psi_{1S}^{(\beta_S^{(0)})}\rangle =0
\end{equation}

\noindent {\it i.e.} $\psi_{2S}$ 
does not mix with $\psi_{1S}$ if it has been chosen variationally to minimize H. 
This argument can be extended; for example, $\psi_{2P}$ 
does not mix with $\psi_{1P}$ if it has been chosen variationally to minimize H.

\newpage

\noindent  {APPENDIX C:  {\bf form factor modifications for ISGW2}}
\bigskip

As described in the text, the formulas of Appendix B of 
ISGW (and of those additional formulas in Refs. [25] and [30] needed when
the lepton mass cannot be neglected)
require modification. 

    First, as discussed in Section III.C, all formulas are affected by the replacement
in eq. (B1) of ISGW shown in eq. (20). In addition, the conversion from the $\tilde f_i^{qm}$
to the $f_i$ described in Section III.A introduces factors of
$( {\bar m_B}/{\tilde m_B})^{n_B(\alpha)} ( {\bar m_X}/{\tilde m_X})^{n_X(\alpha)}$
into each ISGW formula. With both of these changes effected, the factor $F_n$ of eq. (B1) of ISGW
is converted to a factor we denote by $F_n^{(\alpha)}$ since it now depends on the form factor
$\alpha$ under consideration. The powers $n_B(\alpha)$ and $n_X(\alpha)$ required to make these
conversions are given in Table C1. 
Note that in many instances it is a special combination ({\it e.g.}, $f_++f_-$) which has a
simple mass scaling law and not the individual form factors ({\it e.g.}, $f_+$). For this
reason we quote below formulas for these special combinations. To compute a 
particular form factor in such cases, one must apply the methods described here to those 
special combinations and then combine these results.
Section III.A also describes how all
$S$-wave to $S$-wave transition form factors must be modified by
the matching conditions given in Section III.A. These corrections
lead below to the appearance of the factors 
$R^{(\alpha)} \equiv {\bf C}_{ji}(1 + \tilde \beta^{(\alpha)}\alpha_s/ \pi)$
~(or, in the case of $a_++a_-$, to terms proportional 
to ${\bf C}_{ji} \tilde \beta^{(\alpha)}\alpha_s/ \pi$)   which are the coefficients
of $\xi(w)$ in eq. (5). (In practice we use the
``renormalization group improved" matching for all decays except those induced by the $s \rightarrow u$
transition for which we implement ``lowest order matching".)

In the following we employ the notation of ISGW augmented with  $\tilde{w}$ as identified by 
Eq. (3).
As in ISGW, all formulas are
given for the $b \rightarrow c$ transition but can be immediately adapted to any decay; see Ref. [1]
for details including the explicit definition of each form factor.
For an example, see eq. (103) below for the $f$ form factor. The new formulas are:
\begin {enumerate}

\item Eqs. (B8) and (B9) are replaced by the two equations

\begin {eqnarray}
f_{+}+f_{-}&=& 
\left[ 2-\frac{\tilde m_X}{m_q} \left( 1-{m_d m_q\over {2 \mu_+ \tilde m_X}}
{{\beta^2_B}\over {\beta^2_{BX}}} 
\right) \right] F^{(f_{+}+f_{-})}_3 R^{(f_{+}+f_{-})} \\ \cr
\cr
f_{+} - f_{-}&=& 
\frac{\tilde m_B}{m_q} \left( 1-{m_d m_q\over {2 \mu_+ \tilde m_X}}
{{\beta^2_B}\over {\beta^2_{BX}}} 
\right) F^{(f_{+}-f_{-})}_3 R^{(f_{+}-f_{-})}
\end{eqnarray}
which determine both $f_+$ and $f_-$.

\item Eq. (B15)
becomes 

\begin {equation}
f = C_f{\tilde m}_B \left[ (1+\tilde w)+{{m_d (\tilde w-1)}\over
{2\mu_+}}\right]F^{(f)}_3 R^{(f)}.
\end{equation}

The $(1+\tilde w)$ and $(\tilde w-1)$ terms come from the constraints of 
Heavy
Quark Symmetry in leading [4] and next-to-leading order [42-44], 
respectively,
in the
$1/m_Q$ expansion.  $C_f$ is from Eq. (7), and $F_3^{(f)}$ is the modified factor $F_3$
described above, namely

\begin{equation}
F_3^{(f)}=(\bar m_X/\tilde m_X)^{\frac{1}{2}}(\bar m_B/\tilde m_B)^{\frac{1}{2}}\times
(\tilde m_X/\tilde m_B)^{\frac{1}{2}}\left(\frac{\beta_X \beta_B}{\beta_{BX}^2}\right)^{\frac{3}{2}}
[1+{\frac{1}{12}}r_{XB}^2(t_m-t)]^{-2}~~.
\end{equation}

\item Eq. (B16) for $g$ becomes

\begin{equation}
g=\frac{1}{2}
\left[ 
\frac{1}{m_q} - \frac{m_d \beta_B^2}{2 \mu_- \tilde m_X \beta_{BX}^2} 
\right]
F_3^{(g)}R^{(g)}
\end{equation}

\item Eq. (B17) for $a_+$ is replaced by the two equations

\begin {eqnarray}
a_{+}+a_{-}&=&{\bf C}_{ji}\left[ {{m_d}\over {(1+\tilde w)m_qm_b}} 
{\beta^2_X\over
\beta^2_{BX}}
\left( 1-{m_d\over {2\tilde m_B}}{{\beta^2_X}\over {\beta^2_{BX}}} \right) \right. 
\nonumber \\ && \nonumber \\ &&
 \left.     + \tilde \beta^{(a_{+}+a_{-})} \frac {\alpha_s}{\pi}
\tilde m_B \right] F_3^{(a_{+}+a_{-})}\\ \cr
\cr
a_{+} - a_{-}&=&-~{1 \over \tilde m_X}\left[ {\tilde m_B\over
m_b}-{m_d\over {2\mu_+}}{\beta^2_X\over \beta^2_{BX}} \right.
\nonumber \\ && \nonumber \\ && 
 \left. +{{\tilde wm_d\tilde m_B}\over {(\tilde w+1)m_qm_b}} {\beta^2_X\over
\beta^2_{BX}}\left( 1-{m_d\over {2\tilde m_B}}{\beta^2_X\over 
\beta^2_{BX}}
\right) \right] F^{(a_{+} - a_{-})}_3 R^{(a_{+} - a_{-})}~~. 
\end{eqnarray}
These formulas follow from ISGW and Ref. [25] with $\tilde w$-dependence 
dictated by
the constraints of 
Heavy
Quark Symmetry in order $1/ m_Q$ [42-44]. They determine both $a_+$ and $a_-$.

\item Eq. (B23) for $h$ becomes

\begin{equation}
h=\frac{m_d}{2\sqrt{2} \tilde m_B \beta_B}
\left[ 
\frac{1}{m_q} - \frac{m_d \beta_B^2}{2 \mu_- \tilde m_X \beta_{BX}^2} 
\right]
F_5^{(h)}~~.
\end{equation}

\item  Eq. (B24) for  $k$ becomes

\begin{equation}
k=\frac{m_d}{\sqrt{2} \beta_B}
(1+\tilde w)
F_5^{(k)}~~.
\end{equation}

\item Eq. (B25) for $b_+$ is replaced by the two equations

\begin {eqnarray}
b_{+}+b_{-}&=&{{m_d^2 }\over {4 \sqrt{2}m_qm_b\tilde m_B 
\beta_B }} 
{\beta^2_X\over
\beta^2_{BX}}
\left( 1-{m_d\over {2\tilde m_B}}{{\beta^2_X}\over {\beta^2_{BX}}} 
\right)F^{(b_{+}+b_{-})}_5 \\ \cr
\cr
b_{+} - b_{-}&=&-~{ m_d\over \sqrt{2} m_b \tilde m_X \beta_B}
\left[ 1-{m_d m_b\over {2\mu_+ \tilde m_B}}{\beta^2_X \over \beta^2_{BX}} \right.
\nonumber \\ && \nonumber \\ &&
\left.  +{{m_d}\over {4m_q}} {\beta^2_X\over
\beta^2_{BX}}\left( 1-{m_d\over {2\tilde m_B}}{\beta^2_X\over 
\beta^2_{BX}}
\right)  \right] F^{(b_{+} - b_{-})}_5
\end{eqnarray}
which determine both $b_+$ and $b_-$.

\item  Eq. (B31) for $q$ becomes 

\begin{equation}
q=~-\frac{m_d}{2 \tilde m_X \beta_B}
\left({{5+\tilde 
w}\over
6}\right)
F_5^{(q)}~~.
\end{equation} 
See also eqs. (125) and (129) below.

See ref. [23] of ref. [29] for an explanation of the sign change.

\item Eq. (B32) becomes
\begin {equation}
\ell=- \tilde m_B \beta_B \left[ {1\over {\mu_{-}}}+ {{m_d
\tilde m_X(\tilde w-1)}\over
\beta^2_B}\left( {{5+\tilde w}\over {6m_q} }-{1\over
{2\mu_{-}}}{m_d\over {\tilde m_X}} {\beta^2_B\over \beta^2_{BX}}\right) 
\right]F^{(\ell)}_5
\end {equation}
See also eqs. (122) and (126) below.

\item Eq. (B33) for $c_+$ is replaced by the two equations

\begin {eqnarray}
c_{+}+c_{-}&=& - {{m_d \tilde m_X }\over {2 m_q \tilde m_B 
\beta_B }} 
\left( 1-{m_d m_q\over {2\tilde m_X \mu_-}}{{\beta^2_B}\over {\beta^2_{BX}}} 
\right)F^{(c_{+}+c_{-})}_5 \\ \cr
\cr
c_{+} - c_{-}&=&-~{{m_d \tilde m_X }\over {2 m_q \tilde m_B 
\beta_B }} 
\left( \frac{\tilde w+2}{3}-{m_d m_q\over {2\tilde m_X \mu_-}}{{\beta^2_B}\over {\beta^2_{BX}}} 
\right)F^{(c_{+} - c_{-})}_5
\end{eqnarray}
which determine both $c_+$ and $c_-$.
See also eqs. (123), (124), (127),  and (128) below.

\item Eq. (B37) for $u_+$ is replaced by the two equations

\begin {eqnarray}
u_{+}+u_{-}&=& - \sqrt{\frac{2}{3}} {{m_d  }\over { \beta_B }}F^{(u_{+}+u_{-})}_5  \\ \cr
\cr
u_{+} - u_{-}&=& \sqrt{\frac{2}{3}} {{m_d \tilde m_B }\over { \beta_B
\tilde m_X}}F^{(u_{+} - u_{-})}_5
\end{eqnarray}
which determine both $u_+$ and $u_-$.

\item Eq. (B43) for $v$ becomes
\begin {equation}
v=\left[ {{\tilde m_B\beta_B}\over {4\sqrt{2}m_bm_q\tilde 
m_X}}+
{({\tilde w-1})\over 6\sqrt{2}}{{m_d}\over{\tilde m_X\beta_B}}\right]F^{(v)}_5~~.
\end{equation}
See also eqs. (125)  and (129) below.

\item Eq. (B44) for $r$ becomes
\begin {equation}
r={{\tilde m_B\beta_B}\over \sqrt{2}}\left[ {1\over 
\mu_{+}}+
{{m_d\tilde m_X}\over {3m_q\beta^2_B}}{(\tilde w-1)}^2 \right]F^{(r)}_{5}~~.
\end {equation}
See also eqs. (122)  and (126) below.

\item Eq. (B45) for $s_+$ is replaced by the two equations

\begin {eqnarray}
s_{+}+s_{-}&=& {{m_d  }\over {\sqrt{2} \tilde m_B 
\beta_B }} 
\left( 1-\frac{m_d}{m_q}+{m_d \over {2 \mu_+}}{{\beta^2_B}\over {\beta^2_{BX}}} 
\right)F^{(s_{+}+s_{-})}_5 \\ \cr
\cr
s_{+} - s_{-}&=& {{m_d  }\over {\sqrt{2} m_q  
\beta_B }} 
\left( \frac{4-\tilde w}{3}-{m_d m_q\over {2\tilde m_X \mu_+}}{{\beta^2_B}\over {\beta^2_{BX}}} 
\right)F^{(s_{+} - s_{-})}_5
\end{eqnarray}
which determine both $s_+$ and $s_-$.
See also eqs. (123), (124), (127),  and (128) below.

\item Eq. (B49) for $f_+'$ is replaced by the two equations 

\begin {eqnarray}
f'_{+}+f'_{-}&=&\sqrt{3 \over 2}\left[ (1-\frac{m_d}{m_q}) U 
-{m_d\over { m_q }}V \right]
{ F}^{(f'_{+}+f'_{-})}_3 \\ \cr
\cr
f'_{+} - f'_{-}&=&\sqrt{3 \over 2}\frac{\tilde m_B}{m_q} \left[ U 
+{m_d \over {\tilde m_X }}V \right]
{ F}^{(f'_{+}-f'_{-})}_3
\end{eqnarray}
where
\begin{eqnarray}
U&=&
{{\beta^2_B-\beta^2_X}\over {2 \beta^2_{BX}}}+
{{\beta^2_B \tau}\over {3 \beta^2_{BX}}}  \\ \cr
V&=&{{\beta^2_B}\over {6  \beta^2_{BX}}}(1+\frac{m_q}{m_b}) 
\left[7-{{\beta^2_B}\over { \beta^2_{BX}}}(5+\tau) \right]~~.
\end{eqnarray}
and where 
\begin{equation}
\tau \equiv \frac{m_d^2\beta_X^2(\tilde w -1)}{\beta_B^2 \beta_{BX}^2}~~.
\end{equation}
These equations determine both $f'_+$ and $f'_-$.

\item Eq. (B55) for $f^{\prime}$ becomes
\begin{equation}
 f'= C_{f'}\sqrt{\frac{3}{2}} \tilde m_B (1+\tilde w) U F_3^{(f')}
\end{equation}
where $U$ is given above.

\item Eq. (B16) for $g'$ becomes
\begin{equation}
 g'= \frac{1}{2}\sqrt{\frac{3}{2}} \left[ 
(\frac{1}{m_q} - \frac{m_d \beta_B^2}{2 \mu_- \tilde m_X \beta_{BX}^2})U
+\frac{m_d\beta_B^2\beta_X^2}{3\mu_- \tilde m_X \beta_{BX}^4} 
\right]
F_3^{(g')}
\end{equation}
where once again $U$ is given above.
  
\item Eq. (B17) for $a_+'$ is replaced by the two equations

\begin {eqnarray}
a'_{+}+a'_{-}&=&~-\sqrt{2\over 3}{\beta_B^2 \over {m_qm_b \beta_{BX}^2}} 
\left\{ 
{ {7m_d^2\beta^4_X} \over {8 \tilde m_B \beta^4_{BX}} }[1+\frac{1}{7} \tau]
-{ {5m_d \beta^2_X} \over {4   \beta^2_{BX}} }[1+\frac{1}{5} \tau] \right.
\nonumber \\ && \nonumber \\ &&  
\left. -{ {3m_d^2\beta^4_X} \over {8 \tilde m_B \beta_B^2 \beta^2_{BX}} }
+{ {3m_d \beta^2_X} \over {4 \beta^2_{B }} } 
\right\} F^{(a'_{+}+a'_{-})}_3 \\ \cr
\cr
a'_{+} - a'_{-}&=&\sqrt{2\over 3}{ {3\tilde m_B} \over {2m_b \tilde m_X} } 
\left\{1 
-{ {\beta^2_B} \over {   \beta^2_{BX}} }[1+\frac{1}{7} \tau]
-{ {m_d \beta^2_X} \over {2  \tilde m_B \beta^2_{BX}} }
(1-{ 5{\beta^2_B} \over {  3 \beta^2_{BX}} }[1+\frac{1}{5} \tau]) \right.
\nonumber \\ && \nonumber \\ &&   
\left. -{ {7 m_d^2 \beta^2_B \beta^2_X} \over {12 m_q \tilde m_B  \beta^4_{BX}} }
(1-{{\beta^2_X} \over {   \beta^2_{BX}} } +{{\beta^2_B \tau} \over {  7 \beta^2_{BX}} })
\right\}
F^{(a'_{+} - a'_{-})}_3~~. 
\end{eqnarray}
These formulas determine both $a_+'$ and $a_-'$.

\bigskip

\end{enumerate}

For transitions to excited P-wave heavy quark systems, heavy quark 
symmetry
tells us that the $L-S$ coupled states $^3P_1$ and $^1P_1$ which are 
appropriate
to the light $I=1$ and $I=0$ sectors and to states of definite $C$ 
parity
should be replaced by the $j-j$ coupled states with
$s^{\pi_\ell}_\ell = {3 \over 2}^{+}$ and ${1\over 2}^+$.  If we define 
form
factors $\ell_{3\over 2}$,$c_{{+}{3\over 2}}$, $c_{{-}{3\over 2}}$, and
$q_{3\over 2}$ to be the exact analogs of $\ell$, $c_{+}$, $c_{-}$, and 
$q$ of
equations (B26) and (B27) of ISGW but for the $s^{\pi_\ell}_\ell = {3 
\over
2}^{+}$ state with $J^P=1^{+}$, and a parallel set $\ell_{1\over 2}$,
$c_{{+}{1\over 2}}$,$c_{{-}{1\over 2}}$, and $q_{1\over 2}$ for the
$s^{\pi_\ell}_\ell = {1\over 2}^{+}$ state with $J^P=1^{+}$, then

\begin {eqnarray}
\ell_{3\over 2}&=&-~{{2\tilde m_B\beta_B}\over 
{\sqrt{3}}}\left\{{1\over
m_q} \right.
\nonumber \\ && \nonumber \\ &&
\left. +{{\tilde m_Xm_d(\tilde w-1)}\over 2\beta^2_B}\left( {{\tilde 
w+1}\over
{2m_q}}-{{m_d\beta^2_B}\over {2\mu_{-}\tilde m_X\beta^2_{BX}}}\right) 
\right\}F^{(\ell_{3\over 2})}_5 \\
\cr
\cr
c_{{+}{3\over 2}}+c_{{-}{3\over 2}}&=&
-~{{\sqrt{3} m_d}\over {2\beta_B \tilde m_B}}
\left[1-\frac{m_d}{3m_q}-
{{m_d\beta^2_B}\over {3\beta^2_{BX}}}\left({1\over
2\mu_{-}}-{1\over \mu_+}\right) \right]F^{(c_{{+}{3\over 2}}+c_{{-}{3\over 2}})}_5
\\ \cr
\cr
c_{{+}{3\over 2}}-c_{{-}{3\over 2}}&=&-~{{m_d}
\over {2\sqrt{3}\beta_B \tilde m_X}}\left[
{({2-\tilde w})\tilde m_X\over m_q}+{{m_d\beta^2_B}\over
{\beta^2_{BX}}}\left({1\over 2\mu_{-}}-{1\over \mu_+}\right) 
\right]F^{(c_{{+}{3\over 2}}-c_{{-}{3\over 2}})}_5
\\ \cr
\cr
q_{3\over 2}&=&-~{1\over 2\sqrt{3}} \left\{ {{1+\tilde w}\over 2}+
{{\beta^2_B\tilde m_B}\over {2m_dm_qm_b}}\right\} {{m_d}\over
{\beta_b\tilde m_X}}F^{(q_{3\over 2})}_5
\end{eqnarray}
and
\begin{eqnarray} 
\ell_{1\over 2}&=&\sqrt{2\over 3}\tilde m_B\beta_B\left\{ 
{1\over
2m_q} \right.
\nonumber \\ && \nonumber \\ &&
 \left. -{3\over {2m_b}} 
+{{m_d\tilde m_X(\tilde w-1)}\over \beta^2_B} 
\left[
{1\over m_q}-{{m_d\beta^2_B}\over {2\mu_{-}\tilde 
m_X\beta^2_{BX}}}\right]
\right\}F^{(\ell_{1\over 2})}_5
\\ \cr
\cr
c_{{+}{1\over 2}}+c_{{-}{1\over 2}}&=&{{m_d^2\beta^2_{X}}
\over {\sqrt{6}\tilde m_B m_q\beta_B \beta^2_{BX}}}
F^{(c_{{+}{1\over 2}}+c_{{-}{1\over 2}})}_5 
 \\ \cr
\cr
c_{{+}{1\over 2}}-c_{{-}{1\over 2}}&=& -~\sqrt{\frac{2}{3}}
{{m_d}\over {\tilde m_X\beta_B}}\left[ 
1+ {{m_d \beta^2_X}\over {2m_q  
\beta^2_{BX}}} \right]F^{(c_{{+}{1\over 2}}-c_{{-}{1\over 2}})}_5\\ \cr
\cr
q_{1\over 2}&=&\sqrt{1\over 6}\left\{ 1- {{\beta^2_B\tilde m_B}\over
{4m_dm_qm_b}}\right\} {m_d\over {\beta_B\tilde m_X}}F^{(q_{1\over 2})}_5 ~~. 
\end {eqnarray}
We use these latter formulas for decays to all heavy-light final states,
including the kaons.  Of course, to the extent that the  
$s^{\pi_\ell}_\ell = {3
\over 2}^{+}$ and ${1\over 2}^{+}$ multiplets are degenerate there 
formulas will
give total rates to the two $1^{+}$ states that are identical to the 
$^3P_1$
and $^1P_1$ formulas.  However, the latter are needed to predict the 
rates to
individual states, e.g., the rate for $\bar{B} \rightarrow D^{ \left( 
3\over
2 \right)}_1  \ell \bar{\nu}_{\ell}$.
\eject
%%%%%%%%%%%%%%%%%%%%%%%%%%%%%%%%%%%%%%%%%%%%%% END TEXT %%%%%%%%%%%%%%%%%%%%%%%%%%%%%%%%%%%%%%%
%
%%%%%%%%%%%%%%%%%%%%%%%%%%%%%%%%%%%%%%%%%%%%%% TABLES  %%%%%%%%%%%%%%%%%%%%%%%%%%%%%%%%%%%%%%%%%
%%%%%%%%%%%%%%%%%%%%%%%%%%%%%%%%%%%%%%%%%%%%%% TABLES  %%%%%%%%%%%%%%%%%%%%%%%%%%%%%%%%%%%%%%%%%

\draft

\baselineskip=24pt

\begin{table}
 
{\bf Table I}. Relativistic corrections to the $f$ form factor
\begin{tabular}{| ccc |}
       
decay     &$C_f$ for 1S &$C_f$ for 2S \\ \hline
$D \rightarrow \rho, \omega$           &     0.889        &     0.740         \\ 
$D \rightarrow K^*$             &   0.928          &    0.782          \\
$D_s \rightarrow K^*$             &    0.873         &    0.739          \\ 
$D_s \rightarrow \phi$             &     0.911        &       0.773       \\
$\bar B \rightarrow \rho, \omega$             &    0.905         &    0.776          \\ 
$\bar B \rightarrow D^*$             &    0.989         &    0.929          \\
$\bar B_s \rightarrow K^*$             &  0.892           &  0.781            \\ 
$\bar B_s \rightarrow D_s^*$             &  0.984           &  0.924            \\
$\bar B_c \rightarrow \bar D^*$             &  0.868           &  0.779            \\ 
$\bar B_c \rightarrow \psi$             &  0.967           &  0.899           

\end{tabular}

\end{table}

\baselineskip=24pt

\begin{table}

  {\bf Table II}. Exclusive partial widths for the $b {\mbox{$\rightarrow$}} c$ semileptonic
  decays, $ {\mbox{$\bar B$}} {\mbox{$\rightarrow$}} X_{c \bar d} {\mbox{$e \bar \nu_e$}}$, ${\mbox{$\bar B_s$}} {\mbox{$\rightarrow$}} X_{c \bar s} {\mbox{$e \bar \nu_e$}}$ and
  $ \bar B_c \rightarrow  X_{c \bar c} e \bar \nu_e$, in units of $10^{13} |V_{bc}|^2 \sec^{-1}$.
     The Heavy Quark Symmetry notation $n^{s_\ell}L_J$ is used for the final states with unequal mass
quarks~$^{b)}$.
Also included are the physical meson masses used (in GeV), taken
  from Ref. [33] if possible; properties of unobserved or controversial
  states (given in parentheses) are taken from Ref. [35].
  The masses of the decaying particles (in GeV) are 5.28, $5.38^{d)}$, and
  (6.27).

\begin{tabular}{|c|cc|cc|cc|} 
\rule{0ex}{2.5ex} & \multicolumn{2}{c|}{$ {\mbox{$\bar B$}}_d {\mbox{$\rightarrow$}} X_{c \bar d} {\mbox{$e \bar \nu_e$}}$} &
                    \multicolumn{2}{c|}{$ {\mbox{$\bar B_s$}} {\mbox{$\rightarrow$}} X_{c \bar s} {\mbox{$e \bar \nu_e$}}$} &
                    \multicolumn{2}{c|}{$ \bar B_c \rightarrow  X_{c \bar c} e \bar \nu_e$} \\
$X$ & mass & partial width & mass & partial width & mass & partial width \\
\hline 
$1^{1\over 2}S_0$ &  1.87   & 1.42 &  1.97  & 1.31 &  2.98  & 0.99  \\
$1^{1\over 2}S_1$ &  2.01   & 2.81 &  2.11  & 2.49 &  3.10  & 1.57 \\
$1^{3\over 2}P_2$ &  2.46   & 0.10 &  2.57$^{c)}$  & 0.14 & 3.56 & 0.12  \\
$1^{3\over 2}P_1$ &  2.42   & 0.20 & (2.54) & 0.25 &  3.52$^{b)}$  & 0.21  \\
$1^{1\over 2}P_1$ & (2.49)  & 0.04 & (2.57) & 0.05 &  3.51$^{b)}$  & 0.08 \\
$1^{1\over 2}P_0$ & (2.40)  & 0.03 & (2.48) & 0.04 &  3.42  & 0.04 \\
$2^{1\over 2}S_0$ & (2.58)  & 0.00 & (2.67) & 0.01 & (3.62) & 0.07 \\
$2^{1\over 2}S_1$ & (2.64)  & 0.06 & (2.73) & 0.13 &  3.69  & 0.29 \\
\hline 
total      &         & 4.66 &        &4.41 &        & 3.36 

\end{tabular}

\end{table}

\baselineskip=12pt
\noindent $~^{a)}$ The $\Gamma_L/\Gamma_T$ values for these decays are 1.08, 1.03, and 0.88, respectively.

\noindent $~^{b)}$ We list $\bar B_c \rightarrow 1^1P_1$ ({\it i.e.}, $\chi_{c1}$)
under $1^{3 \over 2}P_1$ and $\bar B_c \rightarrow 1^3P_1$ ({\it i.e.}, $h_{c1}$)
under $1^{1 \over 2}P_1$. 

\noindent $~^{c)}$ See Ref. [36].

\noindent $~^{d)}$ See Ref. [37].

\eject

\baselineskip=24pt

\begin{table}
  {\bf Table III}. Exclusive partial widths for the $c {\mbox{$\rightarrow$}} s$ semileptonic
  decays, $ {\mbox{$D$}}  {\mbox{$\rightarrow$}} X_{s \bar u} {\mbox{$e^+ \nu_e$}}$, ${\mbox{$D_s$}}  {\mbox{$\rightarrow$}} X_{s \bar s} {\mbox{$e^+ \nu_e$}}$ and
  ${\mbox{$B_c$}} {\mbox{$\rightarrow$}} X_{s \bar b} {\mbox{$e^+ \nu_e$}}$, in units of $10^{10} |V_{cs}|^2 \sec^{-1}$.
     The Heavy Quark Symmetry notation $n^{s_\ell}L_J$ is used for the final states with unequal mass
quarks~$^{c)}$.
Also included are the physical meson masses used (in GeV), taken
  from Ref.[33] if possible; properties of unobserved or controversial
  states (given in parentheses) are taken from Ref.[35].
  The masses of the decaying particles (in GeV) are 1.87, 1.97, and (6.27),
  respectively.

\begin{tabular}{|c|cc|cc|cc|}  
\rule{0ex}{2.5ex}
      & \multicolumn{2}{c|}{$ {\mbox{$D$}}  {\mbox{$\rightarrow$}} X_{s \bar u} {\mbox{$e^+ \nu_e$}}$} &
        \multicolumn{2}{c|}{$ {\mbox{$D_s$}}  {\mbox{$\rightarrow$}} X_{s \bar s} {\mbox{$e^+ \nu_e$}}$} &
        \multicolumn{2}{c|}{$ {\mbox{$B_c$}}  {\mbox{$\rightarrow$}} X_{s \bar b} {\mbox{$e^+ \nu_e$}}$} \\
$X$ & mass  & partial width & mass & partial width & mass & partial width  \\
\hline 
$1^{1\over 2}S_0$ &  0.49   & 10.5 &  0.55$^{b)}$  & 3.7 &  5.38$^{d)}$ & 2.2 \\
         &         &       &  0.96$^{b)}$  & 3.2 &        &       \\
$1^{1\over 2}S_1$$^{a)}$ &  0.89   & 5.7 &  1.02  &4.8 & (5.45) & 2.7 \\
$1^{3\over 2}P_2$ &  1.43   & 0.00 & 1.53 & 0.00 & (5.88) & 0.00 \\
$1^{3\over 2}P_1$ &  1.27   & 0.34 &  1.38$^{c)}$  & 0.27 & (5.88) & 0.06 \\
$1^{1\over 2}P_1$ & 1.40   & 0.00 &  1.51$^{c)}$  & 0.03 & (5.88) & 0.00\\
$1^{1\over 2}P_0$ &  1.43   & 0.00 &  1.52  & 0.00 & (5.88) & 0.00 \\
$2^{1\over 2}S_0$ & (1.45)  & 0.00 & (1.63) & 0.00 & (5.98) & 0.01 \\
$2^{1\over 2}S_1$ & (1.58)  & 0.00 & (1.69) & 0.01 & (6.01) & 0.01 \\
\hline
total      &         & 16.6 &        & 12.1 &        & 5.0 

\end{tabular}

\end{table}

\baselineskip=12pt

\noindent $~^{a)}$ The $\Gamma_L/\Gamma_T$ values for these decays 
are 0.94, 0.96, and 1.03, respectively.

\noindent $~^{b)}$ We use the approximation of ideal mixing in $I=0$ states in every
   sector except the ground state pseudoscalars where we assume an $\eta-\eta'$
   mixing angle of $-20^{\circ}$. If this mixing angle were changed to 
   $-10^{\circ}$, then the entries in the Table to $\eta$ and $\eta'$ would
   change to 5.6 and 2.4; note that while the individual rates change
   substantially, the total rate to these two states would only increase by about 
   $16\%$.

\noindent $~^{c)}$ For  $D_s \rightarrow X_{s \bar s}$ we list the
rate to the $1^1P_1$ 
under $1^{3 \over 2}P_1$ and that for the $1^3P_1$ state
under $1^{1 \over 2}P_1$. 

\noindent $~^{d)}$ See Ref. [37].

\eject

\baselineskip=24pt

\begin{table}
  {\bf Table IV}. Exclusive partial widths for the $c {\mbox{$\rightarrow$}} d$ semileptonic
  decays ${\mbox{$D$}}^0 {\mbox{$\rightarrow$}} X_{d \bar u} {\mbox{$e^+ \nu_e$}}$ and
  ${\mbox{$D_s$}} {\mbox{$\rightarrow$}} X_{d \bar s} {\mbox{$e^+ \nu_e$}}$, in units of $10^{10} |V_{cd}|^2 \sec^{-1}$.
  Also included are the physical meson masses used (in GeV), taken
  from Ref. [33] if possible; properties of unobserved or controversial
  states (given in parentheses) are taken from Ref. [35].
  The masses of the decaying particles (in GeV) are 1.87 and 1.97, respectively.

\begin{tabular}{|c|cc|cc|}   
      & \multicolumn{2}{c|}{$ {\mbox{$D$}}^0 {\mbox{$\rightarrow$}} X_{d \bar u} {\mbox{$e^+ \nu_e$}}$} &
        \multicolumn{2}{c|}{$ {\mbox{$D_s$}}^+ {\mbox{$\rightarrow$}} X_{d \bar s} {\mbox{$e^+ \nu_e$}}$} \\

$X$ & mass  & partial width &  mass & partial width  \\
\hline
$1^1S_0$ &  0.14   &9.8 &  0.50  & 8.9 \\
$1^3S_1$$~^{a)}$ &  0.77   & 4.9 &  0.89  & 4.4 \\
$1^3P_2$ &  1.32   & 0.01 &  1.43  & 0.01 \\
$1^1P_1$ &  1.23   & 0.52 &  1.27$~^{b)}$  & 1.5 \\
$1^3P_1$ &  1.26   & 0.32 &  1.40$~^{b)}$  & 0.01 \\
$1^3P_0$ &  1.30   & 0.00 &  1.43  & 0.00 \\
$2^1S_0$ &  1.30   & 0.02 & (1.45) & 0.04 \\
$2^3S_1$ & (1.45)  & 0.03 & (1.58) & 0.03 \\
\hline
total      &         & 15.6 &        & 14.9 
\end{tabular}

\end{table}

\baselineskip=12pt

\noindent $~^{a)}$ The $\Gamma_L/\Gamma_T$ values for these decays 
are 0.67  and 0.76, respectively.

\noindent $~^{b)}$We list the
rate to the mainly $1^{3 \over 2}P_1$ state under $1^1P_1$ 
and that for the mainly  $1^{1 \over 2}P_1$ under $1^3P_1$ state.

\eject

\baselineskip=24pt

\begin{table}
  {\bf Table V}. Exclusive partial widths for
  the $c {\mbox{$\rightarrow$}} d$ semileptonic decay ${\mbox{$D$}}^+ {\mbox{$\rightarrow$}} X_{d \bar d} {\mbox{$e^+ \nu_e$}}$
  in units of $10^{10} |V_{cd}|^2 \sec^{-1}$, separated into $I=1$ and $I=0$
  contributions.
  Also included are the physical meson masses used (in GeV), taken
  from Ref. [33] if possible; properties of unobserved or controversial
  states (given in parentheses) are taken from Ref. [35].
  The mass of the decaying particle is 1.87 GeV.

\begin{tabular}{|c|cc|cc|}   \rule{0ex}{2.5ex}
      & \multicolumn{2}{c|}{$ {\mbox{$D$}}^+ {\mbox{$\rightarrow$}} X_{d \bar d} {\mbox{$e^+ \nu_e$}}$} &
        \multicolumn{2}{c|}{$ {\mbox{$D$}}^+ {\mbox{$\rightarrow$}} X_{d \bar d} {\mbox{$e^+ \nu_e$}}$} \\ 
      & \multicolumn{2}{c|}{$I=1$} & \multicolumn{2}{c|}{$I=0$} \\
$X$ & mass & partial width & mass & partial width  \\
\hline
$1^1S_0$ &  0.14   & 4.9 &  0.55$~^{b)}$  &3.0  \\
         &         &       &  0.96$~^{b)}$  & 0.6  \\
$1^3S_1$ $~^{a)}$ &  0.77   & 2.5 &  0.78  & 2.4  \\
$1^3P_2$ &  1.32   & 0.00 &  1.27  & 0.00 \\
$1^1P_1$ &  1.23   & 0.26 &  1.17  & 0.39  \\ 
$1^3P_1$ &  1.26   & 0.16 &  1.28  & 0.13 \\
$1^3P_0$ &  1.30   & 0.00 &  1.30  & 0.00 \\
$2^1S_0$ &  1.30   & 0.01 & (1.44) & 0.00 \\
$2^3S_1$ & (1.45)  & 0.02 & (1.46) & 0.01 \\
\hline
total      &         & 7.8 &        & 6.5 
\end{tabular}

\end{table}
\baselineskip=12pt

\noindent $~^{a)}$ The $\Gamma_L/\Gamma_T$ values for these decays 
are 0.67  and 0.68, respectively.

\noindent $~^{b)}$ We use the approximation of ideal mixing in $I=0$ states in every
   sector except the ground state pseudoscalars where we assume an $\eta-\eta'$
   mixing angle of $-20^{\circ}$. If this mixing angle were changed to 
   $-10^{\circ}$, then the entries in the Table to $\eta$ and $\eta'$ would
   change to 2.2 and 0.9; note that while the individual rates change
   substantially, the total rate to these two states would only decrease by about 
   $14\%$.

\eject

\baselineskip=24pt

\begin{table}
  {\bf Table VI}. Exclusive partial widths for
  the $b {\mbox{$\rightarrow$}} u$ semileptonic decays with a light spectator,
  $ {\mbox{$\bar B$}}^0 {\mbox{$\rightarrow$}} X_{u \bar d} {\mbox{$e \bar \nu_e$}}$, and ${\mbox{$B$}}^- {\mbox{$\rightarrow$}} X_{u \bar u} {\mbox{$e \bar \nu_e$}}$,
  in units of $10^{13} |V_{bu}|^2 \sec^{-1}$.  
  Also included are the physical meson masses used (in GeV), taken
  from Ref. [33] if possible; properties of unobserved or controversial
  states (given in parentheses) are taken from Ref. [35]. (The masses for the $I=1$
final states in $B^- {\mbox{$\rightarrow$}}
  X_{u\bar u} {\mbox{$e \bar \nu_e$}}$ are the same as those for ${\mbox{$\bar B$}}^0 {\mbox{$\rightarrow$}} X_{u \bar d} {\mbox{$e \bar \nu_e$}}$.)
  The mass of the decaying particle is 5.28 GeV.

\begin{tabular}{|c|cc|cccc|}  
\rule{0ex}{2.5ex} & \multicolumn{2}{c|}{$ {\mbox{$\bar B$}}  {\mbox{$\rightarrow$}} X_{u \bar d} {\mbox{$e \bar \nu_e$}}$} &
                    \multicolumn{4}{c|}{$ {\mbox{$B$}}^- {\mbox{$\rightarrow$}} X_{u \bar u} {\mbox{$e \bar \nu_e$}}$} \\
            &  & &  \multicolumn{2}{c}{$I = 1$} & \multicolumn{2}{c|}{$I = 0$}  \\
$X$ & mass  & partial width &mass & partial width & mass & partial width  \\
\hline
$1^1S_0$ &  0.14   & 0.96  &0.14& 0.48 & 0.55$~^{b)}$  & 0.45  \\
         &         &       &&       & 0.96$~^{b)}$  & 0.28  \\
$1^3S_1$$~^{a)}$ &  0.77   & 1.42 &0.77& 0.71 & 0.78  & 0.71  \\
$1^3P_2$ &  1.32   & 0.33 &1.32& 0.16 &  1.27  & 0.18  \\
$1^1P_1$ &  1.23   & 1.09 &1.23& 0.54 &  1.17  & 0.57  \\
$1^3P_1$ &  1.26   & 0.87 &1.26& 0.43 &  1.28  & 0.41 \\
$1^3P_0$ &  1.30   & 0.05 &1.30& 0.02 &  1.30  & 0.03  \\
$2^1S_0$ &  1.30   & 0.17 &1.30& 0.08 & (1.44) & 0.08  \\
$2^3S_1$ & (1.45)  & 0.41 &(1.45)& 0.20 & (1.46) & 0.20  \\
\hline
partial total &      & 5.3 && 2.6 &       & 2.9  
\end{tabular}
\end{table}

\baselineskip=12pt

\noindent $~^{a)}$ The $\Gamma_L/\Gamma_T$ values for these decays 
are 0.30, 0.30, and 0.30 respectively.

\noindent $~^{b)}$ We use the approximation of ideal mixing in $I=0$ states in every
   sector except the ground state pseudoscalars where we assume an $\eta-\eta'$
   mixing angle of $-20^{\circ}$. If this mixing angle were changed to 
   $-10^{\circ}$, then the entries in the Table to $\eta$ and $\eta'$ would
   change to 0.34 and 0.41; note that while the individual rates change
   substantially, the total rate to these two states would only increase by about 
   $3\%$.

\eject

\baselineskip=24pt

\begin{table}
  {\bf Table VII}. Exclusive partial widths for
  the $b {\mbox{$\rightarrow$}} u$ semileptonic decays with a heavy spectator,
  $ {\mbox{$\bar B_s$}} {\mbox{$\rightarrow$}} X_{u \bar s} {\mbox{$e \bar \nu_e$}}$, and $\bar B_c \rightarrow X_{u \bar c} e \bar \nu_e$,
  in units of $10^{13} |V_{bu}|^2 \sec^{-1}$.   The Heavy Quark Symmetry notation $n^{s_\ell}L_J$ is used for the final states with unequal mass
quarks.
 Also included are the physical meson masses used (in GeV), taken
  from Ref. [33] if possible; properties of unobserved or controversial
  states (given in parentheses) are taken from Ref. [35].
   The masses of the decaying particles (in GeV) are  5.38$^{b)}$ and (6.27), respectively..

\begin{tabular}{|c|cc|cc|}  
\rule{0ex}{2.5ex} & \multicolumn{2}{c|}{$ {\mbox{$\bar B_s$}} {\mbox{$\rightarrow$}} X_{u \bar s} {\mbox{$e \bar \nu_e$}}$} &
                    \multicolumn{2}{c|}{$\bar B_c \rightarrow X_{u \bar c} e \bar \nu_e$} \\

$X$ & mass  & partial width & mass & partial width  \\
\hline
$1^{1\over 2}S_0$    &  0.49   & 0.85 &  1.87  & 0.30 \\
$1^{1\over 2}S_1$$~^{a)}$    &  0.89   & 1.14 &  2.01  & 0.62 \\
$1^{3\over 2}P_2$    &  1.43   & 0.28 &  2.46  & 0.06 \\
$1^{3\over 2}P_1$    &  1.27   & 1.72 &  2.42  & 0.62 \\
$1^{1\over 2}P_1$    &  1.40   & 0.08 & (2.49) & 0.04 \\
$1^{1\over 2}P_0$    &  1.43   & 0.04 & (2.40) & 0.01 \\
$2^{1\over 2}S_0$    & (1.45)  & 0.45 & (2.58) & 0.46 \\
$2^{1\over 2}S_1$    & (1.58)  & 0.54 & (2.64) & 0.40 \\
\hline
partial total &         & 5.1 &        & 2.5 
\end{tabular}
\end{table}

\baselineskip=12pt

\noindent $~^{a)}$ The $\Gamma_L/\Gamma_T$ values for these decays 
are 0.45  and 0.61, respectively.

\noindent $~^{b)}$see Ref. [37]

\eject

\baselineskip=24pt

\begin{table}
  {\bf Table VIII}. Comparison at zero recoil of the ISGW2 meson form factors 
without perturbative matching corrections
to those of leading order Heavy Quark Symmetry and Heavy Quark Symmetry including the
  $\cal O$$({1 / {m_Q}})$ corrections predicted by ISGW2. 
  Note that these ISGW2 form factors {\it cannot be
directly compared to experiment}: they are just the $\tilde f_{\alpha}^{qm}$
scaled by factors of $( {\bar m_B}/{m_B})^{n_B(\alpha)}( {\bar m_X}/{m_X})^{n_X(\alpha)}$
with the exponents of Table C1. The form factors are those for mesons corresponding to
  a light ($\bar u$ or $\bar d$)
  spectator.

\begin{tabular}{|lc|cccccc|}  \rule{0ex}{2.7ex}
    && $\tilde{f}_+$ & $\tilde{f}_-$ & $\tilde{g}$ & $\tilde{f}$ &
     $(\tilde{a}_+ + \tilde{a}_-) $ & $(\tilde{a}_+ - \tilde{a}_-) $ \\ 
\hline
\rule{0ex}{2.5ex}
ISGW2-no matching & $b {\mbox{$\rightarrow$}} c {\mbox{$e \bar \nu_e$}}$ & 1.00 & -0.03 & 1.11 & 0.97 & -0.08 & 1.04 \\
ISGW2-no matching & $c {\mbox{$\rightarrow$}} s {\mbox{$e^+ \nu_e$}}$ & 0.98 & -0.00 & 1.28 & 0.84 & -0.23 & 1.18 \\
\hline  
\rule{0ex}{2.5ex}
HQS     & $b {\mbox{$\rightarrow$}} c {\mbox{$e \bar \nu_e$}}$ & 1 & 0 & 1 & 1  & 0  & 1  \\
HQS     & $c {\mbox{$\rightarrow$}} s {\mbox{$e \bar \nu_e$}}$ & 1 & 0 & 1 & 1 & 0 & 1 \\
\hline
\rule{0ex}{2.5ex}
HQS+$\cal O$$({1 \over {m_Q}})$ & $b {\mbox{$\rightarrow$}} c {\mbox{$e \bar \nu_e$}}$ & 1    & -0.06& 1.12 & 1 & -0.09  &1.03 \\
HQS+$\cal O$$({1 \over {m_Q}})$ & $c {\mbox{$\rightarrow$}} s {\mbox{$e^+ \nu_e$}}$ & 1    &-0.21& 1.39 & 1 & -0.30  &1.09 
\end{tabular}
\end{table}

\eject

\baselineskip=24pt

\begin{table}
 {\bf Table IX}.  Predictions for the six form factors for 
$\bar B \rightarrow D \ell \bar \nu_{\ell}$ and $\bar B \rightarrow D^* \ell \bar \nu_{\ell}$.
The HQET column shows the effect of QCD radiative corrections {\it alone} to the HQS symmetry 
limit column.

 \begin{tabular}{|c|cccc|} 
         &ISGW2 & ISGW & HQS & HQET \\ \hline
   $\tilde f_+(t_m)$   & 1.00 & 1.01  & 1   &   1.00 \\
   $\tilde f_-(t_m)$   & -0.09  &  -0.05   &  0  & -0.02  \\
   $\tilde g(t_m)$ & 1.17 & 1.12 & 1 & 1.06 \\
   $\tilde f(t_m) $  & 0.91 & 1.00 & 1 & 0.93 \\
   $2\tilde a_+(t_m) $       & 0.83 &   0.95               & 1 &0.88 \\
   $2\tilde a_-(t_m) $     & -1.19 &    -1.11              & -1 & -1.08
\end{tabular}
\end{table}

\eject

\baselineskip=24pt

\begin{table}
 {\bf Table X}.  Comparison of the form factors for ${\mbox{$D$}} {\mbox{$\rightarrow$}} \bar K{\mbox{$e^+ \nu_e$}}$ 
and ${\mbox{$D$}} {\mbox{$\rightarrow$}} \bar K^*{\mbox{$e^+ \nu_e$}}$  with experiment.  We
 have used the $t$ dependence assumed in the fits to data
to extrapolate the experimental form factors to $t = t_m$
 from $t = 0$.

 \begin{tabular}{|c|ccc|} 
         & experiment [23] & ISGW2 & ISGW \\ \hline
   $f_+(t_m)$     & $1.42 \pm 0.25$  & 1.23 & 1.16 \\      
   $f(t_m) $ (GeV)& $2.21 \pm 0.19$ & 1.92 & 2.76 \\
   $g(t_m) $ (GeV$^{-1}$) & $0.55 \pm 0.08$ & 0.55 & 0.47 \\  
   $a_+(t_m) $ (GeV$^{-1}$)    & $-0.21 \pm 0.04$            & -0.34 & -0.37 
\end{tabular}
\end{table}

\eject

\baselineskip=24pt

\begin{table}
 
{\bf Table A1}. quark model parameters
\begin{tabular}{| ccc |}
       
parameter    &ISGW2&ISGW\\ \hline
$b$           &     0.18 GeV$^2$        &     0.18 GeV$^2$          \\ 
$c$             &   -0.81 GeV          &    -0.84 GeV          \\
$\alpha_s$      &$0.60 \rightarrow 0.30$  (see text)    & $0.50 \rightarrow 0.30$ (see ISGW)      \\
$m_u=m_d$             &    0.33 GeV       &    0.33 GeV          \\ 
$m_s$             &     0.55 GeV        &       0.55 GeV       \\
$m_c$             &   1.82 GeV         &    1.82 GeV          \\ 
$m_b$             &    5.20 GeV         &    5.12 GeV         \\
$a$             &  2.8          & not applicable           
\end{tabular}

\end{table}

\eject
\baselineskip=24pt

\begin{table}
 
{\bf Table A2}. the masses and $\beta$ values in GeV for variational solutions of
the hyperfine-corrected Coulomb plus linear problem in the $1S$, $1P$, $2S$ basis
\begin{tabular}{| c c c c c c c c c c c c|}
        
          &meson flavor: & $u\bar d$ &    $u\bar s$   
                      &   $s\bar s$ &  $c\bar u$ &  $c\bar s$ 
                      &  $u\bar b$ &  $s\bar b$  & $c\bar c$            &$b \bar c$&\\ 
\hline
          &     &      &      &      &      &      &      &      &      &    &\\ 
          & $m$ & 0.35 & 0.55 & 0.62 & 1.86 & 1.94 & 5.27 & 5.33 & 2.95 & 6.33   &\\
$1^1S_0$  &     &      &      &      &      &      &      &      &      &    &\\ 
      & $\beta$ & 0.41 & 0.44 & 0.53 & 0.45 & 0.56 & 0.43 & 0.54 & 0.88 & 0.92   &\\
          &     &      &      &      &      &      &      &      &      &    &\\
\hline 
          &     &      &      &      &      &      &      &      &      &    &\\
          & $m$ & 0.74 & 0.87 & 0.97 & 2.01 & 2.10 & 5.33 & 5.40 & 3.13 & 6.42   &\\
$1^3S_1$  &     &      &      &      &      &      &      &      &      &    &\\ 
      & $\beta$ & 0.30 & 0.33 & 0.37 & 0.38 & 0.44 & 0.40 & 0.49 & 0.62 & 0.75   &\\
          &     &      &      &      &      &      &      &      &      &    &\\
\hline 
          &     &      &      &      &      &      &      &      &      &    &\\
          & $m$ & 1.24 & 1.35 & 1.42 & 2.48 & 2.53 & 5.81 & 5.84 & 3.51 & 6.79   &\\
$1^1P  $  &     &      &      &      &      &      &      &      &      &    &\\ 
      & $\beta$ & 0.28 & 0.30 & 0.33 & 0.33 & 0.38 & 0.35 & 0.41 & 0.52 & 0.60   &\\
          &     &      &      &      &      &      &      &      &      &    &

\end{tabular}

\end{table}
\eject
%%%%%%%%%%%%%%%%%%%%%%%%%%%%%%%%%%%%Table C1 %%%%%%%%%%%%%%%%%%%%%%%%%%%%%%%%%%%%%%%%%%%%%%%%%%%%
%%%%%%%%%%%%%%%%%%%%%%%%%%%%%%%%%%%%Table C1 %%%%%%%%%%%%%%%%%%%%%%%%%%%%%%%%%%%%%%%%%%%%%%%%%%%%
Table C1:  The factors $F_n$ of eq. $(B1)$ of ISGW are to be replaced by
$F_N^{(\alpha)}$ which have the modification shown in eq. (27) of Section III.C
and are multiplied by the $\alpha$-dependent factor $\left( {\bar m_B} \over
{ \tilde m_B} \right)^{n_B ( \alpha )} \left( { \bar m_X} \over {\tilde
m_X} \right)^{n_X( \alpha )}$ with $n_B ( \alpha)$ and $n_X( \alpha )$ given
here.
\begin{table}
[tbhp]
  \protect\caption[Table C1:  The factors $F_n$ of eq. (B1) ISGW are to be replaced by 
$F^{\alpha}_n$ which have the modification shown in eq. (27) of Section III.C and
are multiplied by the $\alpha$-dependent factors $\left(\overline 
m_B\over\tilde_B\right)^{n_B)\alpha}$]{
  \label{tbl:bcpw} }
\begin{center}
\begin{tabular}{|c|c|c|}  \hline \hline
{form factors}&&\\
%\rule{0ex}{2.5ex} 
{($\alpha$)}& {$n_{B}(\alpha)$} &{$n_x(\alpha)$}\\
\hline
$f_+ + f_-, f^{\prime}_+ + f^{\prime}_-$ & $-{1/2}$ & $+{1/2}$\\
$f_+-f_-, f_+^\prime -f_-^\prime$& $+{1/2}$ & $-{1/2}$\\
$g,g^\prime$ & $-{1/2}$ & $-{1/2}$\\
$f, f^\prime$ &  $+{1/2}$  & $ +{1/2}$\\
$a_++a_-, a_+^\prime+a_-^\prime$  & $ -{3/2}$  & $ + {1/2}$\\
$a_+-a_-, a_+^\prime-a_-^\prime$ & $-1/2$ & $-1/2$\\
$h$  & $-3/2$  &  $-1/2$\\
$k$  &  $-1/2$  & $ +1/2$\\
$b_++b_-$  &  $-5/2$  &  $+1/2$\\
$b_+-b_-$  &  $-3/2$  &  $-1/2$\\
$\ell,r,\ell_{3\over2},\ell_{1\over 2}$  &  $+1/2$  &  $+1/2$\\
$c_++c_-,s_++s_-,c_{+{3\over2}}+c_{-{3\over2}},c_{+{1\over2}}+c_{-{1\over2}}$&$-3/2$&$+1/2$\\
$c_+-c_-,s_+-s_-,c_{+{3\over2}}-c_{-{3\over2}},c_{+{1\over2}}-c_{-{1\over2}}$&$-1/2$&$-1/2$\\
$q,v,q_{3 \over 2},q_{1 \over 2}$ & $-1/2$ & $-1/2$\\
$u_++u_-$&$-1/2$&$+1/2$\\
$u_+-u_-$&$+1/2$&$-1/2$\\
\hline
   &       &       \\
\hline \hline
\end{tabular}
\end{center}
\end{table}
\eject
%End Table C1
%%%%%%%%%%%%%%%%%%%%%%%%%%%%%%%%%%%%%%%%%%%%%%%%%%End of Table C1 %%%%%%%%%%%%%%%%%%%%%%%%%%%%%%%%%%
\end{document}